\documentclass[12pt,preprint]{aastex}
 
\shorttitle{Circumstellar Gas and Dust in Nearby Edge-On Disks}
\shortauthors{Redfield et al.}

\begin{document}
 
\newcommand{\php}[0]{\phantom{--}}
\newcommand{\kms}[0]{km~s$^{-1}$}
 
\title{{\it Spitzer} Limits On Dust Emission and Optical Gas
Absorption Variability Around Nearby Stars with Edge-On Circumstellar
Disk Signatures}
 
\author{Seth Redfield\altaffilmark{1}, Jacqueline E. Kessler-Silacci\altaffilmark{2}, and Lucas A. Cieza} 
\affil{Department of Astronomy and McDonald Observatory, University of Texas, Austin, TX, 78712}
\email{sredfield@astro.as.utexas.edu}
\altaffiltext{1}{Hubble Fellow.}
\altaffiltext{2}{Spitzer Fellow.}
\setcounter{footnote}{2}

\begin{abstract}

We present {\it Spitzer Space Telescope} infrared spectroscopic and
photometric observations and McDonald Observatory Smith Telescope and
Anglo-Australian Telescope high spectral resolution optical
observations of 4 nearby stars with variable or anomalous optical
absorption, likely caused by circumstellar material.  The optical
observations of \ion{Ca}{2} and \ion{Na}{1} cover a 2.8 year baseline,
and extend the long term monitoring of these systems by previous
researchers.  In addition, mini-surveys of the local interstellar
medium (LISM) around our primary targets provide a reconstruction of
the intervening LISM along the line of sight.  We confirm that the
anomalous absorption detected toward $\alpha$ Oph is not due to
circumstellar material, but to a small filamentary cloud $<$14.3\,pc
from the Sun.  The three other primary targets, $\beta$ Car, HD85905,
and HR10 show both short and long term variability, and little of the
observed absorption can be attributed to the LISM along the line of
sight.  The {\it Spitzer Space Telescope} photometry and spectroscopy
did not detect infrared excesses.  We fit the maximum hypothetical
infrared excess that is consistent with observed upper limits.  We are
able to place upper limits on any possible fractional infrared
luminosity, which range from $L_{IR}/L_{\star} < 2$--$5 \times
10^{-6}$, for our three disk stars.  These fractional luminosities are
significantly less than that found toward $\beta$ Pic, but comparable
to other nearby debris disks.  No stable gas absorption component
centered at the radial velocity of the star is detected for any of our
targets, consistent with no infrared excess detections.  Based on
simple assumptions of the variable gas absorption component, we
estimate limits on the circumstellar gas mass causing the variable
absorption, which range from 0.4--20 $\times 10^{-8} M_{\oplus}$.
These multiwavelength observations place strong limits on any possible
circumstellar dust, while confirming variable circumstellar gas
absorption, and therefore are interesting targets to explore the
origins and evolution of variable circumstellar gas.

\end{abstract}
 

\section{Introduction}

The presence of circumstellar gas around mature stars presents an
interesting diagnostic of the formation and evolution of stars and
their immediate environments.  A collection of ``shell'' stars,
including stars that exhibit strong and sharp absorption features
(e.g., in \ion{Ca}{2}) and sometimes hydrogen emission lines (i.e., Be
stars), include many examples of stars in which gas from the stellar
atmosphere is deposited in the circumstellar environment via winds
\citep{slettebak88,porter03}.  \citet{slettebak75} identified $\beta$
Pic as an object that displayed strong and sharp absorption in
\ion{Ca}{2} that they postulated might be circumstellar.  Indeed,
observations with the {\it Infrared Astronomical Satellite} ({\it
IRAS}) led to the discovery of dust disks around $\beta$~Pic ($L_{\rm
dust}/L_{\star} \sim 3 \times 10^{-3}$; \citealt{backman93}), and
other nearby stars, including Vega, $\epsilon$~Eri, and $\alpha$~PsA
\citep{aumann85}.  The gas and dust in the circumstellar environment
of $\beta$~Pic have been shown to be distributed in an edge-on disk by
\citet{brandeker04} and \citet{smith84}, respectively.  Although
$\beta$ Pic shares some diagnostic characteristics of ``shell'' stars,
it has become clear that its observed gas and dust are the secondary
products left from the final stages of stellar formation, which has
resulted in a debris disk.  Therefore, two possible mechanisms exist
for depositing substantial amounts of circumstellar gas, one is
associated with debris disks and stellar formation and the other with
shell stars and stellar winds.

In the process of studying the properties of the circumstellar gas in
the dusty edge-on disk surrounding $\beta$ Pic, high resolution
optical (\ion{Ca}{2} and \ion{Na}{1}) spectra showed night-to-night
absorption line variability, evidence of a variable circumstellar gas
component located close to the star \citep{hobbs85, vidalmadjar86}.
Of the four major dust disks discovered by {\it IRAS}, (i.e.,
$\beta$~Pic, Vega, $\alpha$~PsA, and $\epsilon$~Eri), only $\beta$~Pic
shows \ion{Ca}{2} absorption line variability.  $\beta$~Pic is also
the only one to have an edge-on orientation, which allows for a
detectable optical depth along the line of sight through the midplane
of the circumstellar disk, and may explain why it is the only one of
the four major dust disks to show gas phase absorption.  Indeed, if
the three-dimensional density profile of \ion{Na}{1} in the disk
around $\beta$~Pic, derived by \citet{brandeker04}, is observed from
inclinations consistent with the other three major dust disks, the
resulting observable \ion{Na}{1} column density falls well below the
threshold of detectability.

The formation of stars and planets appears to necessarily include the
construction (and subsequent dispersal) of disks comprised of gas and
dust.  Therefore, it is likely that every star has experienced a
transitory phase in which the secondary gas and dust products of
stellar and planetary formation lead to a debris disk.  Understanding
the process of disk formation, evolution, and dissipation is critical
to placing known stellar and planetary systems, including our own,
into context.  Observations with the {\it Spitzer Space Telescope}
({\it Spitzer}) are finding much fainter debris disks than observed
toward $\beta$ Pic, around much older stars.  {\it Spitzer} has
detected dusty material in stars similar in age to the Sun, with
$L_{\rm dust}/L_{\star} \sim 3 \times 10^{-6}$
\citep[e.g.,][]{chen05,silverstone06,beichman06,chen06}, whereas our
solar system zodiacal dust emits $\sim$10$^{-7} L_{\odot}$.  In
particular, \citet{su06} find that such disks are quite common, with
$\sim$1/3rd of a sample of $\sim$160 A stars showing infrared (IR)
excesses due to a debris disk.

If we use $\beta$ Pic as a prototypical debris disk, the structure of
dust and gas in the circumstellar disk can be characterized by three
distinct components.  (1) The large-scale bulk dust disk, which causes
an IR excess \citep[e.g.,][]{aumann85} and scattered light
emission \citep[e.g.,][]{smith84}.  (2) The large-scale bulk gas disk,
which causes a stable gas absorption feature \citep{brandeker04}.  (3)
The variable gas component of the disk, which is located very close to
the star and causes gas absorption line variability over short (e.g.,
night-to-night) timescales \citep[e.g.,][]{petterson99}.  For reviews
of the various properties of all the components of the $\beta$ Pic
debris disk, see \citet{zuckerman01}, \citet*{lagrange00}, and
\citet*{vidalmadjar98}.

The source of the variable gas absorption component of the $\beta$ Pic
disk has received particular attention.  Long \ion{Ca}{2} monitoring
campaigns of $\beta$~Pic \citep[e.g.,][]{petterson99}, support the
theory that the short-term spectral variability is due to gas clouds
caused by evaporating, star-grazing, kilometer-sized, cometary-like
bodies, simply referred to as, Falling Evaporating Bodies
\citep[FEB's;][]{thebault01, beust94}.  The dynamics and frequency of
these events, potentially originating from a mean-motion resonance
with a giant planet, can constrain the structure of the disk and even
the geometry of a young planetary system \citep{beust00}.

No detailed study, comparable to the work on $\beta$ Pic, on the
variability of gas absorption in an edge-on disk toward a shell or Be
star has been made to date.  Despite the fact that gas may be
deposited irregularly in the circumstellar environment of rapidly
rotating, early type stars via weak stellar winds, like a scaled down
version of the disks that form around Be stars.  \citet*{abt97}
describe three epochs of observations toward shell stars taken over a
20 year baseline, and note the long term variations in the strength of
the gas absorption.  In order to understand the origins and structure
of circumstellar gas around mature stars it is critical to increase
the sample of known edge-on circumstellar gas absorption systems and
measure as comprehensively as possible the properties of gas and any
dust in the circumstellar environment.

Researchers have tried to find other circumstellar gas disks or
$\beta$~Pic-like systems (i.e., edge-on debris disks) through IR
excesses from {\it IRAS} \citep{cheng92} and \ion{Ca}{2} variability
(\citeauthor{lagrangehenri90} \citeyear{lagrangehenri90};
\citeauthor*{holweger99} \citeyear{holweger99}).  To date, several
other main sequence edge-on circumstellar disk systems have been
studied, including $\beta$ Car \citep{lagrangehenri90,hempel03},
HD85905 \citep{welsh98}, HR10 \citep{lagrangehenri90hr10,welsh98}, AU
Mic (\citeauthor*{kalas04} \citeyear{kalas04}; \citeauthor{roberge05}
\citeyear{roberge05}), and HD32297 (\citeauthor*{schneider05}
\citeyear{schneider05}, \citeauthor{redfield07hd32297}
\citeyear{redfield07hd32297}).

We selected three of the \ion{Ca}{2} and \ion{Na}{1} variable objects
(i.e., $\beta$ Car, HD85905, and HR10) together with $\alpha$ Oph,
which has an anomalously high absorption signature in these ions
\citep{crawford01}, to study any gas and dust in their circumstellar
environments.  We monitored their optical absorption properties to
probe the stable and variable components of the gas disk (see
Section~\ref{sec:var}).  Observations of their IR emission were also
made in order to look for any excess due to a dusty debris disk (see
Section~\ref{sec:ddd}).

In addition, we conducted mini-surveys of a handful of stars in close
angular proximity to our program stars to look for absorption due to
intervening interstellar gas.  Measuring the interstellar medium (ISM)
along the line of sight and in the locality directly surrounding a
circumstellar disk candidate, is critical to reconstructing the
distribution of possible ``contaminating'' ISM absorption
\citep{crawford01,redfield07hd32297}.  In particular, the Sun resides
in a large scale structure known as the Local Bubble, whose boundary
at $\sim$100\,pc is defined by a significant quantity of interstellar
material \citep{lallement03}.  These mini-surveys allow us to
differentiate between a stable circumstellar absorption component and
an interstellar absorption feature (see Section~\ref{sec:nab}).

Our program stars are all rapidly rotating, and therefore likely close
to edge-on \citep[i.e., their $v \sin i$ places them on the high
velocity tail of the distribution of predicted equatorial rotational
velocities, and therefore $\sin i \sim 1$;][]{abt95}.  They are
relatively mature systems, with ages of several hundreds of millions
of years, based on isochronal fitting (see Section~\ref{sec:ddd}).
Although they are older than $\beta$ Pic, which is $\sim$12 Myr, they
are comparable in age to other stars with debris disks, such as Vega
and $\alpha$ PsA \citep{barradoynavascues98}, as well as our solar
system during the Late Heavy Bombardment \citep{gomes05}.  Therefore,
with evidence that these systems have an edge-on orientation, and ages
consistent with the final stages of planetary system formation, our
program stars are inviting targets with which to makes observations of
secondary gas and dust products that may still reside in their
circumstellar environments.

\section{Observations}

Our sample was determined with the intent to investigate $\beta$ Pic
like systems, that is, edge-on circumstellar disks that are in the
evolutionary transition period of clearing their dusty debris disks.
These edge-on transitional systems provide an opportunity to probe
properties of both the dust, via IR spectral energy distributions
(SEDs), and gas, through atomic absorption lines.  We selected 4
systems, see Table~\ref{tab:basics}, that were suspected of having gas
disks, from absorption line variability on timescales of days to
years, or anomalous \ion{Ca}{2} absorption features that have been
difficult to attribute solely to local interstellar medium (LISM)
absorption.  Three of the four targets (HR10, HD85905, and
$\beta$~Car) show \ion{Ca}{2} and/or \ion{Na}{1} absorption
variability \citep{lagrangehenri90hr10,welsh98,hempel03}.  The fourth
target, $\alpha$~Oph, has an anomalously large \ion{Ca}{2} column
density compared to observations of other nearby stars
\citep{redfield02}.  \citet{crawford01} observed 8 angularly close
stars and detected \ion{Ca}{2} absorption in 2 stars which were
significantly more distant (120--211 pc) than $\alpha$ Oph (14.3 pc),
leaving open the possibility that the absorption toward $\alpha$ Oph
is circumstellar in origin.  None of our targets had significant IR
excess detections with {\it IRAS}, although these observations were
not sensitive enough to reach the stellar photospheres.

Our observational strategy included (1) continued high resolution
optical spectroscopy of our primary targets to monitor the short term
variability of atomic absorption lines, Section~\ref{sec:var}, (2)
observations of several stars close in angle and distance to our
primary targets in order to reconstruct the LISM absorption profile
along the line of sight, and to be able to distinguish between
interstellar absorption and a stable circumstellar feature,
Section~\ref{sec:nab}, and (3) {\it Spitzer} observations of the IR
SED to search for excess emission from dust and bulk gas emission
lines in a circumstellar disk, Sections~\ref{sec:ddd} and
\ref{sec:ddg}.

\subsection{Optical Spectroscopy}

High resolution optical spectra were obtained using the Coud\'{e}
Spectrometers on the 2.7m Harlan J. Smith Telescope at McDonald
Observatory and the Ultra High Resolution Facility (UHRF) on the 3.9m
Anglo-Australian Telescope (AAT) at the Anglo-Australian Observatory
(AAO).  Observations began in October 2003 and continued until July
2006, a temporal baseline of 2.8 years.  The observational parameters
for our primary targets are given in Table~\ref{tab:optobs} and stars
proximate to our primary targets in Table~\ref{tab:optobsnab}.  During
this interval, repeated observations of our primary targets monitored
absorption variability on timescales from days to years, and
observations of close neighbors surveyed the spatial and radial
variations in the interstellar medium around our primary targets. Two
atomic doublets were monitored: \ion{Ca}{2} H and K (3968.4673 and
3933.6614\,\AA, respectively) and \ion{Na}{1} D$_1$ and D$_2$
(5895.9242 and 5889.9510\,\AA, respectively).  These are among the
strongest transitions in the optical wavelength band, appropriate for
observing absorption toward nearby stars \citep{redfield06}.

The McDonald spectra were obtained with a range of resolving powers.
High resolution spectra ($R\,\equiv\,\lambda/\Delta\lambda\,
\sim\,$400,000) were obtained using the CS12 double-pass configuration
\citep{tull72}.  The detector was TK4, a 1024 $\times$ 1024 Tektronix
CCD chip, with 24 $\mu$m pixels.  The resolution was confirmed using
the HeNe laser line at 6328\,\AA, at $R\,=\,$520,000.  Only a single
order falls on the detector, and therefore the spectral range in this
configuration is very small, $\sim$1.4\,\AA\ near \ion{Ca}{2} at
3934\,\AA, and $\sim$2.0\,\AA\ near \ion{Na}{1} at 5896\,\AA, too
small to observe both transitions in the doublet simultaneously.  We
also utilized the 2dcoud\'{e} Spectrograph \citep{tull95} in both the
CS21 configuration ($R\,\sim\,$240,000) and the CS23 configuration
($R\,\sim\,$60,000).  The detector was TK3, a 2048 $\times$ 2048
Tektronix CCD chip, with 24 $\mu$m pixels.  The resolutions were
confirmed using the HeNe laser line at 6328\,\AA, at $R\,=\,$210,000
for CS21 and $R\,=\,$70,000 for CS23.  The spectral range for CS21
near the \ion{Ca}{2} doublet (3934 and 3968\,\AA) is $\sim$570\,\AA\
with $\sim$30\,\AA\ gaps between orders, and near the \ion{Na}{1}
doublet (5890 and 5896\,\AA) is $\sim$2800\,\AA\ with $\sim$130\,\AA\
gaps between orders.  In either configuration both lines of the
doublet (of either \ion{Ca}{2} or \ion{Na}{1}) can be observed
simultaneously.

The AAO spectra were obtained with the highest resolving power
available ($R\,\sim\,$940,000), using the UHRF spectrograph
\citep{diego95}.  The detector was EEV2, a 2048 $\times$ 4096 CCD
chip, with 15 $\mu$m pixels.  The resolution was confirmed using the
HeNe laser line at 6328\,\AA, at $R\,=\,$1,090,000.  Only a single
order falls on the detector, but due to the large chip size, the
spectral range is $\sim$4.8\,\AA\ near \ion{Ca}{2} at 3934\,\AA, and
$\sim$7.2\,\AA\ near \ion{Na}{1} at 5896\,\AA, although again, too
small to easily observe both transitions in either the \ion{Ca}{2} or
\ion{Na}{1} doublets simultaneously.  By utilizing an image slicer
\citep{diego93}, the throughput is significantly better than single
slit high resolution spectrographs.

The data were reduced using Image Reduction and Analysis Facility
\citep[IRAF;][]{tody93} and Interactive Data Language (IDL) routines
to subtract the bias, flat field the images, remove scattered light
and cosmic ray contamination, extract the echelle orders, calibrate
the wavelength solution, and convert to heliocentric velocities.
Wavelength calibration images were taken using a Th-Ar hollow cathode
before and after each scientific target.

The extracted one-dimensional spectra were then normalized using fits
of low order polynomials to regions of the continuum free of
interstellar and telluric absorption lines.  Numerous water vapor
lines are commonly present in spectra around the \ion{Na}{1} doublet.
Although the telluric H$_2$O lines are relatively weak, they need to
be modeled and removed from the spectrum, in order to measure an
accurate \ion{Na}{1} absorption profile, particularly for observations
toward nearby stars which may be expected to exhibit weak interstellar
(or circumstellar) absorption.  The traditional telluric subtraction
technique of observing a nearby, rapidly rotating, early type star at
a similar airmass in order to divide out an empirically derived
telluric spectrum is not feasible for observations of our nearby
targets.  It is precisely our primary targets, which are nearby,
rapidly rotating, early type stars and likely candidates themselves to
be used as a telluric standard, that we want to scrutinize for
interstellar or circumstellar absorption.  Instead, we use a forward
modeling technique demonstrated by \citet{lallement93} to remove
telluric line contamination in the vicinity of the \ion{Na}{1} D
lines.  We use a relatively simple model of terrestrial atmospheric
transmission (AT - Atmospheric Transmission program, from Airhead
Software, Boulder, CO) developed by Erich Grossman to fit and remove
the telluric water vapor lines.  Observing both transitions of the
\ion{Na}{1} doublet is an important confirmation that the telluric
subtraction is successful.  With two independent measurements of
\ion{Na}{1} absorption at the same projected velocity, it is easy to
identify contaminating telluric absorption.  No telluric features fall
near the \ion{Ca}{2} H \& K lines.

Atmospheric sodium absorption was occasionally detected, particularly
in high signal-to-noise ($S/N$) spectra.  This absorption is easily
identified at the Doppler shift of the projected velocity of the
Earth's atmosphere in the heliocentric rest frame.  This projected
velocity is given in Table~\ref{tab:optobs} for all \ion{Na}{1}
observations.  For the vast majority of \ion{Na}{1} observations, the
location of an atmospheric absorption line is well separated from any
astrophysical absorption features.

\subsection{Infrared}

The IR observations were obtained with {\it Spitzer} \citep{werner04},
from 2004 September through 2005 September.  Table~\ref{tab:obs} lists
dates and astronomical observation request (AOR) numbers.  Near-IR
photometry from 3.6 to 8.0 $\mu$m was obtained with the Infrared Array
Camera \citep[IRAC;][]{fazio04}, and mid- to far-IR photometry was
obtained with the Multiband Imaging Photometer for {\emph Spitzer}
\citep[MIPS;][]{rieke04}.  Moderate resolution spectroscopy
($R\,\approx160$--600) in the 10--37 $\mu$m range was obtained with
Short-High (SH) and and Long-High (LH) modules of the Infrared
Spectrograph \citep[IRS;][]{houck04}, while low resolution
($R\approx15$--25) spectroscopy in the 55--95 $\mu$m region was
obtained with the SED mode of MIPS.

\subsubsection{{\it Spitzer} Photometry}

The IRAC and MIPS 24 $\mu$m data were processed using the the c2d
mosaicking/source extraction software, c2dphot, \citep{harvey04} which
is based on the mosaicking program, Astronomical Point Source
Extraction (APEX), developed by the \emph{Spitzer} Science Center
(SSC) and the source extractor Dophot \citep*{schechter93}. While the
photometric measurement uncertainties are small ($\sim$5$\%$ and
$\sim$9$\%$) for IRAC and MIPS 24 $\mu$m sources with good $S/N$, the
absolute calibration uncertainty is estimated to be 10$\%$ (Evans et
al. 2006).

We used the Mosaicking and Point Source Extraction (MOPEX) software
package version 030106\footnotemark \citep{mopex}, to create 70 and
160 $\mu$m mosaics starting from the basic calibrated data (BCD)
processed by the Spitzer Science Center (SSC) through the S14.4
pipeline \citep{mipsdatahandbook}.  We used the median-filtered BCDs
provided by the SSC, which are optimized for photometry of point
sources.  Only two of our sources, $\alpha$ Oph and $\beta$ Car, are
clearly detected at 70 $\mu$m. We obtained 70 $\mu$m fluxes and
uncertainties for $\alpha$ Oph and $\beta$ Car using MOPEX
point-source fitting from half-pixel (i.e., 4$\arcsec$) re-sampled
mosiacs.  We obtained 70 $\mu$m upper limits for HR 10 and HD 85905
through aperture photometry from mosaics re-sampled at the original
pixel scale. We use an aperture with a radius of 16$\arcsec$ and a sky
annulus with an inner and an outer radius of 48 and 80$\arcsec$,
respectively. Based on high $S/N$ 70 $\mu$m point sources identified
in the BCD mosaics of the \emph{Spitzer} c2d Legacy project (program
identification (PID) = 173), we derived a multiplicative aperture
correction ($AC$) of 1.6. Thus, we compute the observed 70 $\mu$m
flux, $F_{70} = FA_{70} \times AC$, where $FA_{70}$ is the flux within
the aperture minus the contribution from the sky. We estimate a
1$\sigma$ photometric uncertainty, $\sigma =AC \times RMS_{\rm sky}
\times n^{1/2}$, where $RMS_{\rm sky}$ is the root mean square (RMS)
of the pixels in the sky annulus, and $n$ is the number of pixels in
our aperture. An absolute calibration uncertainty, estimated to be
20$\%$ (Evans et al. 2006), is added in quadrature to the photometric
uncertainties for MIPS 70 $\mu$m and 160 $\mu$m.  $\alpha$ Oph and
$\beta$ Car are detected with $S/N \sim 30$, while HD85905 and HR10
are very close to the 70 $\mu$m detection limit (e.g., $F_{70} \sim
3\times \sigma$).  We note that the absolute calibration uncertainty
is estimated at the 20$\%$ level for 70 $\mu$m observations
\citep{mipsdatahandbook}, and becomes the dominant source of error for
moderate and high $S/N$ observations.  \footnotetext{MOPEX is
available for distribution at {\url
http://ssc.spitzer.caltech.edu/postbcd/}.}

None of our four sources are detected at 160 $\mu$m. The MIPS 160
$\mu$m channel has a short-wavelength filter leak in which stray light
in the wavelength range of 1--1.6 $\mu$m produces a ghost image offset
$\sim$40 arcsecs from the true 160 $\mu$m image.  The leak is only
detectable above the confusion limit for sources brighter than $J \sim
5.5$ \citep{mipsdatahandbook}. Given the brightness of $\alpha$ Oph
and $\beta$ Car ($J = 1.75$ and 1.55, respectively), their MIPS 160
$\mu$m images are severely affected by this leak.  For our two fainter
targets, HR10 and HD85905 ($J = 5.85$ and 6.05, respectively), the 160
$\mu$m signal produced by the near-IR leak should be just below the
160 $\mu$m confusion limit expected from extragalactic sources
\citep{dole04}.  In this case, we use an aperture 32$\arcsec$ in
radius and a sky annulus with an inner and an outer radius of 40 and
80$\arcsec$, respectively.  We adopt an aperture correction of 2.0,
appropriate for the size of our aperture and sky annulus\footnotemark.
Similar to the 70 $\mu$m upper limits, we calculate the 1$\sigma$
uncertainty, $\sigma =AC \times RMS_{\rm sky} \times n^{1/2}$, from
mosaics re-sampled at the original pixel scale.  The 160 $\mu$m flux
at the position of all four sources is affected by the
short-wavelength leak.  Therefore, we adopt a conservative 3$\sigma$
upper limit by adding 3$\sigma$ to the flux measured at the source
position.  \footnotetext{A discussion of aperture corrections applied
to MIPS data can be viewed at {\url
http://ssc.spitzer.caltech.edu/mips/apercorr/}.}

\subsubsection{{\it Spitzer} Spectroscopy}

The IRS spectra were extracted via the c2d Interactive Analysis
(c2dia) reduction environment (F. Lahuis et al. 2006, in
preparation)\footnotemark.  Prior to extraction, the dither positions
are combined, reducing noise and adding to spectral stability.  The
spectra are extracted using an optimal point spread function (PSF)
extraction, in which an analytical cross dispersion PSF, defined from
high $S/N$ observations of a calibrator, is fit to the collapsed order
data by varying the offset for the trace and width of the source
profile. The observed signal is assumed to be that of a point source
plus a uniform zero level (representing extended/sky emission).  The
amplitude of the zero level is determined via the profile fitting
method and used to make sky corrections.  The optimal PSF extraction
uses a fit to the good pixels only, therefore direct bad/hot pixel
corrections are not required.  The final spectra are defringed using
the IRSFRINGE package developed by the c2d team.  \footnotetext{A
description of c2dia is also available in the documentation for the
final c2d Legacy data delivery, which is available at {\url
http://ssc.spitzer.caltech.edu/legacy/all.html}.}

MIPS SED spectra were calibrated and coadded using MOPEX.  The spectra
were then extracted from the coadded images via standard methods using
the IRAF apall tool within the National Optical Astronomy
Observatories (NOAO) TWODSPEC package.  Wavelength calibration and
aperture corrections were performed post extraction by using the
``MIPS70 SED sample calibration'' file provided by the SSC with the
MOPEX package (version 030106).  

Two of the four sources (HR10 and HD85905) were not detected in the
IRS LH module, and 3$\sigma$ upper limits are used together with our
photometric measurements to constrain the estimates of the excess
emission (Section~\ref{sec:ddd}).  Additionally, no gas-phase atomic
or molecular lines or solid-state silicate or polycyclic aromatic
hydrocarbon (PAH) bands were detected in the {\it Spitzer} spectra
(Section~\ref{sec:ddg}).

\section{Detections and Constraints on Variable Circumstellar Gas}

Gas phase absorption due to \ion{Ca}{2} and \ion{Na}{1} gas phase
atoms was monitored in spectra of all four targets.
Table~\ref{tab:optobs} details the observational parameters.  All
targets were observed over short ($\sim$few nights) and long
($\sim$few months) timescales, spanning 2.75 years.  Examples of
observed spectra are shown in Figure~\ref{fig:examspec}.  Two epochs
are shown for each star to emphasize any variability and demonstrate
the use of different spectrographs and spectral resolutions.  The
$\alpha$ Oph absorption, which shows no evidence for variability,
exemplifies the consistency in data collection and reduction over 7
months while using two different instruments (CS21 and UHRF).  HD85905
and HR10 show clear evidence for variability over these two epochs,
while only the \ion{Na}{1} profiles of $\beta$ Car varied
significantly over the two epochs shown.  The temporal variability of
the entire dataset will be discussed in detail in
Section~\ref{sec:var}.

Figure~\ref{fig:examspec} also shows quite clearly that the observed
\ion{Na}{1} column density is significantly lower relative to the
observed \ion{Ca}{2} absorption.  In fact, we detect only very weak
ISM or circumstellar \ion{Na}{1} absorption toward HR10, despite it
being our most distant sightline and therefore likely to traverse
significant ISM material.  It does show quite strong circumstellar
absorption in \ion{Ca}{2}.  The constant \ion{Na}{1} absorption in
HD85905, despite the variability in \ion{Ca}{2} presumably from a
circumstellar gas disk, may be a signature of the intervening ISM
along this line of sight, rather than the circumstellar disk.  The
``contamination'' of interstellar absorption on our observations of
gas in circumstellar disks will be discussed in Section~\ref{sec:nab}
where spectra of stars in close angular proximity to our targets are
presented.

\subsection{Temporal Variability \label{sec:var}}

\begin{figure}
\begin{center}
\epsscale{.8}
\plotone{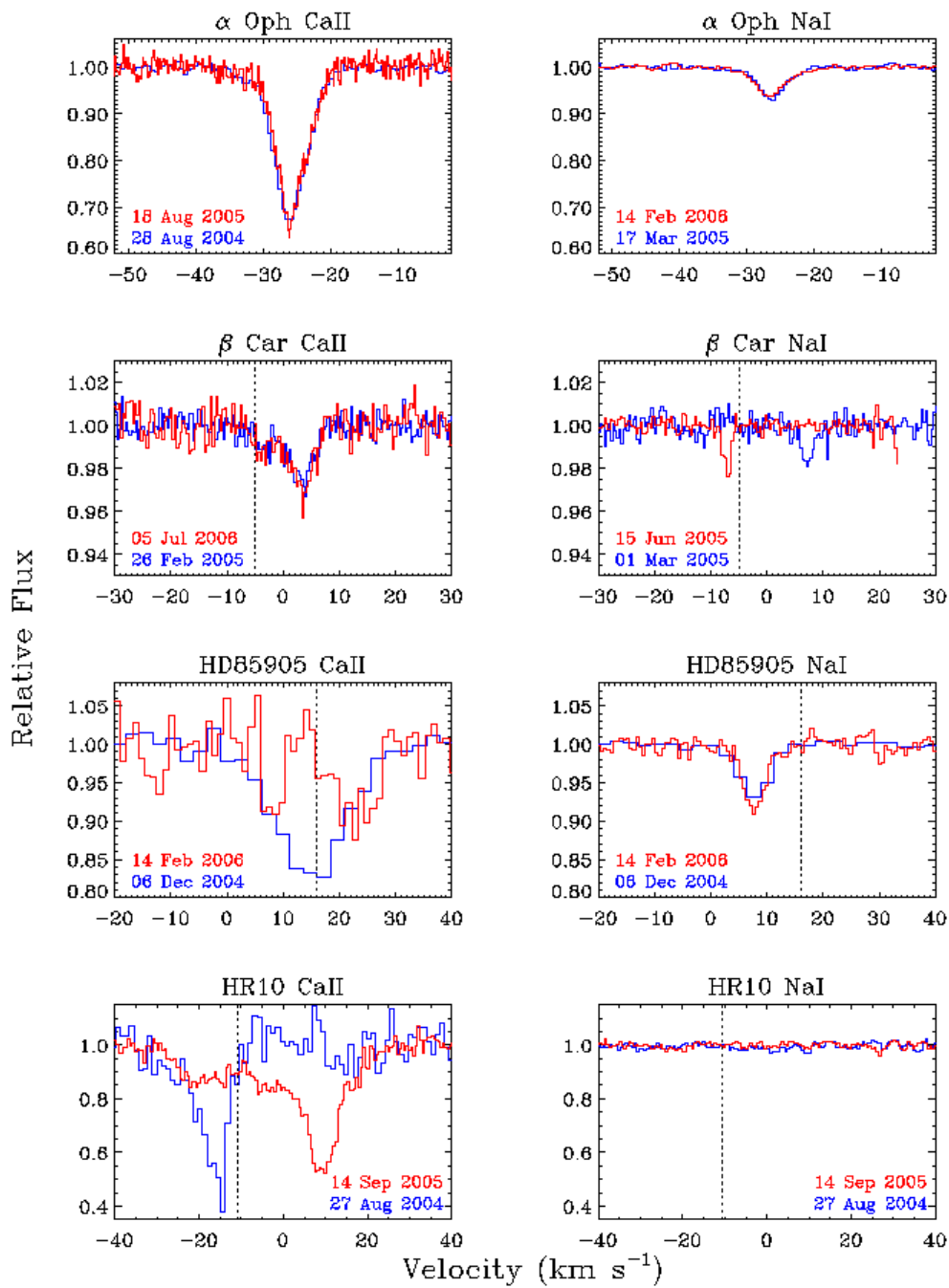}
\end{center}
\caption{Examples of typical \ion{Ca}{2} and \ion{Na}{1} spectra taken
at two different epochs for all 4 primary targets.  The vertical
dotted line indicates the stellar radial velocity.  Both ions are
shown on the same scale, and it is clear that the \ion{Ca}{2}
absorption is much stronger than \ion{Na}{1} in all cases.  Temporal
variability is immediately obvious in \ion{Ca}{2} observations of
HD85905 and HR10, and in \ion{Na}{1} of $\beta$ Car.  These examples
include spectra taken at various spectral resolution from both the
McDonald Observatory H.J. Smith Telescope and the AAT.  Note the
success of telluric subtraction in the \ion{Na}{1} spectra.
\label{fig:examspec}}
\end{figure}

Figure~\ref{fig:examspec} shows that variability is detected in 3 of
our 4 targets.  In order to characterize the absorption profile we use
the apparent optical depth (AOD) method \citep{savage91} to calculate
the observed column density in each velocity bin.  An alternative
characterization would be to model the absorption profile with a
series of Gaussian components, as is often done in high resolution ISM
absorption line analysis \citep*[e.g.,][]{welty94,crawford95}, and is
particularly straightforward in observations of the simple absorption
profiles seen toward nearby stars, where only 1--3 components are
detected \citep{redfield01}.  Component fitting has also been used
successfully to characterize variable absorption due to circumstellar
material \citep[e.g.,][]{welsh98}.  However, the ability to attach
physical properties with the parameters used to fit the series of
Gaussian absorption profiles is highly sensitive to the physical
distinctiveness in projected velocity of the absorbing medium and the
resolving power of the spectrograph.  In other words, a one-to-one
correspondence must exist between an absorbing structure and an
observed absorption component in order to make meaningful physical
measurements.  For example, physical properties (e.g., temperature)
can be derived from a Gaussian fit to the line width for ISM
absorption toward the nearest stars, since 1 absorption component is
observed and the path length is so short that only 1 absorbing cloud
is traversed \citep{redfield04tt}.  However, the circumstellar
environment giving rise to the variable absorption profiles is likely
too complicated (e.g., coincident projected velocities from different
absorbing sites) to allow for a straightforward correspondence with a
series of Gaussian components.  Therefore, we employ the AOD technique
to characterize the observed absorption.

The AOD method is well described in \citet{savage91}, and has been
used extensively to model absorption profiles
\citep[e.g.,][]{jenkins01,roberge02}.  In brief, the apparent optical
depth ($\tau_a$) in velocity ($v$) space is
\begin{equation}
\label{eq:taua}
\tau_a(v) = \ln\frac{I_0(v)}{I_{\rm obs}(v)}\;,
\end{equation}
where $I_{\rm obs}(v)$ is the observed spectrum, and $I_0(v)$ is the
continuum spectrum expected if no interstellar or circumstellar
absorption were present.  In a normalized spectrum, as shown in
Figure~\ref{fig:examspec}, the stellar background intensities have
already been divided out, and $I_0(v)\,=\,1$.  Equation~\ref{eq:taua}
does not describe the true optical depth, since the instrumental line
spread function (LSF) is folded into our observed spectrum ($I_{\rm
obs}(v)$).

The column density ($N_a$) in each velocity bin can be calculated from
the apparent optical depth,
\begin{equation}
\label{eq:na}
N_a(v) = \frac{m_e c}{\pi e^2} \frac{\tau_a(v)}{f \lambda} \;\;[{\rm cm}^{-2}]\;,
\end{equation}
where $f$ is the oscillator strength of the transition, and $\lambda$
is the wavelength of the velocity bin.  The total column density can
be calculated by integrating $N_a(v)$ over the velocity range of
interest.  Equation~\ref{eq:na} provides accurate total column
densities provided the absorption is unsaturated.  Since both
\ion{Ca}{2} and \ion{Na}{1} are doublets, we have two independent
measurements of the absorbed column density in transitions with
different oscillator strengths.  A comparison of $N_a(v)$ for the two
transitions in each doublet confirms the absorption is unsaturated.  A
final $N_a(v)$ is calculated for each ion by taking the weighted mean
of $N_a(v)$ calculated for each transition.  By utilizing the
information from both transitions in the doublet, the impact of
numerous systematic uncertainties (e.g., continuum placement, telluric
line subtraction, wavelength calibration) is greatly reduced.

\begin{figure}
\epsscale{1.}
\plottwo{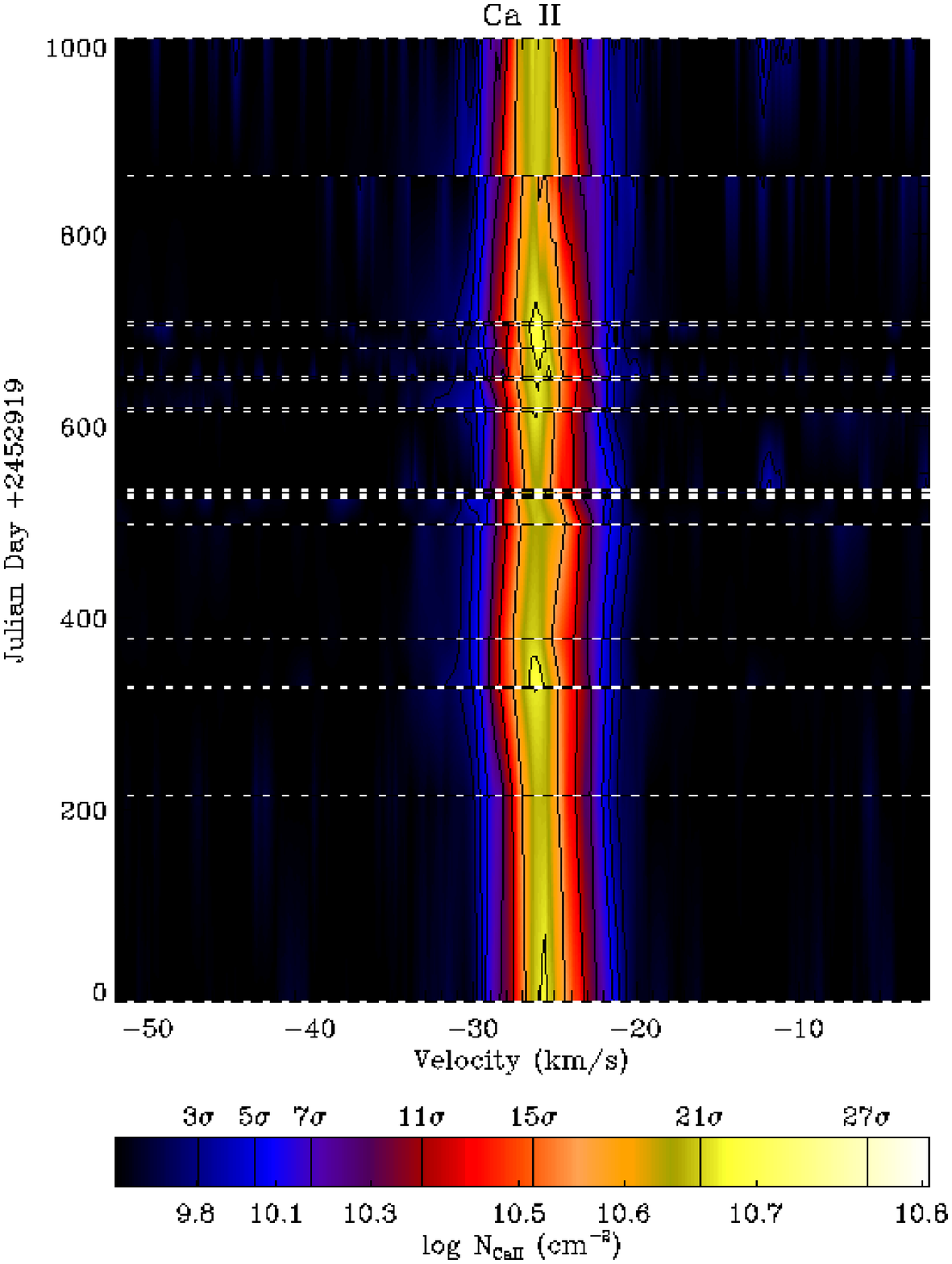}{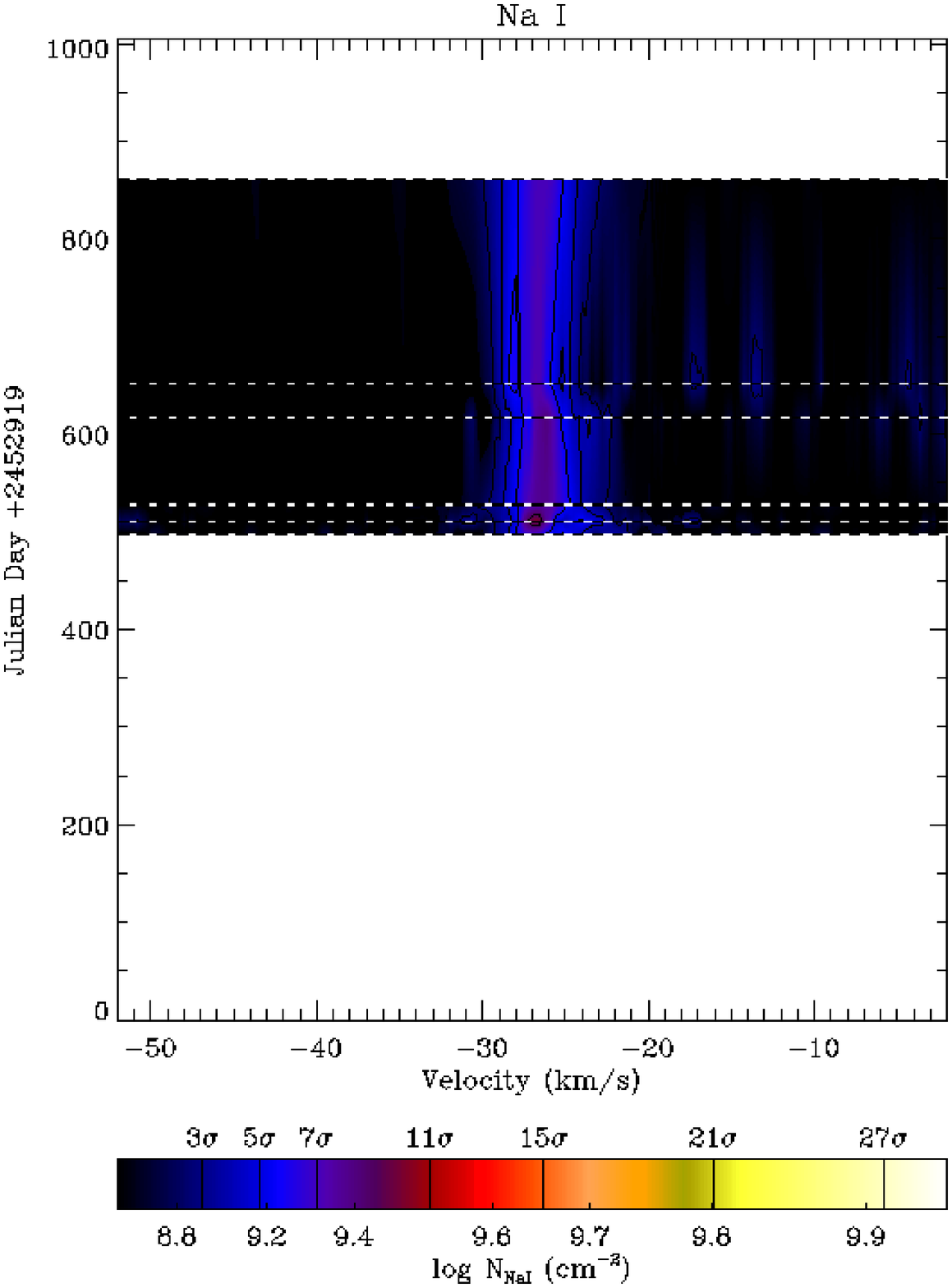}
\caption{Three dimensional data cubes of the observed column density,
signified by the colored contours, in \ion{Ca}{2} ({\it left}) and
\ion{Na}{1} (right), toward $\alpha$ Oph.  The data are shown as a
function of projected velocity and date of observation.  Although the
colored contours are displayed as continuous through time,
observations are sporadic and indicated by the horizontal hatched
lines.  A simple interpolation between observations is used to create
the continuous data cube.  The color-coding is normalized in $S/N$
between \ion{Ca}{2} and \ion{Na}{1}, so that it is clear that the
\ion{Ca}{2} absorption is much stronger compared to \ion{Na}{1}.
Little, if any, variation is detected in the absorption toward
$\alpha$ Oph.
\label{fig:aodaoph}}
\end{figure}

\begin{figure}
\epsscale{1.}
\plottwo{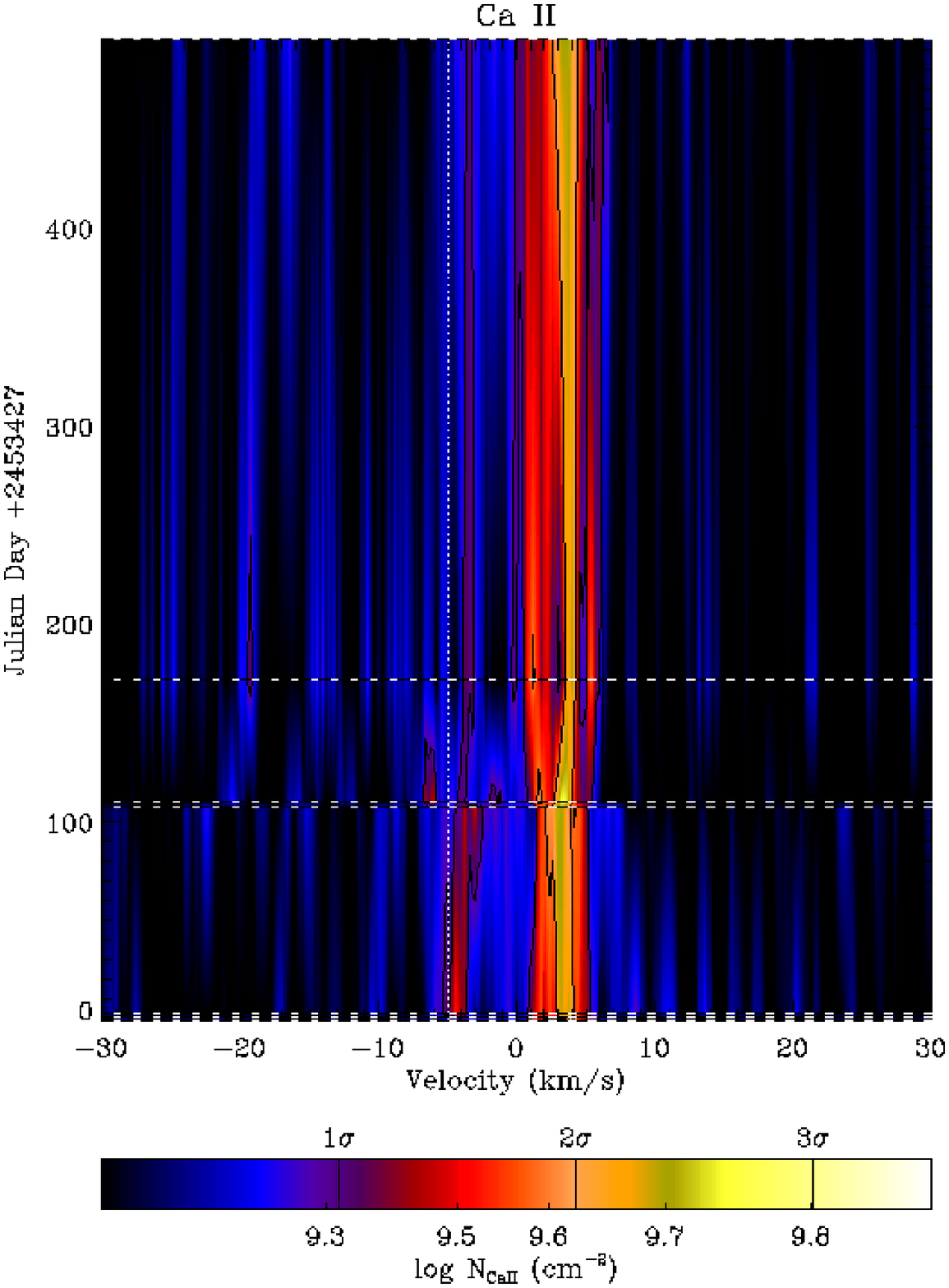}{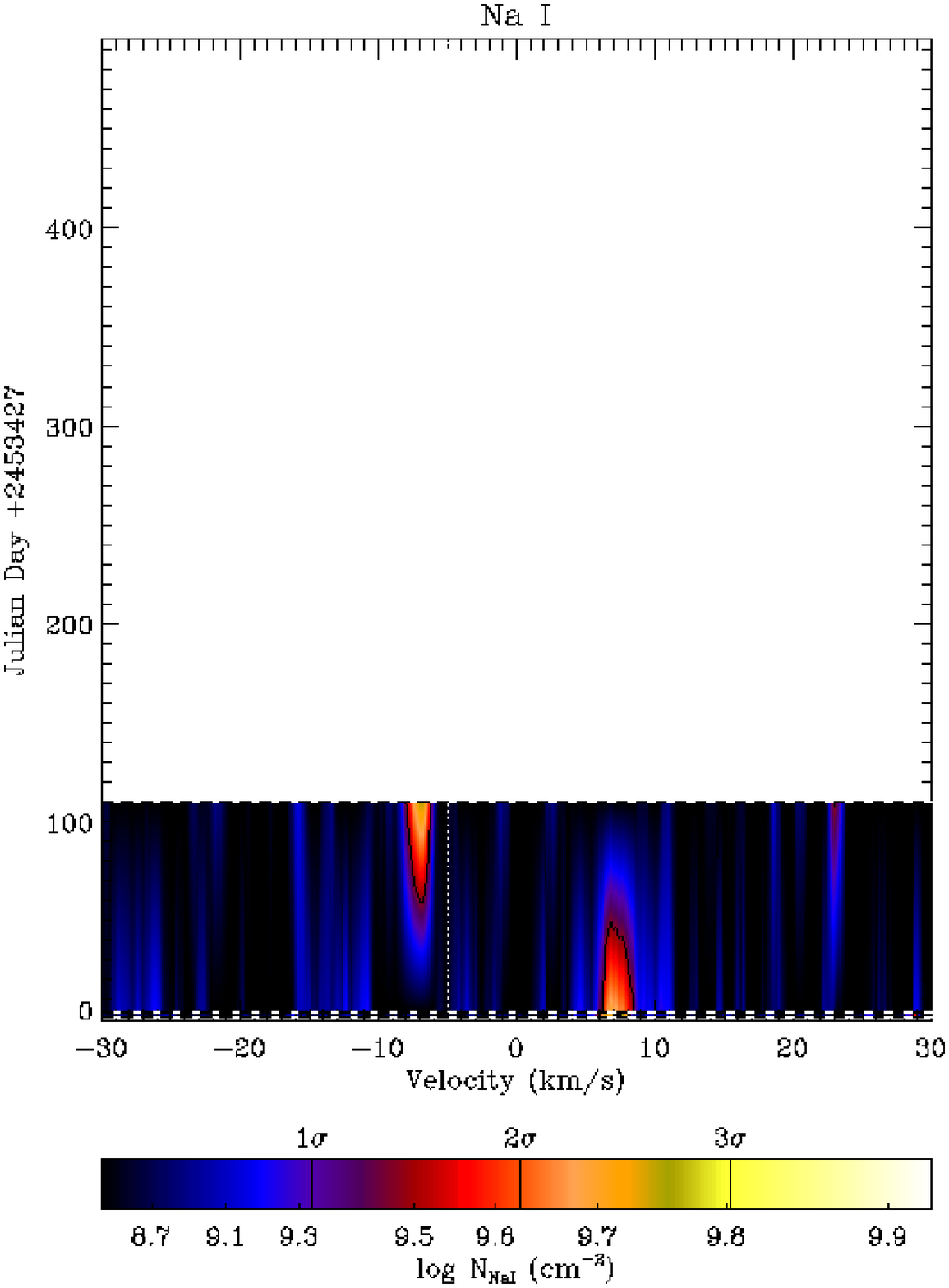}
\caption{Same as Figure~\ref{fig:aodaoph} for absorption toward
$\beta$ Car.  The vertical dotted line indicates the stellar radial
velocity.  Some slight variation is detected in \ion{Ca}{2}, and clear
temporal variability is seen in \ion{Na}{1}.
\label{fig:aodbcar}}
\end{figure}

\begin{figure}
\epsscale{1.}
\plottwo{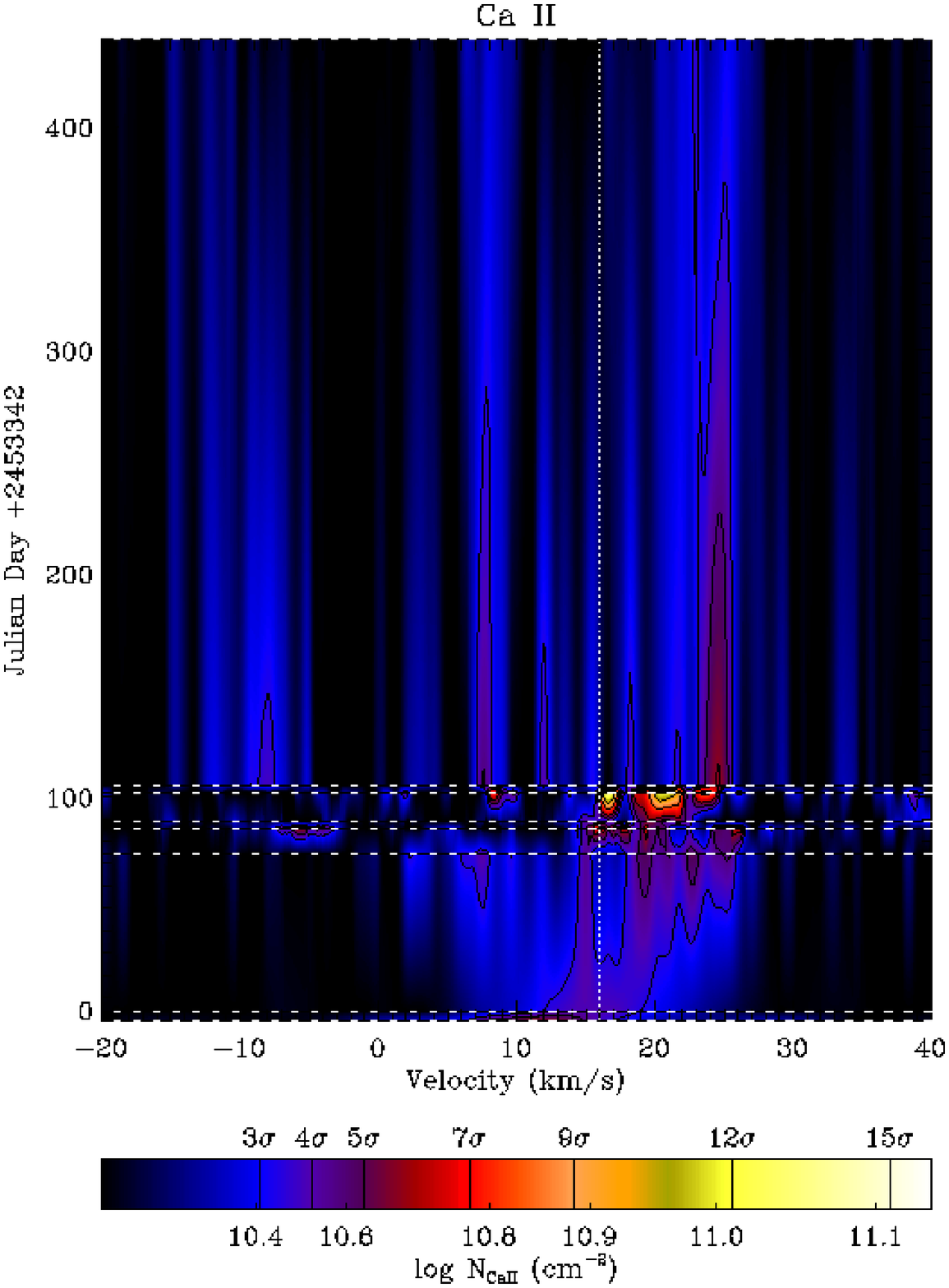}{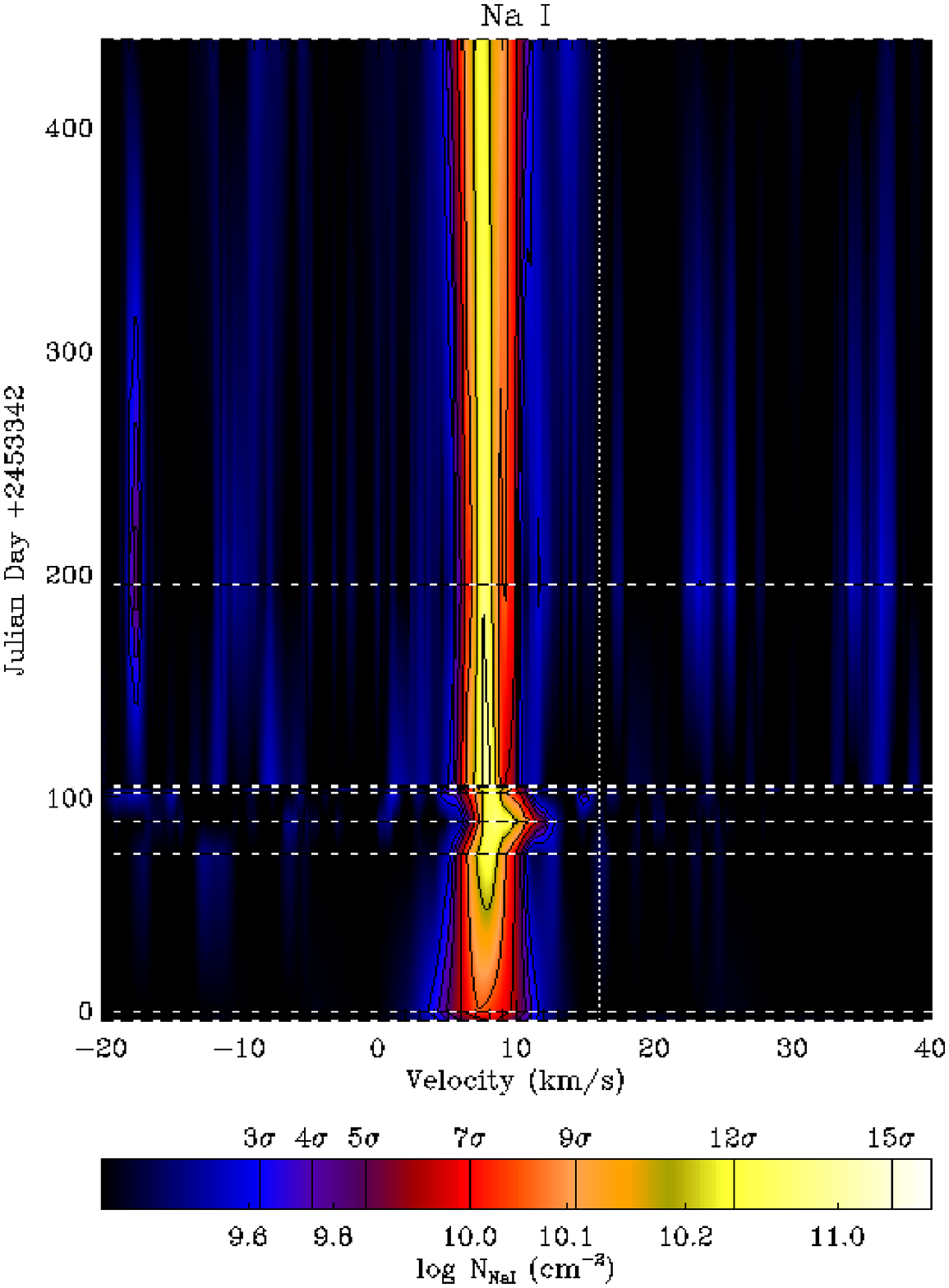}
\caption{Same as Figure~\ref{fig:aodaoph} for absorption toward
HD85905.  The vertical dotted line indicates the stellar radial
velocity.  Temporal variation is detected in \ion{Ca}{2}, while a
relatively constant feature, likely interstellar, is seen in
\ion{Na}{1}.
\label{fig:aodhd85905}}
\end{figure}

\begin{figure}
\epsscale{1.}
\plottwo{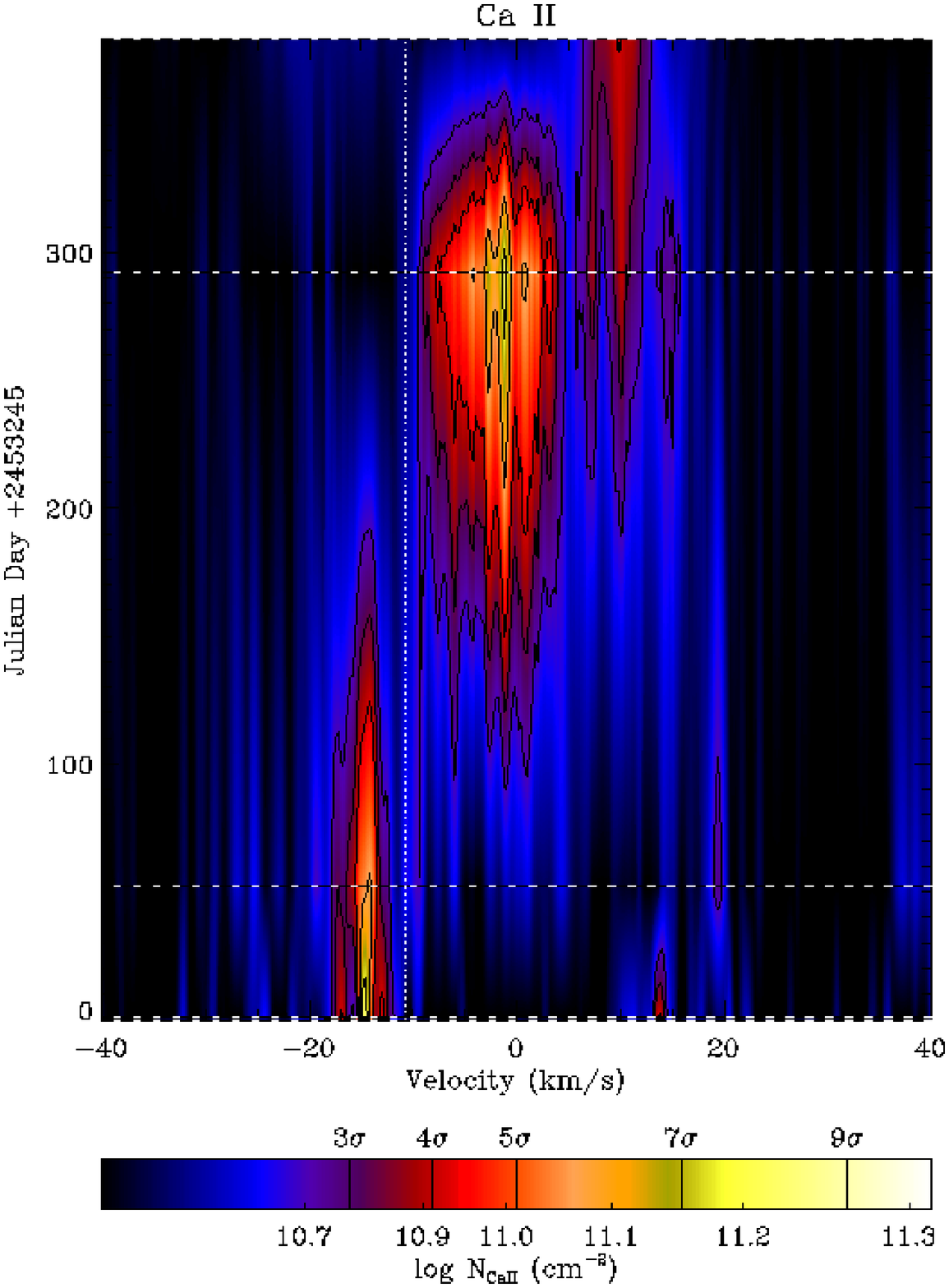}{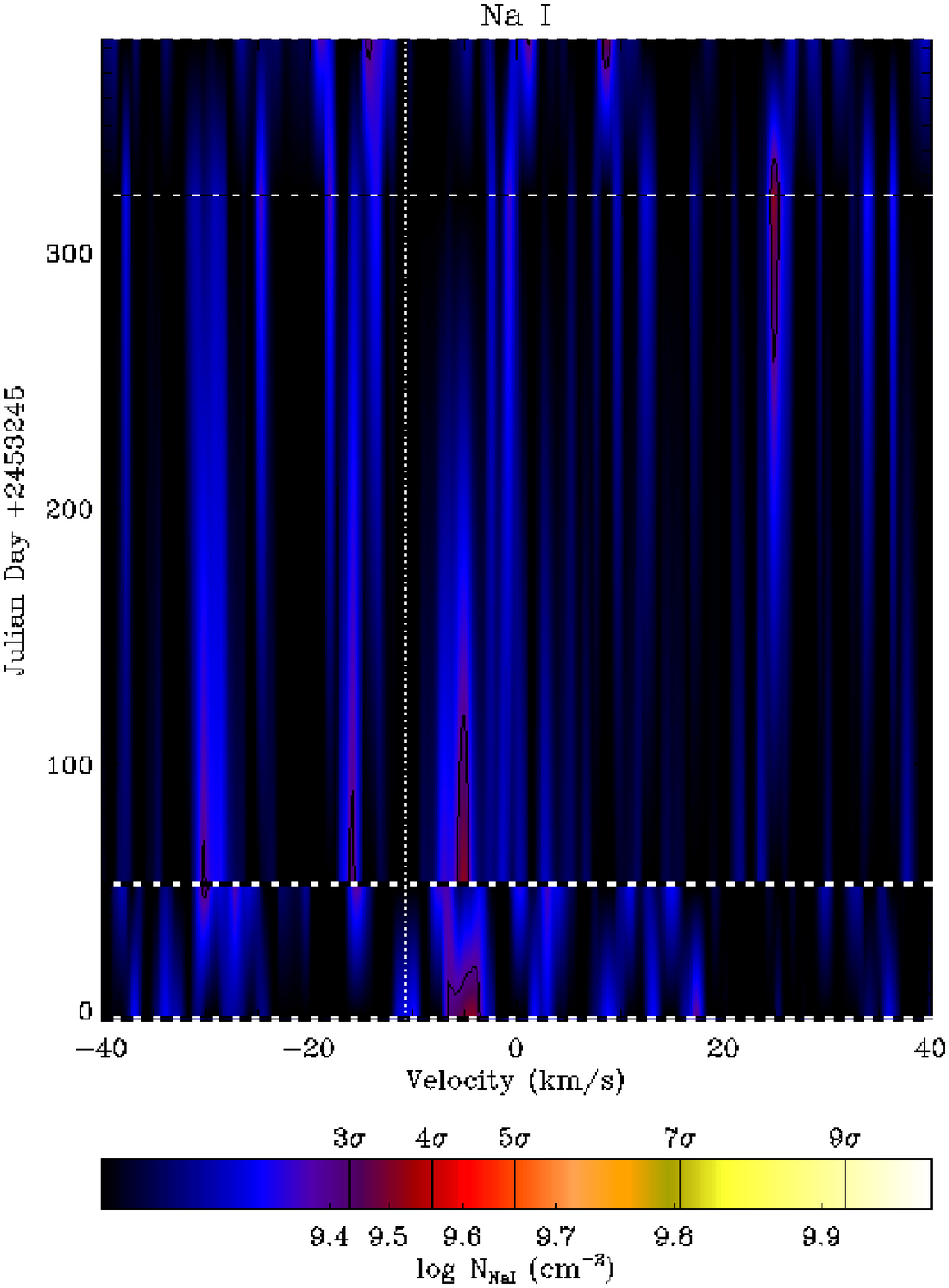}
\caption{Same as Figure~\ref{fig:aodaoph} for absorption toward HR10.
The vertical dotted line indicates the stellar radial velocity.
Strong temporal variation is detected in \ion{Ca}{2}, while only very
weak absorption is detected in \ion{Na}{1}.
\label{fig:aodhr10}}
\end{figure}

Another challenge of evaluating the temporal variability of absorption
profiles is amalgamating the data of a long monitoring campaign.  Even
comparing the spectra of only two epochs, as in
Figure~\ref{fig:examspec}, can be confusing.  Comparing spectra
directly in such a way for as many as 26 epochs is impractical.  Each
observation results in an array of measurements of the column density
as a function of velocity (derived from the normalized flux as a
function of wavelength as described above).  Therefore, we have a
sporadic data cube.  Figures~\ref{fig:aodaoph}--\ref{fig:aodhr10} are
three-dimensional contour plots of observed column density as a
function of velocity as a function of time, for all 4 targets.  It is
important to note that the observations are sporadic and not
continuous in time.  The date of each observation is highlighted with
a hatched line, and the contours between epochs are simple
interpolations between the two observations.  Occasionally closely
spaced observations, (e.g., $<$1 week apart), cannot be distinguished
in Figures~\ref{fig:aodaoph}--\ref{fig:aodhr10} and can be associated
with an apparent discontinuity.  It is likely that subtle changes,
such as in \ion{Ca}{2} toward $\alpha$ Oph in Figure~\ref{fig:aodaoph}
or \ion{Na}{1} toward HD85905 in Figure~\ref{fig:aodhd85905} are
caused by systematic effects, whereas obvious circumstellar
variability is seen for example in \ion{Ca}{2} toward HR10 in
Figure~\ref{fig:aodhr10}.  The short and long-term temporal
variability is discussed in detail below.  For each target, the color
coding is normalized between \ion{Ca}{2} and \ion{Na}{1} such that
column density measurements of comparable $S/N$ are displayed with the
same color.  For example, the normalization of $S/N$ makes it clear in
Figure~\ref{fig:aodaoph}, that the \ion{Na}{1} feature is clearly
weaker when compared to the strong \ion{Ca}{2} absorption.

\begin{figure}
\epsscale{.8}
\plotone{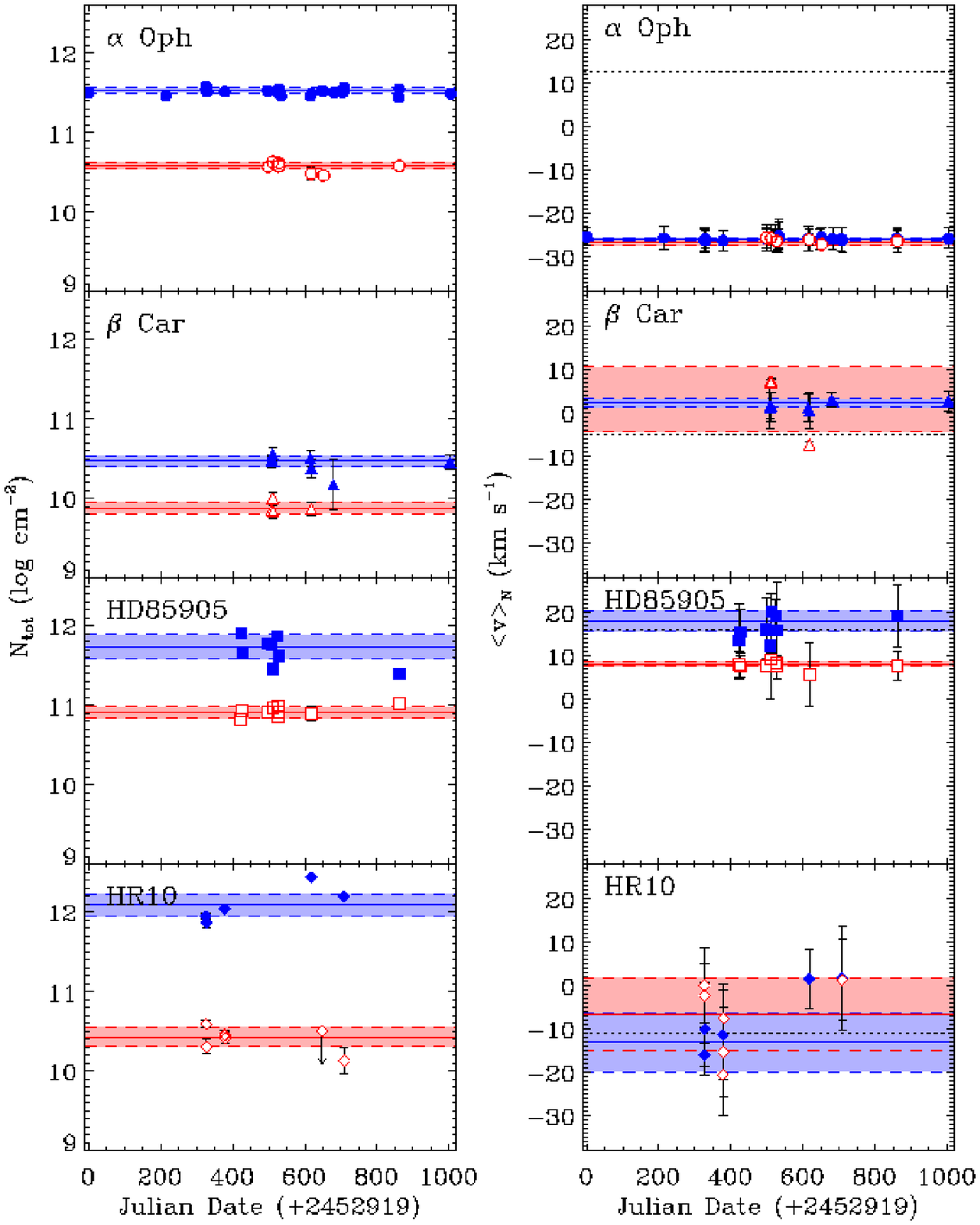}
\caption{Total column density ({\it left}) and column density weighted
velocity ({\it right}) measurements of absorption toward our primary
targets as a function of time.  The filled symbols indicate
\ion{Ca}{2} observations, while open symbols indicate \ion{Na}{1}.
The error bars are the weighted average variance, and in the case of
$\langle v \rangle_N$ indicate the range of observed velocities rather
than the error in measuring the central velocity of absorption.  The
shaded regions (blue corresponds to \ion{Ca}{2}, and red with
\ion{Na}{1}) indicate the weighted mean and weighted average variance
of the absorption detected in the primary target, presented in
Figure~\ref{fig:tarvar}.  The horizontal dotted line in the right
panels indicates the radial velocity of the star.  Three of our four
targets show some signs of temporal variation.  Only the spectra
toward $\alpha$ Oph shows no hint of variation.  The data is also
presented in Table~\ref{tab:longterm}.
\label{fig:tarvar}}
\end{figure}

Some basic attributes of the temporal variability data cubes are
summarized in Figure~\ref{fig:tarvar}.  For each observation, the
total column density, $N_{\rm tot} = \int^{v_2}_{v_1} N_a(v) dv$, and
the column density weighted velocity, $\langle v \rangle_N =
\sum\limits_{i=v_1}^{v_2} v_i N_i / N_{\rm tot}$ are plotted, where
$v_1$ and $v_2$ indicate the range of velocities over which the
absorption is detected.  The error bars on the total column density
are often smaller than the symbol size.  The ``error bars'' shown in
the column density weighted velocity are the weighted average variance
\citep{bevington92}.  Therefore, the ``errors'' shown for $\langle v
\rangle_N$ are not the error in determining the central velocity of
absorption, which for these high resolution spectra range from
0.1--2.0 km~s$^{-1}$, but the range of velocities with significant
absorption.  For example, in the case of the \ion{Ca}{2} spectra shown
for HR10 in Figure~\ref{fig:examspec}, for the 27 Aug 2004 spectrum,
$\langle v \rangle_{\rm CaII} = -15.9$ km~s$^{-1}$, and a weighted
average variance of only 2.6 km~s$^{-1}$.  The fact that the weighted
average variance is relatively small matches the fact that the
absorption profile spans a narrow range of velocities around --16
km~s$^{-1}$.  The 14 Sep 2005 spectrum, on the other hand, shows a
narrow absorption component around 9 km~s$^{-1}$, as well as
significant weak absorption that ranges from --25 to 5 km~s$^{-1}$.
Since the absorption covers a wide range of velocities and is
asymmetric, $\langle v \rangle_{\rm CaII}$ is not centered exactly on
the narrow component but at 1.8 km~s$^{-1}$, and the weighted average
variance is relatively large, 12.1 km~s$^{-1}$, since absorption is
detected over a wide range of velocities.  The total column density
and the column density weighted velocity of each observation is listed
in Table~\ref{tab:longterm}.

Figures~\ref{fig:examspec}--\ref{fig:tarvar} indicate that no temporal
variability is detected toward $\alpha$ Oph in either observed ion.
The \ion{Ca}{2} absorption observed toward $\beta$ Car is relatively
constant, but variation is seen in \ion{Na}{1}.  HD85905 shows some
\ion{Ca}{2} variability, but relatively constant \ion{Na}{1}
absorption, likely dominated by absorption from interstellar material
in the Local Bubble shell, which will be discussed in
Section~\ref{sec:nab}.  HR10 shows dramatic \ion{Ca}{2} temporal
variability, but little to no absorption, interstellar or
circumstellar, is detected in \ion{Na}{1}.

\subsubsection{Search for Short Term Variability}

Short-term variability, on time scales of nights or hours, is detected
in two of our targets: HD85905 and HR10.  Night-to-night measurements
are provided in Table~\ref{tab:longterm}.  Such short temporal
variations have been detected in $\beta$ Pic \citep*{ferlet87} and in
these two stars by \citep{welsh98}.  Column density variations from
night-to-night can reach factors $\geq$2, while shifts in velocity of
$\geq$10 km\,s$^{-1}$ are detected.  However, the magnitude and
frequency of short variations in our targets remains lower than
detected toward the prototypical edge-on debris disk, $\beta$ Pic,
where single feature night-to-night variations in column density can
exceed a factor of 10, and 20 km~s$^{-1}$ in radial velocity
\citep{petterson99}.

\subsubsection{Comparison of Contemporaneous \ion{Ca}{2} and \ion{Na}{1} Observations}

Contemporaneous observations of both \ion{Ca}{2} and \ion{Na}{1}
absorption toward circumstellar disk stars, even $\beta$ Pic, are relatively
rare.  There is a strong preference to observe the \ion{Ca}{2} lines
rather than the \ion{Na}{1} lines, because \ion{Na}{1} is
significantly less abundant (see Section~\ref{sec:ca2na1}) and it is
difficult to model and remove the telluric lines that populate the
spectral region near \ion{Na}{1}.  \citet{welsh97,welsh98} monitor
both \ion{Ca}{2} and \ion{Na}{1} for several edge-on circumstellar disks,
including $\beta$ Pic, HD85905, and HR10.  In the case of $\beta$ Pic,
only the strong component at the rest frame of the star is detected in
both ions, while no time variable absorbers have ever been detected in
\ion{Na}{1}.  Toward HD85905 and HR10, \citet{welsh98} detect
absorption in both ions, often with little one-to-one correspondence
in the velocity of the absorption between \ion{Ca}{2} and \ion{Na}{1}.
Although, it is important to note that the ions were not observed
during the same night, but on adjacent nights.

Although rarely simultaneous, many of our observations of \ion{Ca}{2}
and \ion{Na}{1} were taken one after the other during the same night.
The nightly measurements, given in Table~\ref{tab:longterm}, can be
used to compare such contemporaneous absorption measurements of
\ion{Ca}{2} and \ion{Na}{1}.  Both ions show the same absorption
feature toward $\alpha$ Oph, although, as discussed in
Section~\ref{sec:lismaoph}, this feature is not circumstellar in
origin, but a result of interstellar absorption along the line of
sight.  Toward $\beta$ Car, the \ion{Ca}{2} and \ion{Na}{1} absorption
features do not match in velocity.  The \ion{Ca}{2} is relatively
constant and distinct from the narrow and weak \ion{Na}{1} absorption.
It is possible that we are sampling two different collections of
material, and the relatively constant \ion{Ca}{2} feature is part of
the extended disk, while \ion{Na}{1} is found in the variable gas
component.  Toward HD85905, a comparison between contemporaneous
\ion{Ca}{2} and \ion{Na}{1} observations shows little correspondence,
although there is the possibility that the \ion{Na}{1} absorption is
due to the interstellar medium, as discussed in
Section~\ref{sec:3lism}.  The contemporaneous observations toward HR10
are typically consistent between \ion{Ca}{2} and \ion{Na}{1}, such as
the 2005 September and 2004 October observations.  However, the 2004
August observations of \ion{Ca}{2} and \ion{Na}{1} are not consistent
in velocity, similar to the earlier observations by \citet{welsh98},
where \ion{Ca}{2} and \ion{Na}{1} absorption features differ between
adjacent nights.

\subsubsection{Distribution of Red- vs. Blue-shifted Features}

The velocity distribution of the variable circumstellar absorption
features relative to the rest frame of the host star is an important
constraint on the dynamics of the absorbing gas.  Toward $\beta$ Pic,
the majority of variable absorption features are redshifted relative
to the rest frame of the star, although blueshifted absorption is not
particularly rare \citep*{crawford98,petterson99}.  Although we do not
have the temporal sampling of some of the monitoring campaigns of
$\beta$ Pic \citep[e.g.,][]{petterson99}, we are able to quantify the
frequency of redshifted versus blueshifted absorption features.  Using
the radial velocities listed in Table~\ref{tab:basics}, we calculated
the fraction of the total absorption that was redshifted relative to
the radial velocity of the star, i.e., $f_{\rm red} =
\int^{\infty}_{v_{\rm R}} N(v) / \int^{\infty}_{-\infty} N(v)$.  For
$\beta$ Car, $f_{\rm red}($\ion{Ca}{2}$)$ ranged from 80--100\%, while
$f_{\rm red}($\ion{Na}{1}$)$ ranged from 4--100\%, indicating the
stability of the \ion{Ca}{2} feature in contrast to \ion{Na}{1}.
Toward HD85905, $f_{\rm red}($\ion{Ca}{2}$)$ ranged from 25--79\%,
while $f_{\rm red}($\ion{Na}{1}$)$ ranged from 0--2\%, although as
discussed in Section~\ref{sec:3lism}, there is a possibility that the
\ion{Na}{1} feature is caused by the interstellar medium.  Toward
HR10, $f_{\rm red}($\ion{Ca}{2}$)$ ranged from 0--98\%, while $f_{\rm
red}($\ion{Na}{1}$)$ ranged from 44--100\%.  In contrast to $\beta$
Pic, the velocity of variable absorption toward our three
circumstellar disk stars did not have a distribution dominated by
redshifted radial velocities.

\subsubsection{Search for Very Long Term Variability}

The primary targets were selected based on previous evidence of
anomalous or variable gas phase atomic absorption.  This work builds
on that of previous studies, and presents an opportunity to search for
very long term variability, on the timescale of decades.
Table~\ref{tab:longterm} details the absorption characteristics
observed during the monitoring campaign presented in this work, as
well as past measurements by other researchers.  We have limited the
literature search to relatively high spectral resolution observations
($R \geq 60$,000).

Our measurements of \ion{Ca}{2} and \ion{Na}{1} toward $\alpha$ Oph
are constant and consistent with past observations.  Some of the total
column density measurements are slightly lower than the present values
\citep[e.g.,][]{hempel03}.  However, the central velocity has remained
steady throughout at approximately $-25.8$ km~s$^{-1}$, and systematic
errors are expected due to the different instruments and analysis
techniques employed.  For example, a new analysis by
\citet{crawford01} of identical data from \citet{crawford95} led to a
slight increase in column density, from $\log N_{\rm CaII} = 11.38$ to
11.54, which matches the mean value of the present monitoring
campaign, which is $\log N_{\rm CaII} = 11.53$.

The variation we see in the total column density of \ion{Ca}{2} toward
$\beta$ Car, $\log N_{\rm CaII} = 10.18$--$10.56$, is similar to the
variation detected by \citet{hempel03}, $\log N_{\rm CaII} =
10.14$--$10.50$.  The previous nondetection of \ion{Na}{1} toward
$\beta$ Car of $\log N_{\rm NaI} < 10.23$ by \citet{welsh94} is
consistent with the relatively weak absorption features that are
detected in the present campaign, which range from $\log N_{\rm NaI} =
9.87$--$10.15$.  The radial velocity of the \ion{Na}{1} features
varied by 14.4 km~s$^{-1}$ over the two epochs, from $v_{\rm NaI} =
-7.1$ to $+7.3$ km~s$^{-1}$.

Variation in velocity ($v_{\rm CaII} = 12.4$ to $20.2$ km~s$^{-1}$) and
column density ($\log N_{\rm CaII} = 11.39$--$11.91$) is detected in
\ion{Ca}{2} absorption toward HD85905, just as it has been in a
previous campaigns \citep[$v_{\rm CaII} = -11.5$ to $+7.0$ km~s$^{-1}$,
$\log N_{\rm CaII} = 11.97$--$12.24$;][]{welsh98}.  We do not see as
dramatic velocity shifts, nor quite as large total column densities,
but the magnitude of variability in column density is comparable.  We
see a relatively stable \ion{Na}{1} component in velocity ($v_{\rm
NaI} \sim 8.2$ km~s$^{-1}$), which shows subtle column density
variations ($\log N_{\rm NaI} = 10.82$--$11.02$).  This is similar to
the 1997 November observations by \citet{welsh94}, $v_{\rm NaI} = 9.2$
and $\log N_{\rm NaI} = 11.22$.  However, over a temporal baseline of
1.2 years, we see no dramatic variation, while in two epochs spaced by
$\sim$10 months, \citet{welsh94} observed a dramatic shift in velocity
to $v_{\rm NaI} = -10.8$ and a subtle weakening in column density,
$\log N_{\rm NaI} = 11.02$.

The dramatic \ion{Ca}{2} absorption variability toward HR10 ($v_{\rm
CaII} = -15.9$ to $+1.8$ km~s$^{-1}$, $\log N_{\rm CaII} =
11.87$--$12.44$) is similar to variations seen in previous
observations in velocity ($v_{\rm CaII} = -6.8$ to $+6.1$ km~s$^{-1}$)
and total column densities ($\log N_{\rm CaII} = 11.87$--$12.62$).
The \ion{Na}{1} columns ($\log N_{\rm NaI} = 10.13$--$10.59$) are
significantly smaller for this campaign than observed by
\citet{welsh98}, $\log N_{\rm NaI} = 11.09$--$11.49$.  The upper
limits by \citet{hobbs86} ($\log N_{\rm NaI} < 11.0$) and
\citet{lagrangehenri90} ($\log N_{\rm NaI} < 10.5$)are consistent with
the present campaign's low column densities, and indicates that the
high column density absorption epoch detected by \citet{welsh98} was
short-lived.

\subsection{Comparison With Proximate Targets \label{sec:nab}}

\begin{figure}
\epsscale{1.}
\plotone{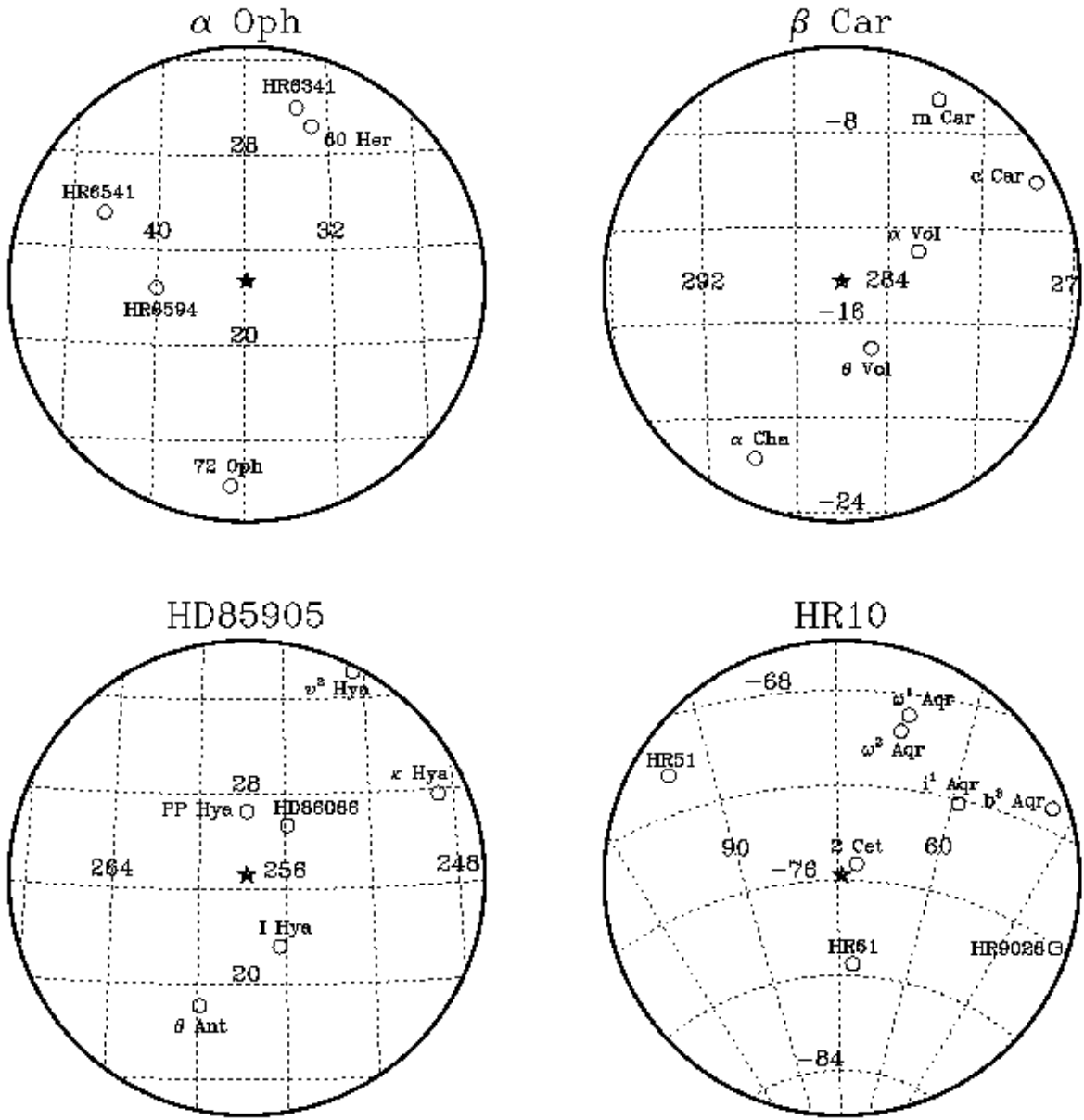}
\caption{Location of target stars and proximate neighbors in Galactic
coordinates.  The field of view is 10 degrees radius, centered on the
primary target.  The distances of the proximate stars bracket the
distance of the primary target, but are generally at shorter distances
in order to reconstruct the three-dimensional morphology of the LISM
in that direction.  The stellar properties and coordinates of the
proximate stars can be found in
Tables~\ref{tab:basicaophnab}--\ref{tab:basichr10nab}.\label{fig:nabloc}}
\end{figure}

Due to the similarity of circumstellar and interstellar absorption
signatures, it is important to understand the distribution of
interstellar material in the vicinity of our primary targets.  For
this reason, we observed several stars in close angular proximity and
at a range of distances, in order to establish the three-dimensional
structure of the ISM in the direction of our primary targets.
Tables~\ref{tab:basicaophnab}--\ref{tab:basichr10nab} provide the
basic stellar parameters of stars proximate to our primary targets,
and Figure~\ref{fig:nabloc} shows the location of the primary and
neighboring stars in Galactic coordinates.  If an absorption feature
is detected toward both our primary target and a proximate star, it
must be located between the Earth and the nearer of the two stars.  In
particular, the last two columns of
Tables~\ref{tab:basicaophnab}--\ref{tab:basichr10nab} give the
separation of the proximate target from the primary target in angle,
$\Delta\theta$, and in distance in the plane of the sky (POS), $\Delta
r_{\rm POS} = d_{\rm ISM} \tan (\Delta\theta)$.  The distance to the
absorbing material, $d_{\rm ISM}$, is rarely known, so the distance to
the closer of the two stars is used as an upper limit, $d_{\rm ISM}
\leq d_{\star}$.  Given the values of $\Delta r_{\rm POS}$ in our
proximate star surveys, we may be probing structure on physical scales
significantly less than 1 parsec, if the absorbing material is even
closer than the nearest of the observed stars.  The observational
parameters for the proximate stars are given in
Table~\ref{tab:optobsnab}.

The LISM is an interstellar environment filled with warm ($T \sim
7000$~K), partially ionized, moderately dense ($n \sim 0.3$~cm$^{-3}$)
material, surrounded by a volume of hot ($T \sim 10^6$~K), rarefied
($n \sim 0.005$~cm$^{-3}$) gas known as the Local Bubble.  It is
relatively common to observe \ion{Ca}{2} from the warm partially
ionized clouds in the LISM \citep{redfield02}, while \ion{Na}{1} is
rarely detected within the Local Bubble, but is clearly observed in
the shell of dense gas that defines its boundary \citep{lallement03}.
Since the Local Bubble shell is a large scale interstellar structure,
it is reasonable to expect that absorption from the Local Bubble will
be present in all sightlines that extend beyond its boundary, at
$\sim$100 pc.  We have used the three-dimensional model of the
morphology of the Local Bubble by \citet{lallement03}, to more
carefully estimate the distance to the edge of the Local Bubble in the
direction of our 4 primary targets.  This is given in 10th column of
Table~\ref{tab:basics}.  Two of our primary targets, $\alpha$ Oph and
$\beta$ Car are well within the Local Bubble and therefore are
unlikely to show signatures of absorption from its shell.  HD85905 is
located just outside the LB shell and therefore, the \ion{Na}{1}
spectrum is likely to show evidence of ISM absorption due to the LB
shell.  HR10 is located in the direction of the southern Galactic
pole.  Due to the lack of dense material in directions perpendicular
to the Galactic plane, the Local Bubble is relatively unconfined at
its poles \citep{welsh99,lallement03}.  Therefore, it is unlikely that
Local Bubble material will be detected, even in the distant sightlines
of HR10 and its proximate stars.

\begin{figure}
\epsscale{1.2}
\plottwo{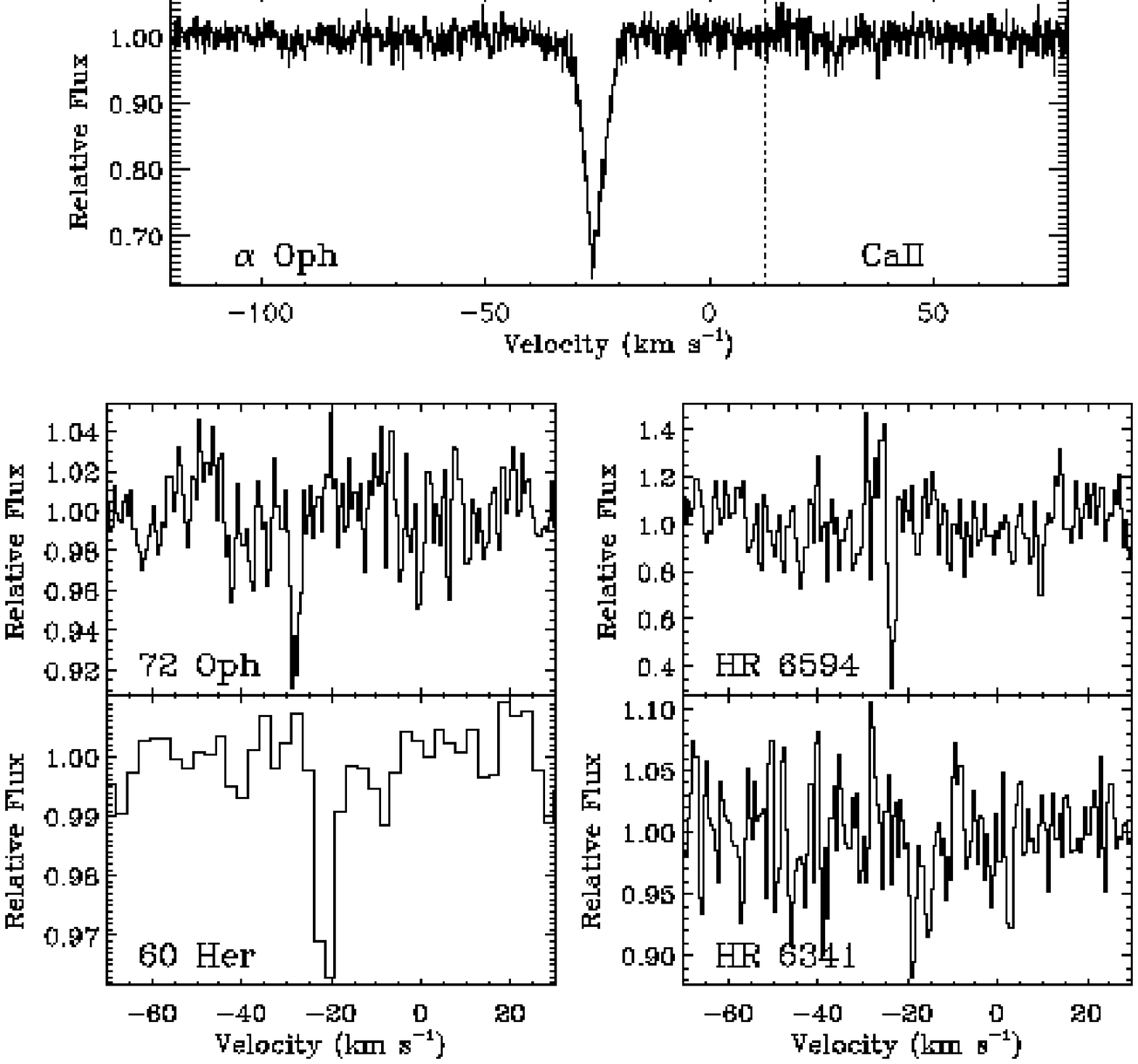}{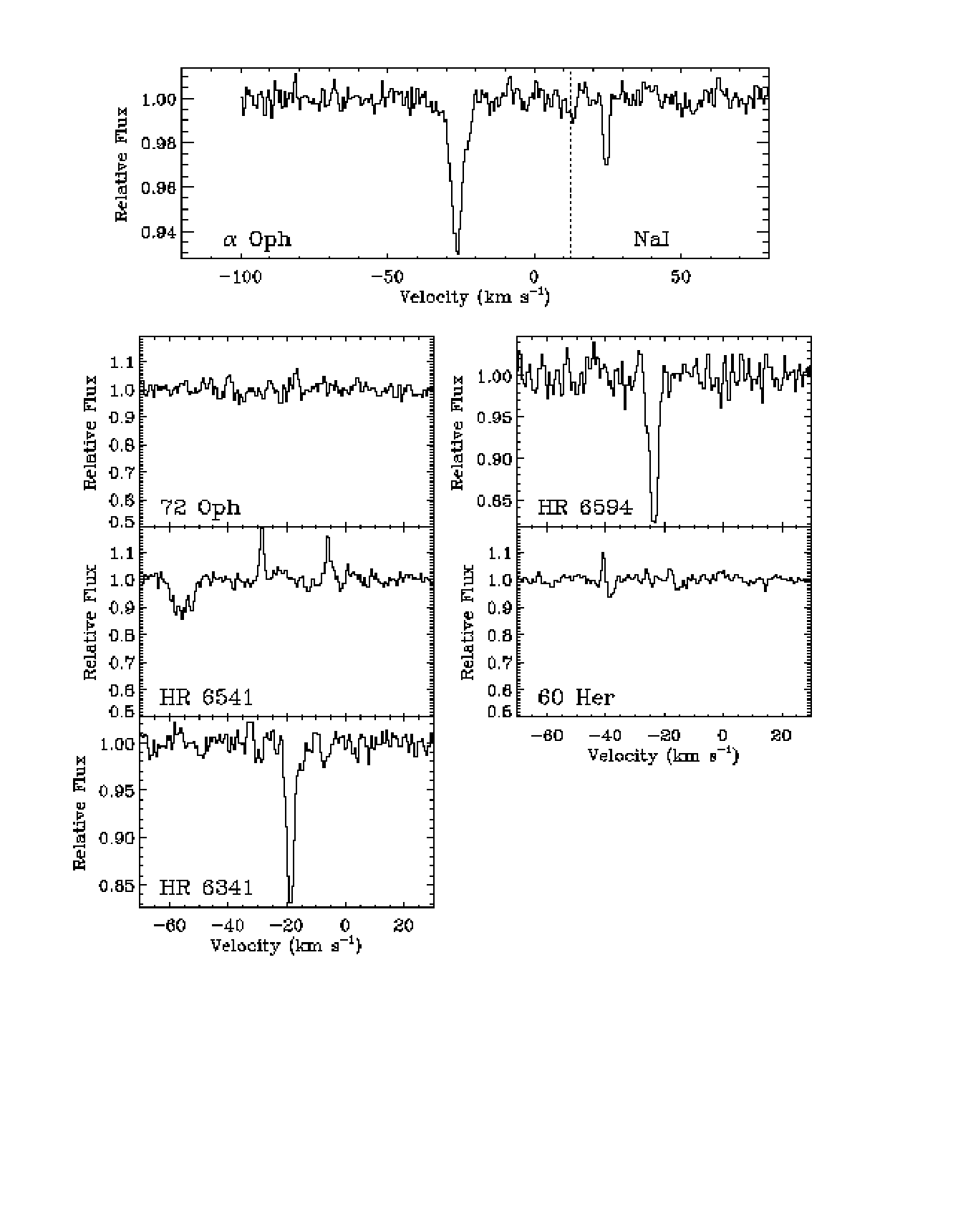}
\caption{\ion{Ca}{2} ({\it left}) and \ion{Na}{1} ({\it right})
absorption lines observed towards stars in close angular proximity to
$\alpha$ Oph.  Representative spectra of $\alpha$ Oph are provided at
the top on the same scale, but with an expanded range.  The vertical
dotted line indicates the stellar radial velocity.  Note the weak
atmospheric \ion{Na}{1} absorption in the $\alpha$ Oph spectrum near
24.2 km~s$^{-1}$, well separated from the strong main absorption near
--26.6 km~s$^{-1}$.  It is clear that several (4 of 4 in \ion{Ca}{2}
and 2 of 5 in \ion{Na}{1}) proximate targets show absorption at a
similar velocity as the main component in $\alpha$ Oph.  The
\ion{Na}{1} spectrum of HR6541 shows signs of the difficulties in
measuring weak absorption in this spectral region, including artifacts
from the incomplete subtraction of the stellar photospheric
\ion{Na}{1} line at --55 km~s$^{-1}$, and overestimated telluric
subtraction leading to ``emission'' spikes at --28 and --6
km~s$^{-1}$.
\label{fig:nabaoph}}
\end{figure}

\begin{figure}
\epsscale{1.2}
\plottwo{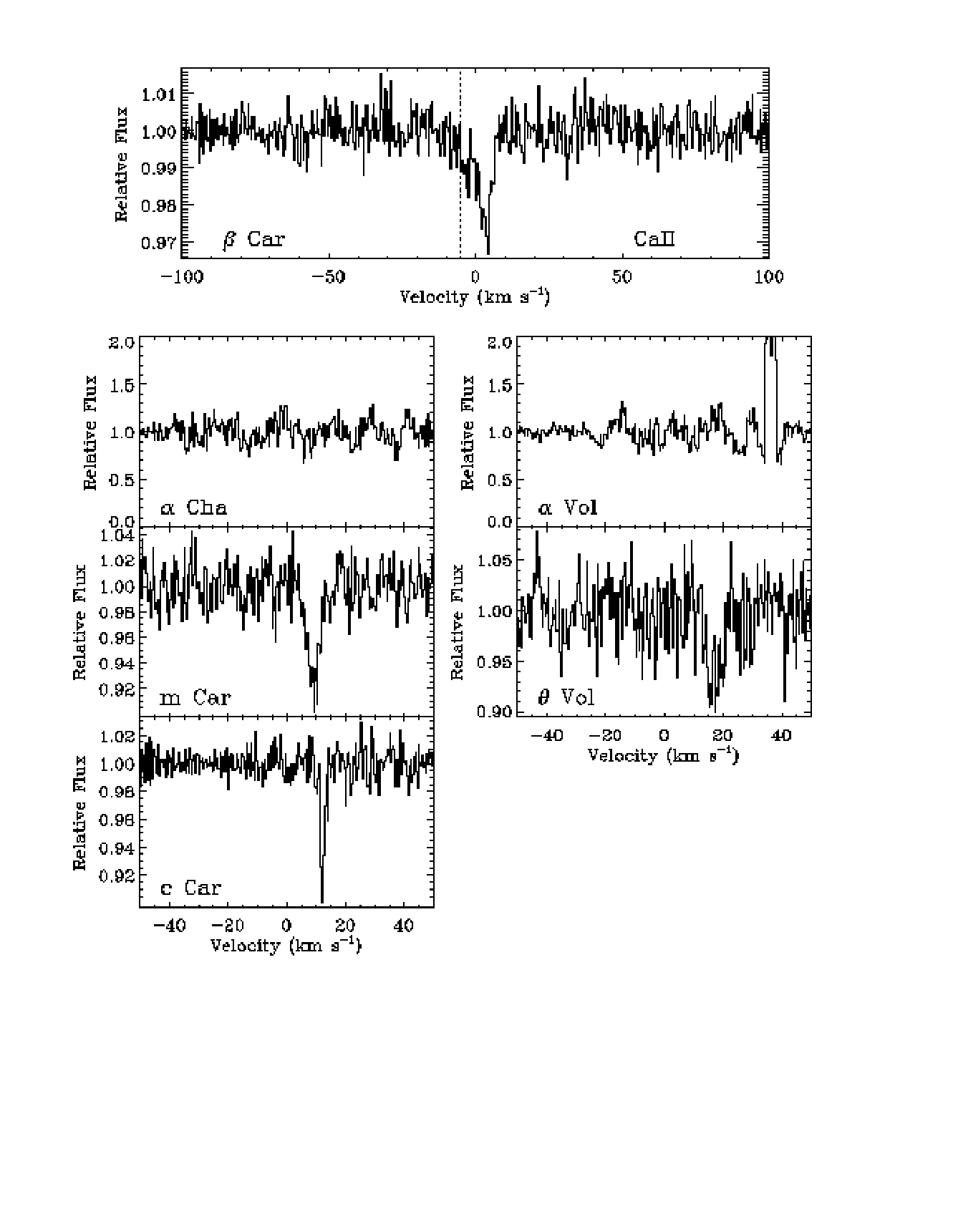}{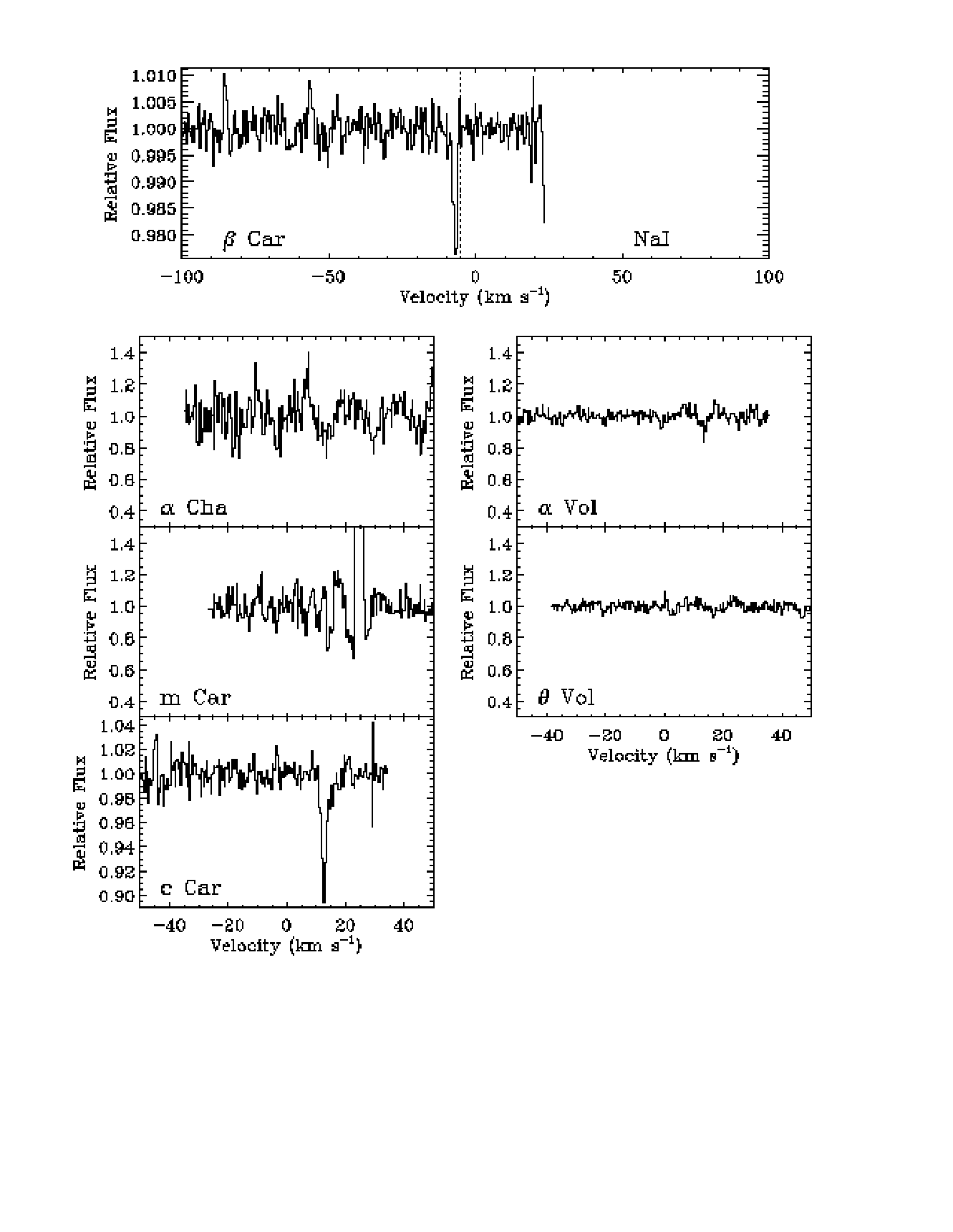}
\caption{Same as Figure~\ref{fig:nabaoph} for stars proximate to
$\beta$ Car.  In this direction, interstellar \ion{Ca}{2} is detected
in 3 of 5 targets, and \ion{Na}{1} detected in only 1 of 5 targets.
However the interstellar absorption is consistently redshifted with
respect to the absorption detected toward $\beta$ Car.  Indeed, the
closest star to $\beta$ Car, $\theta$ Vol, shows the largest projected
velocity difference in comparison to the absorption toward $\beta$
Car.
\label{fig:nabbcar}}
\end{figure}

\begin{figure}
\epsscale{1.2}
\plottwo{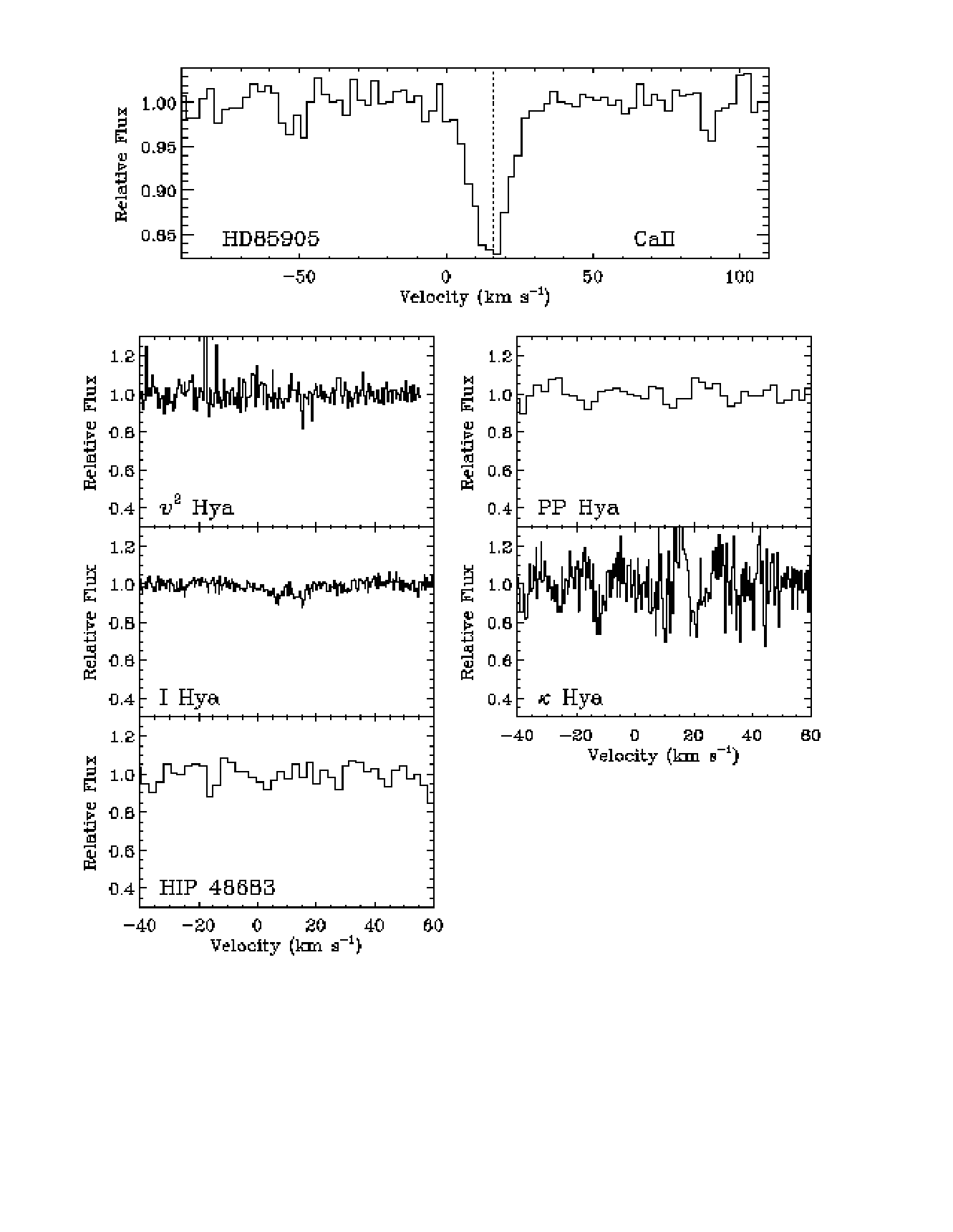}{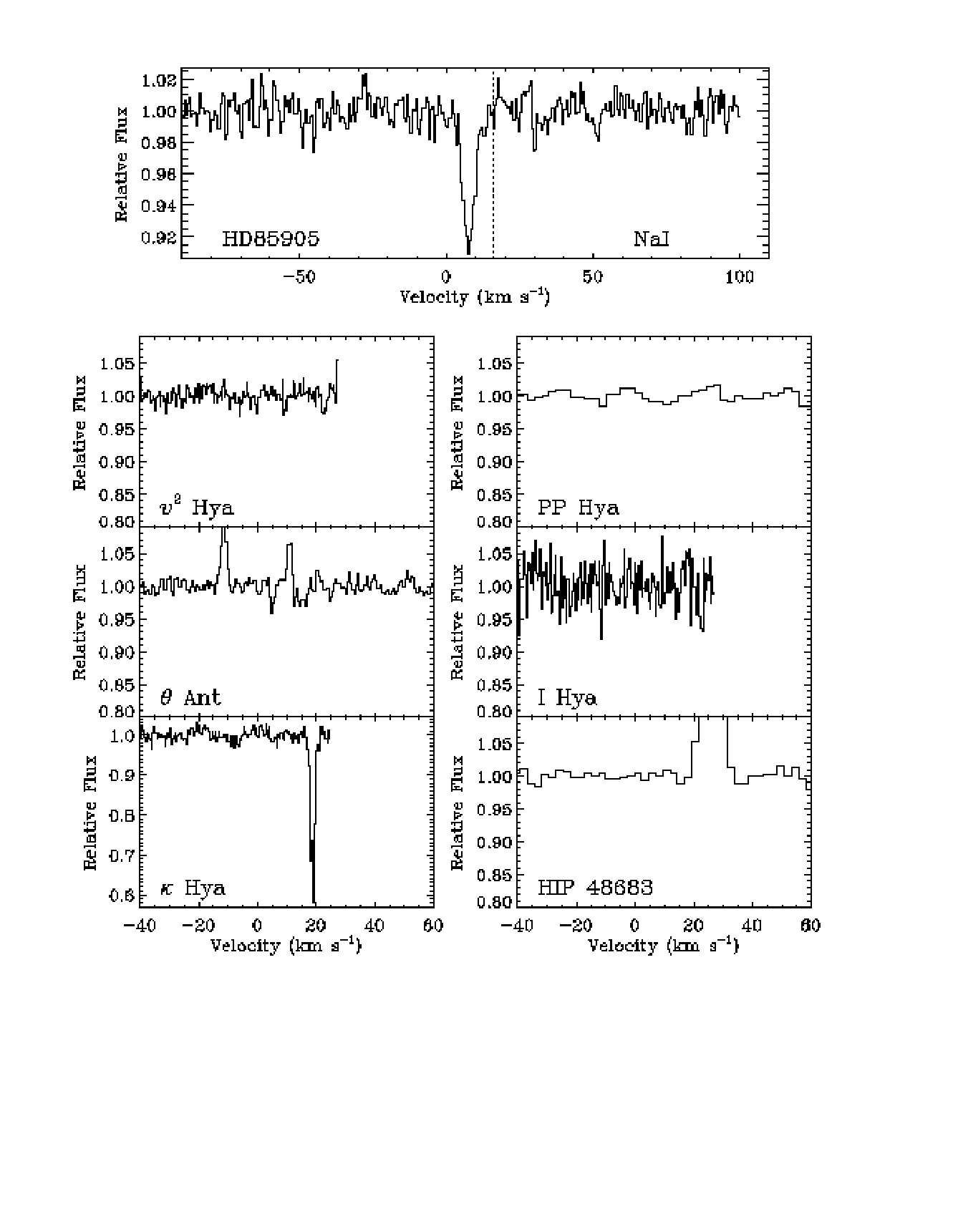}
\caption{Same as Figure~\ref{fig:nabaoph} for stars proximate to
HD85905.  \ion{Ca}{2} absorption is not detected toward any of the 5
targets, while \ion{Na}{1} is detected in 1 of 6.  It is likely that
the \ion{Na}{1} absorption observed toward HD85905 and $\kappa$ Hya is
due to interstellar absorption, even though they show a significant
projected velocity difference, due to the large projected separation
($\Delta r_{\rm POS} \leq 22$ pc) between these stars.
\label{fig:nabhd85905}}
\end{figure}

\begin{figure}
\epsscale{1.2}
\plottwo{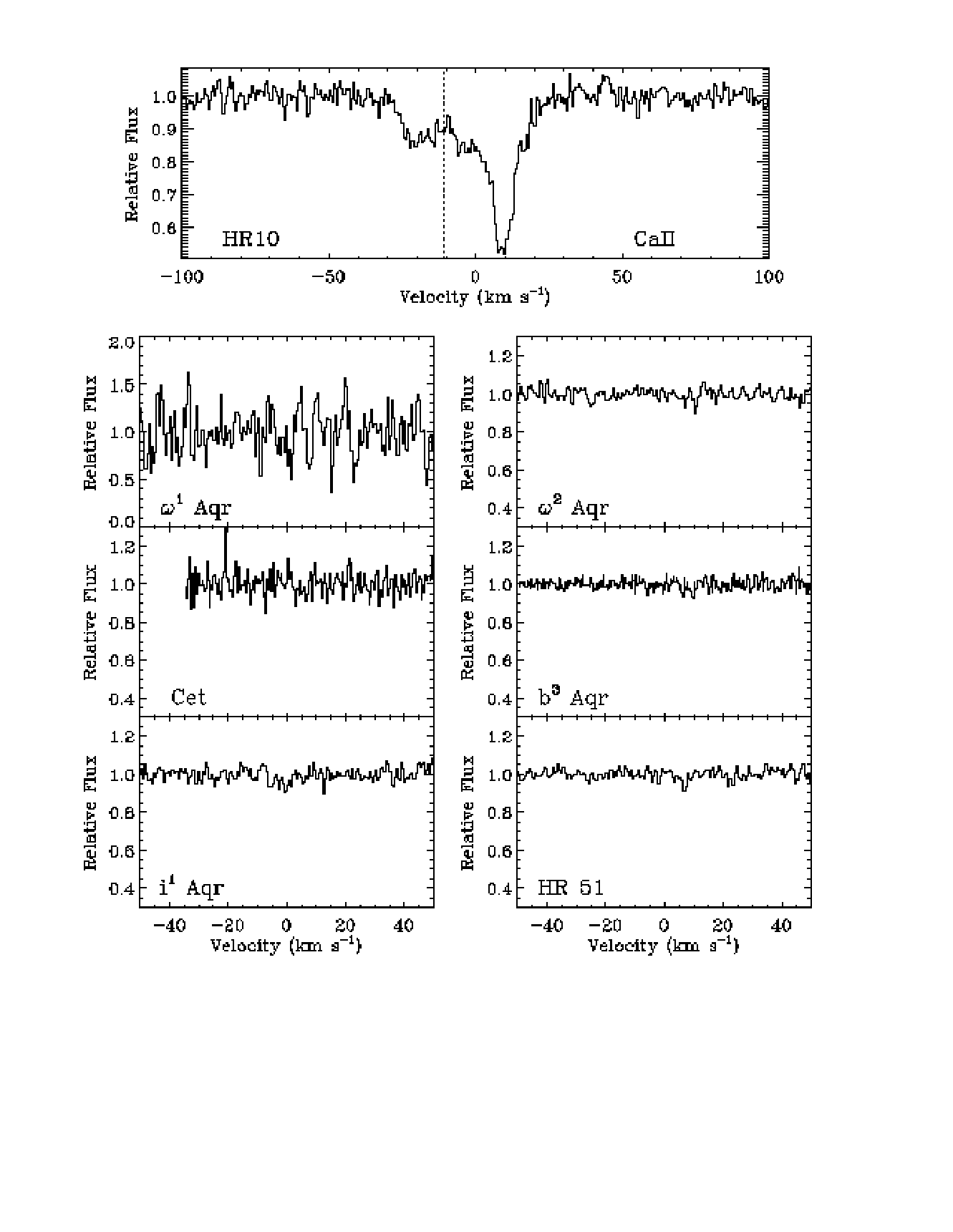}{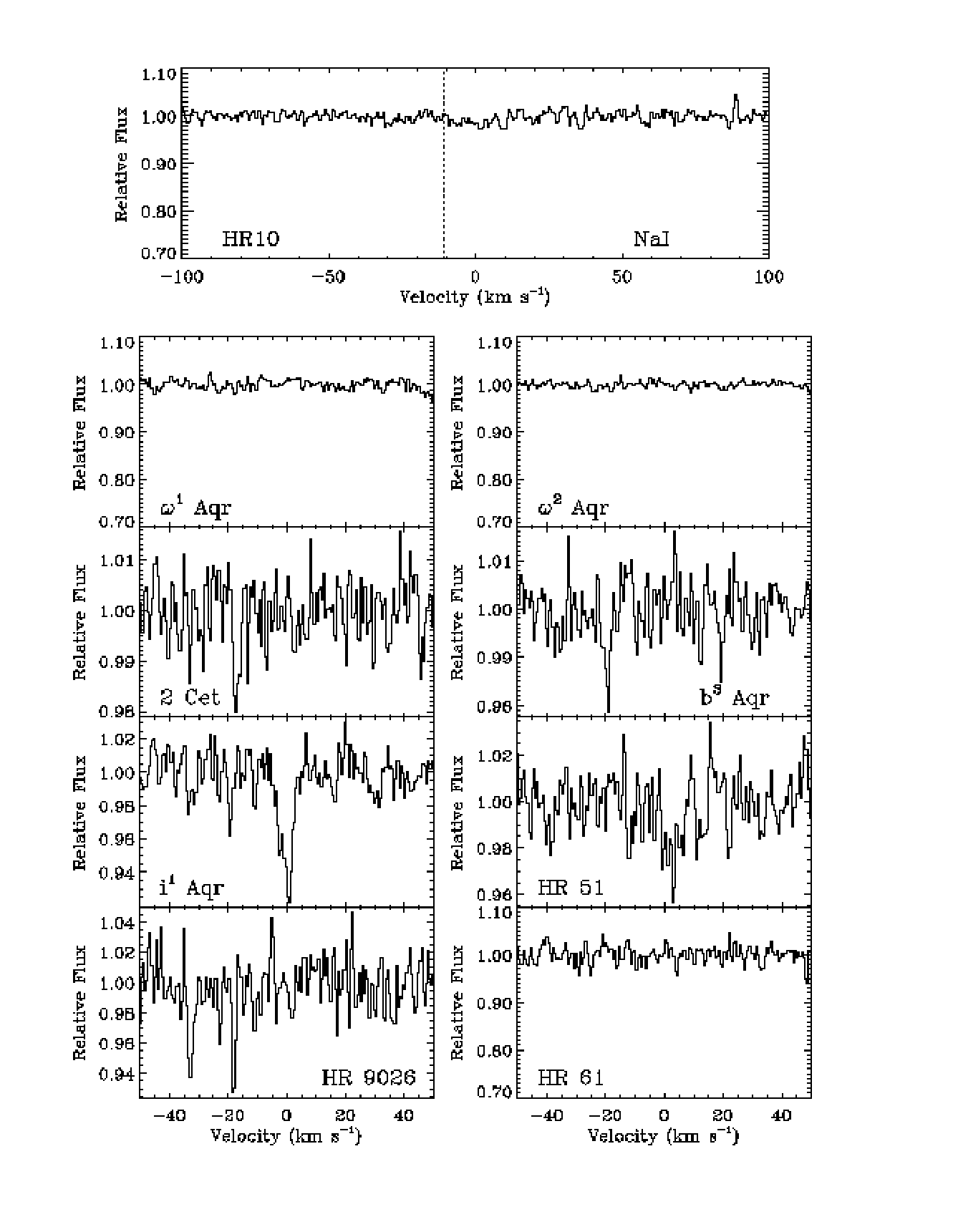}
\caption{Same as Figure~\ref{fig:nabaoph} for stars proximate to HR10.
No \ion{Ca}{2} absorption is detected toward any of 6 targets, while
\ion{Na}{1} is detected in 5 of 8 targets at two different velocities:
near --20 km~s$^{-2}$ and near 0 km~s$^{-2}$.  Our detection limits in
\ion{Ca}{2} are significantly lower than the detected absorption in
\ion{Ca}{2} toward HR10, emphasizing its circumstellar origin.
\label{fig:nabhr10}}
\end{figure}

\begin{figure}
\epsscale{0.8}
\plotone{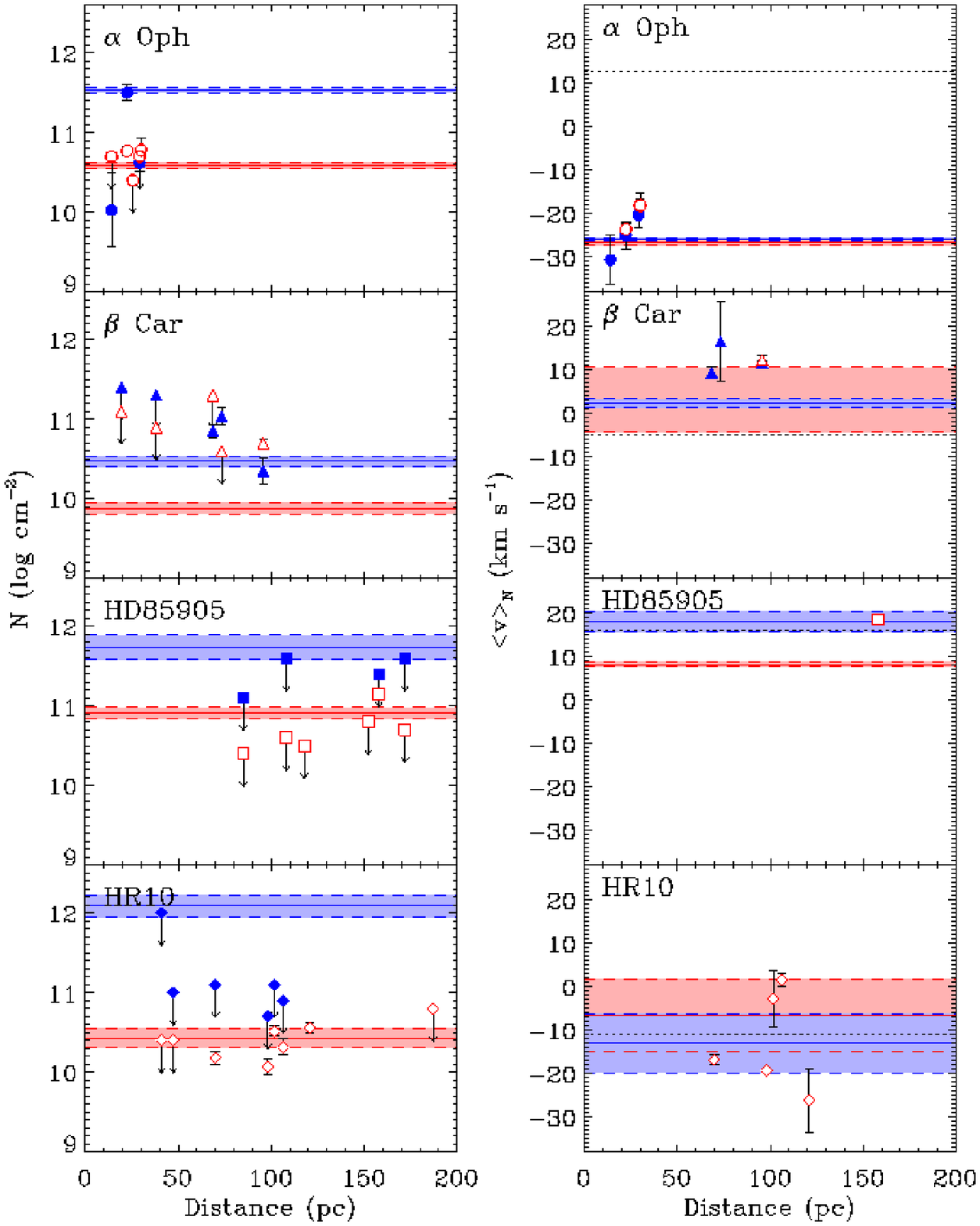}
\caption{Total column density ({\it left}) and column density weighted
velocity ({\it right}) measurements of absorption toward proximate
stars as a function of their distance.  The filled symbols indicate
\ion{Ca}{2} observations, while open symbols indicate \ion{Na}{1}.
The error bars are the weighted average variance, and in the case of
$\langle v \rangle_N$ indicate the range of observed velocities rather
than the error in measuring the central velocity of absorption.  The
horizontal dotted line in the right panels indicates the radial
velocity of the star.  The shaded regions indicate the weighted mean
and weighted average variance of the absorption detected in the
primary target, as presented in Figure~\ref{fig:tarvar} where blue
corresponds to \ion{Ca}{2}, and red with \ion{Na}{1}.  The data is
also presented in Table~\ref{tab:aodmeas}.
\label{fig:nabvar}}
\end{figure}

The \ion{Ca}{2} and \ion{Na}{1} spectral regions for proximate stars
are shown in Figures~\ref{fig:nabaoph}--\ref{fig:nabhr10}.  In order
to maintain small angular distances from our primary targets and to
sample a range of pertinent distances, we were severely limited in
choice of targets.  Often we had to push toward fainter and cooler
stars, resulting in lower $S/N$ observations and some contamination by
narrow stellar atmospheric lines, respectively.  Nonetheless, LISM
absorption is clearly detected in several (12/23, 52\%) of the
targets.  The total observed column density and column density
weighted velocity for the proximate stars are shown in
Figure~\ref{fig:nabvar}.  The weighted average values of our {\it
primary} stars are plotted as in Figure~\ref{fig:tarvar} in order to
make a direct comparison.

\subsubsection{The LISM Toward $\alpha$ Oph \label{sec:lismaoph}}

Inspection of Figures~\ref{fig:nabaoph} and \ref{fig:nabvar} and
Table~\ref{tab:aodmeas} indicate quite clearly that stars in close
proximity to $\alpha$ Oph show absorption in both ions at a similar
velocity and strength.  Indeed, the nearest star to $\alpha$ Oph,
HR6594, only 3.8$^{\circ}$ away, matches the anomalously high
\ion{Ca}{2} and \ion{Na}{1} column densities seen toward $\alpha$ Oph.
The projected distance between $\alpha$ Oph and HR6594 is only 0.9 pc,
at the distance of $\alpha$ Oph, which sets the maximum projected
distance between the location where these two sightlines probe the
absorbing medium.  If the absorbing medium is not as distant as
$\alpha$ Oph, the projected distance between the sightlines will be
even smaller.  The other proximate stars that show interstellar
absorption, albeit weaker than either $\alpha$ Oph or HR6594, range in
projected distance from $\alpha$ Oph from 1.8--2.1 pc in the
north-south direction.  Because no absorption is detected toward
HR6541, 6.7$^{\circ}$ or 1.7\,pc to the northeast, it appears that the
morphology of this cloud is quite elongated in the north-south
direction, similar to other clouds in the LISM (S. Redfield \&
J. Linsky 2006, in preparation).  A more detailed search would be
required to fully delineate the contours of this interesting cloud.

\citet{crawford01} presented a similar survey around $\alpha$ Oph, and
detected absorption at the same projected velocity as $\alpha$ Oph in
2 of their targets.  However, these targets were located at distances
of 120-211\,pc, a significantly greater distance than $\alpha$ Oph at
14.3\,pc.  In his ultra high resolution survey, severe restrictions in
the brightness of the background star resulted in a survey of targets
ranging in distance from 24.1--201 pc, and in angular separation from
0.6--13.2 degrees.  Due to the complex morphology of the interstellar
medium, particularly at the Local Bubble shell and beyond, which lies
only $\sim$55 pc in the direction of $\alpha$ Oph, the observed
spectra of distant targets is dominated by distant material, and one
becomes in effect ``confusion-limited'' in terms of identifying weak
absorption features, or separating overlapping absorption at coincident
projected velocity.  We limited the proximate neighbors to distances
$<$100 pc, with only one target beyond the Local Bubble, HR6341.
Interestingly, only this distant target, and HR6594, the closest star
to $\alpha$ Oph, show \ion{Na}{1} absorption.  Although the absorption
toward HR6341 is of comparable strength, it is likely caused by the
Local Bubble shell because it is significantly redshifted relative to
the strong absorption toward $\alpha$ Oph and HR6594.  Our survey,
using spectra at lower resolution (but higher $S/N$) than
\citet{crawford01}, was able to retain fainter targets close in both
angle and distance to $\alpha$ Oph.  Although the LISM absorption
toward $\alpha$ Oph remains an outlier in comparison to other nearby
stars, our mini-survey of the ISM along its line of sight indicates
that the absorption is due to intervening interstellar material
$<$14.3 pc from the Sun, rather than circumstellar material
surrounding $\alpha$ Oph.

\subsubsection{The LISM Toward $\beta$ Car, HD85905, and HR10\label{sec:3lism}}

The mini-surveys of the LISM near to our remaining 3 targets, $\beta$
Car, HD85905, and HR10, reveal that the majority of absorption
observed toward these targets is unlikely to be interstellar.

In the proximity of $\beta$ Car, \ion{Na}{1} is detected in only one
target, c Car, 95.7\,pc, which is also the only target predicted to be
beyond the Local Bubble shell located at $\sim$85 pc
\citet{lallement03}.  The two shortest sightlines, $\alpha$ Cha and
$\alpha$ Vol show no absorption in either ion, but are not sensitive
to the low column densities that are detected toward $\beta$ Car.  The
star nearest in projected distance to $\beta$ Car is $\theta$ Vol.
Absorption is detected toward this star, but the central velocity
(16.5 km~s$^{-1}$) is significantly different from that observed
toward $\beta$ Car (2.4 km~s$^{-1}$).  Indeed, all absorption
detections in proximate neighbors are redshifted by 7--14.1
km~s$^{-1}$.  Figure~\ref{fig:nabloc} indicates that since common
absorption is detected toward m Car and $\theta$ Vol, which bracket
$\beta$ Car and $\alpha$ Vol, it is likely that the interstellar
material responsible for the absorption is located 38--69 pc away,
between $\alpha$ Vol and m Car.  Observations of stars proximate to
$\beta$ Car indicate that the absorption observed toward $\beta$ Car
is not caused by the LISM.

A significant \ion{Ca}{2} column density is detected toward HD85905.
None of the observed neighboring stars show any \ion{Ca}{2}
absorption, despite that the column density upper limits are 1.4--8.7
times lower than the column observed toward HD85905.  The same is true
for \ion{Na}{1}, where the upper limits are 1.3--3.2 times lower.  One
target, $\kappa$ Hya at a distance of 158 pc, does show \ion{Na}{1}
absorption.  Half of the neighboring stars, and HD85905 itself, are
located beyond the predicted Local Bubble shell, which is located at
$\sim$120 pc in this direction \citep{lallement03}.  The two distant
neighboring stars that are closest in angular distance from HD85905, I
Hya and HIP48683, do not show any indication of Local Bubble shell
absorption.  However, HD85905 itself, has a constant \ion{Na}{1}
feature, at a velocity significantly different than the \ion{Ca}{2}
absorption.  It is possible this absorption is due to the Local Bubble
shell.  However, this requires a patchy morphology of Local Bubble
shell material in this direction because HD85905's nearest neighbors
show no \ion{Na}{1} absorption and the absorption toward the third
distant neighbor is at a different velocity than observed toward
HD85905.  Regardless of the exact nature of the \ion{Na}{1}
absorption, it is clear that the \ion{Ca}{2} absorption cannot be
explained by ISM absorption.
 
Similar to HD85905, the large \ion{Ca}{2} column density observed
toward HR10, is not detected in the proximate stars, despite column
density upper limits 1.2--25 times lower than observed toward HR10.
Very weak \ion{Na}{1} absorption is detect both in HR10 and in
neighboring stars.  Since HR10 is in the direction of the south
Galactic pole, and the Local Bubble is relatively unconstrained in
this direction \citep{lallement03}, no strong Local Bubble absorption
is expected.  We do start to detect weak absorption at distances $>$70
pc, but the velocity and strength of absorption varies across the 10
degree radius survey area.  Three targets show absorption near --20
km~s$^{-1}$ (2 Cet, b$^3$ Aqr, and HR9026) and two show absorption
near 0 km~s$^{-1}$ (i$^1$ Aqr and HR51).  Some of the \ion{Na}{1}
detected toward HR10 may be interstellar, but the variability of the
\ion{Na}{1} absorption toward HR10 indicates much of it is probably
circumstellar.  Again, it is clear that the \ion{Ca}{2} absorption
detected toward HR10 is not caused by the ISM along the line of sight.

\subsection{\ion{Ca}{2} to \ion{Na}{1} Ratio \label{sec:ca2na1}}

The ratio of \ion{Ca}{2} to \ion{Na}{1} has been used as a means of
discriminating between interstellar and circumstellar material
\citep[e.g.,][]{lagrangehenri90}.  ISM values are typically low, while
those observed toward $\beta$ Pic are much higher, such as
$N($\ion{Ca}{2}$)/N($\ion{Na}{1}$) = 38.9 \pm 10.9$, as measured by
\citet{hobbs85}, or even $>$100 as seen by \citet{welsh97}.  However,
other than for extremely high values (i.e., $>$50), it is difficult to
differentiate between circumstellar and interstellar material based on
this abundance ratio alone.  In the ISM, we see a wide range of
\ion{Ca}{2} to \ion{Na}{1} ratios, some approaching the ``high''
values seen toward $\beta$ Pic.  \citet*{welty96} compile a large
sample of ISM measurements, which are dominated by distant sightlines
(out to 1 kpc), and find a wide range of \ion{Ca}{2} to \ion{Na}{1}
ratios, from $\sim$0.003 to $\sim$50.  Even locally, a wide range of
values are measured.  For example, \citet{bertin93} found 8 stars
within 50 pc that showed both \ion{Ca}{2} to \ion{Na}{1} absorption.
Excluding $\alpha$ Oph, the ratio of \ion{Ca}{2} to \ion{Na}{1} ranges
from 2.2--11.9.  In general, calcium appears to be more strongly
effected by depletion onto dust grains than sodium \citep{savage96}.
Long sightlines likely sample a wide range of interstellar
environments, from cold, dense regions where a significant amount of
calcium will be depleted onto dust grains, leading to very low
\ion{Ca}{2} to \ion{Na}{1} ratios, to warm, shocked regions, in which
much of the calcium is maintained in the gas phase, and the abundance
ratio can be quite high.

Table~\ref{tab:aodmeas} includes the
$N($\ion{Ca}{2}$)/N($\ion{Na}{1}$)$ ratios for all our circumstellar
and interstellar observations.  For the three interstellar sightlines
that have both \ion{Ca}{2} and \ion{Na}{1} detections (HR6594, HR6341,
and c Car), the ratio ranges from 0.4--5.4.  Our circumstellar disk
candidates range from 3.9--46.  HR10, clearly a variable absorption
edge-on disk, has an abundance ratio
$N($\ion{Ca}{2}$)/N($\ion{Na}{1}$) = 46^{+23}_{-17}$ at the very high
end of the range, comparable with $\beta$ Pic.  However, the other
three edge-on disk candidates fall well within that found for LISM
sightlines \citep{bertin93}.  $\alpha$ Oph, which we argue is not an
edge-on disk, but a particularly small, high column density local
cloud, has a moderately high \ion{Ca}{2} to \ion{Na}{1} ratio, but is
consistent with other LISM sightlines.  While our other two edge-on
disk stars, HD85905 and HR10, which show variability and little
indication from neighboring sightlines that they are significantly
contaminated by ISM absorption, have relatively high \ion{Ca}{2} to
\ion{Na}{1} ratios, but not extreme enough to clearly differentiate
from the general ISM, on the basis of the ratio of \ion{Ca}{2} to
\ion{Na}{1} alone.  Note that due to the possibility that some of the
constant \ion{Na}{1} absorption observed toward HD85905 may be
interstellar, the \ion{Ca}{2} to \ion{Na}{1} ratio given in
Table~\ref{tab:aodmeas} should be considered a lower limit to the
circumstellar ratio of these two ions.

\subsection{Estimates of Physical Properties of the Variable Circumstellar Gas}

The observed temporal variability and lack of comparable absorption in
proximate neighbors demonstrates that most of the absorption detected
toward our 3 edge-on disk targets ($\beta$ Car, HD85905, and HR10) is
due to circumstellar material.  These targets have high $v\sin i$
velocities, and are likely viewed edge-on.  Since no circumstellar
absorption is detected in rapidly rotating intermediate inclination or
pole-on debris disks (e.g., Vega and $\alpha$ PsA;
\citeauthor{hobbs86} \citeyear{hobbs86}), it is likely that the
circumstellar gas is distributed in an edge-on disk.  An edge-on disk
morphology is confirmed in infrared and scattered light observations
of $\beta$ Pic \citep{smith84,heap00}.  However, the short and
long-term temporal variability in our 3 edge-on disk targets
demonstrate that the distribution of material in the gas disk is
clumpy.

Our observations are unable to constrain independently the physical
size of the absorbing gas structure or its density, other than the
absorbing material is presumably very close to the host star in order
to cause the observed short term temporal variability.  Since we don't
see a particularly stable component centered at the radial velocity of
the star, it is unlikely that the gas is smoothly distributed in any
extended disk structure, as observed in the stable component of
$\beta$ Pic by \citet{brandeker04}.  Instead, it is likely that the
absorbing gas is located between approximately 0.3--1.0\,AU
\citep{lagrange00}, and the maximum pathlength is on the order of
$\sim$1 AU.  If the gas absorption is caused by star-grazing families
of evaporating bodies as in the FEB model \citep{beust94}, the
pathlength through a gaseous coma-like structure could be
significantly less.  Although comet comae can reach sizes approaching
$\sim$1 AU \citep{jones00}, observations of nonblack saturated
variable absorption lines toward $\beta$ Pic indicate that the
absorbing material does not cover the entire stellar surface and is
likely to have a pathlength significantly less than 1 AU
\citep{vidalmadjar94}.  An upper limit to the amount of variable
absorbing gas around $\beta$ Car, HD85905, and HR10 can be estimated
if we assume it is distributed in a disk with an inner radius $l_1 =
0.3$\,AU and an outer radius of $l_2 = 1.0$\,AU.  The inner radius
($l_1$) is calculated at approximately 0.3 AU, due to the sublimation
of most types of grains at distances closer to the host star
\citep{mann06,vidalmadjar86}.

In order to convert our observable, $N_{\rm CaII}$, to a hydrogen
column density, we use the abundances measured for the stable
component of the disk around $\beta$ Pic \citep{roberge06}, where the
ratio $N($\ion{H}{1}$)/N($\ion{Ca}{2}$) \lesssim 2.4 \times 10^6$ is based on
$\beta$ Pic \ion{Ca}{2} measurements by \citet{crawford94} and
\ion{H}{1} limits by \citet{freudling95}.  The observed \ion{Ca}{2}
column density is assumed to be caused by circumstellar material only.
In this crude upper limit estimate, we assume the largest and simplest
configuration of gas closest to the star causing the variable gas
absorption.  The precise distribution of hydrogen gas in the
circumstellar disk is still highly uncertain \citep{brandeker04}.

If we assume the morphology of the disk is roughly cylindrical, the
total mass in the gas disk can be calculated from,
\begin{equation}
M_{\rm gas} \sim m_{\rm H} \frac{N_{\rm CaII}}{(l_2 - l_1)}~\frac{N_{\rm HI}}{N_{\rm CaII}}~\pi h (l^2_2 - l^2_1) \;\; [{\rm g}]\;,
\end{equation}
where $m_{\rm H}$ is the mass of a hydrogen atom, and $h$ is the
height of the disk and assumed to be equal to $0.2l_2$
\citep{hobbs85}.  Given the assumptions above, we calculate an upper
limit to the total gas mass of the variable component toward $\beta$
Car of $M_{\rm gas} \lesssim 4 \times 10^{-9} M_{\oplus}$, HD85905 of
$M_{\rm gas} \lesssim 7 \times 10^{-8} M_{\oplus}$, and HR10 of
$M_{\rm gas} \lesssim 2 \times 10^{-7} M_{\oplus}$.  In units of
$M_{\rm Halley} = 10^{17}$\,g \citep{whipple87}, the upper limits on
the variable gas component mass would be 9000, 4000, 200 $M_{\rm
Halley}$, respectively.  Note that due to the lack of constraints on
the distribution of the absorbing material, these are likely upper
limits to the total amount of variable component gas surrounding these
stars.


\section{Detections and Constraints on Circumstellar Disk Dust \label{sec:ddd}}

\begin{figure}
\includegraphics[angle=90,width=6.9in]{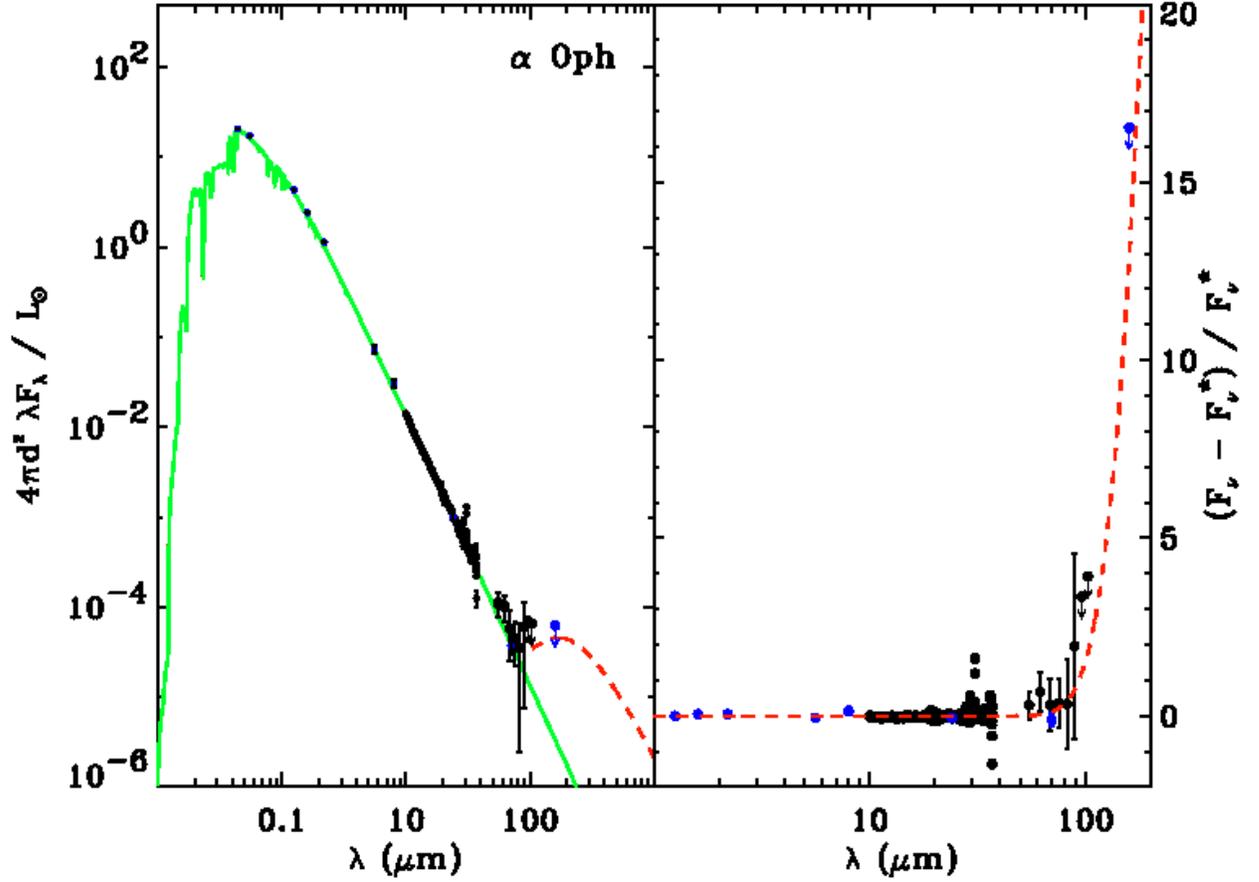}
\caption{SED ({\it left}) and (SED-star)/star ({\it right}) for
$\alpha$ Oph. Photometry from $B$, $V$, $J$, $H$, and $K$ bands and
IRAC and MIPS are shown as blue points, with black error bars
overlaid. IRS and MIPSSED spectra are shown in black.  NextGen models
of stellar photospheres specified by spectral type are overlaid in
green.  Blackbody fits to the maximum hypothetical excess still
allowed by our observations are depicted by red dashed lines.
\label{fig:sed_alphaoph}}
\end{figure}

\begin{figure}
\includegraphics[angle=90,width=6.9in]{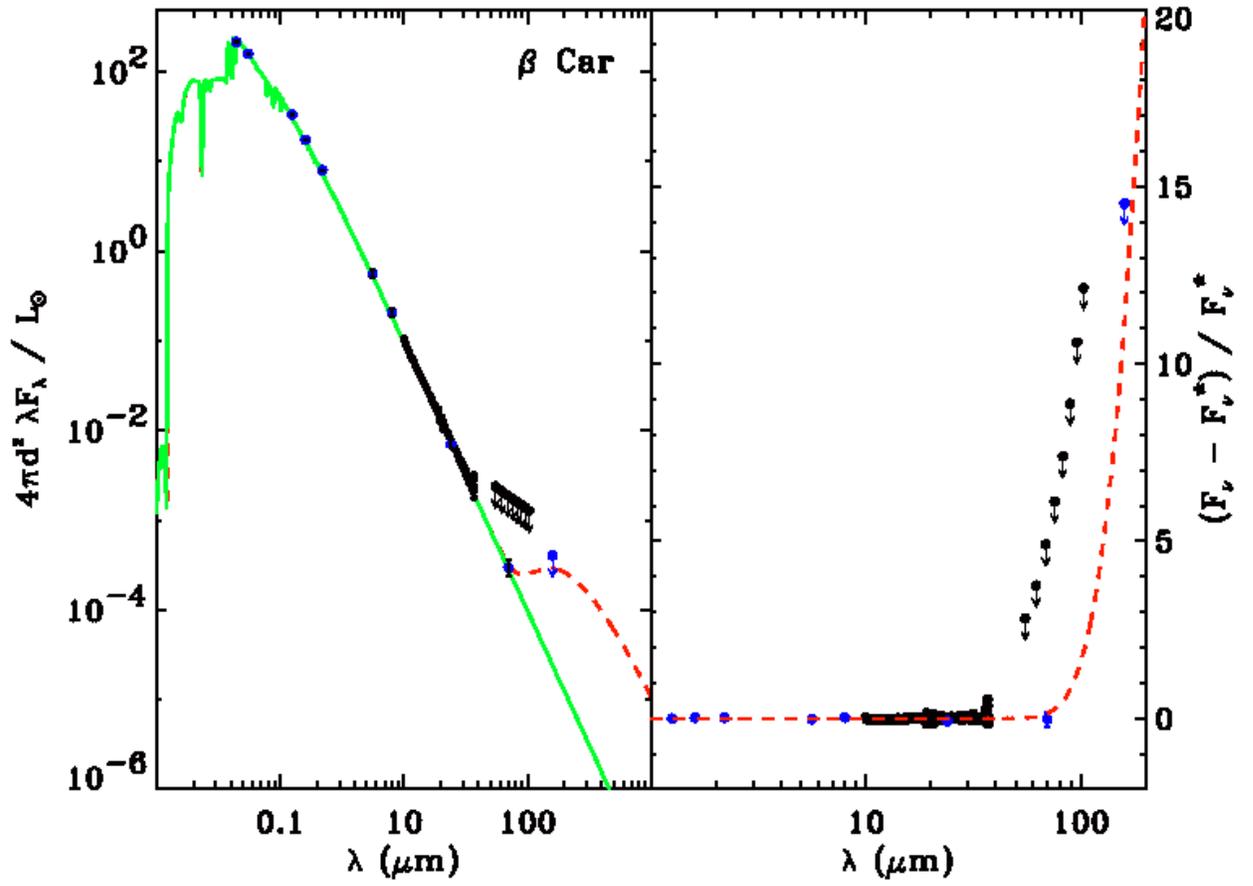}
\caption{SED ({\it left}) and (SED-star)/star ({\it right}) for $\beta$
Car. Symbols are the same as in Figure~\ref{fig:sed_alphaoph}.
\label{fig:sed_betacar} }
\end{figure}

\begin{figure}
\includegraphics[angle=90,width=6.9in]{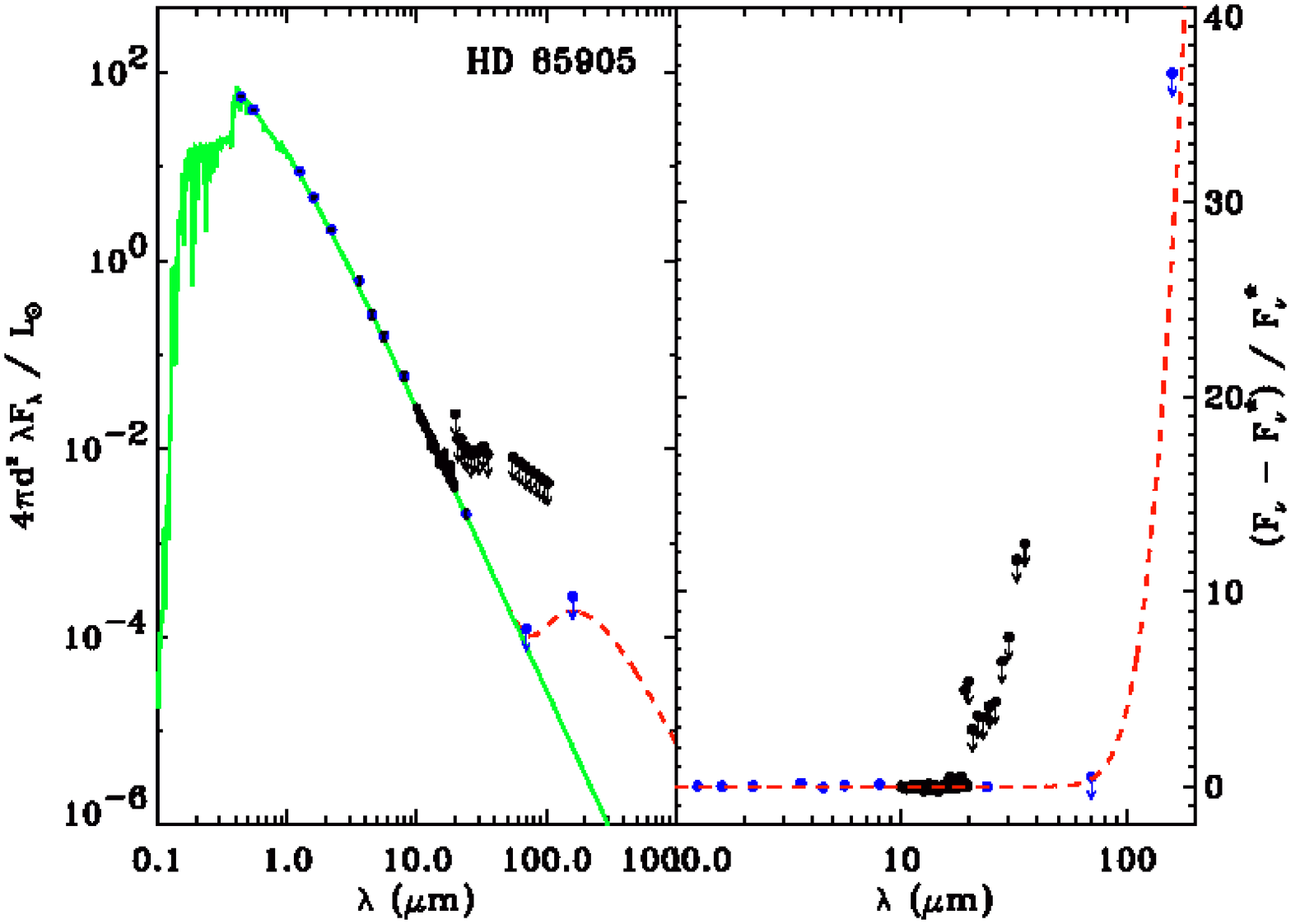}
\caption{SED ({\it left}) and (SED-star)/star ({\it right}) for
HD85905. Symbols are the same as in Figure~\ref{fig:sed_alphaoph}. 
IRS LH spectrum was not detected and a single upper limit equal to 
3$\times$RMS is plotted for each order. MIPS SED data are outside
the plotted region in the right panel.
\label{fig:sed_hd85905} }
\end{figure}

\begin{figure}
\includegraphics[angle=90,width=6.9in]{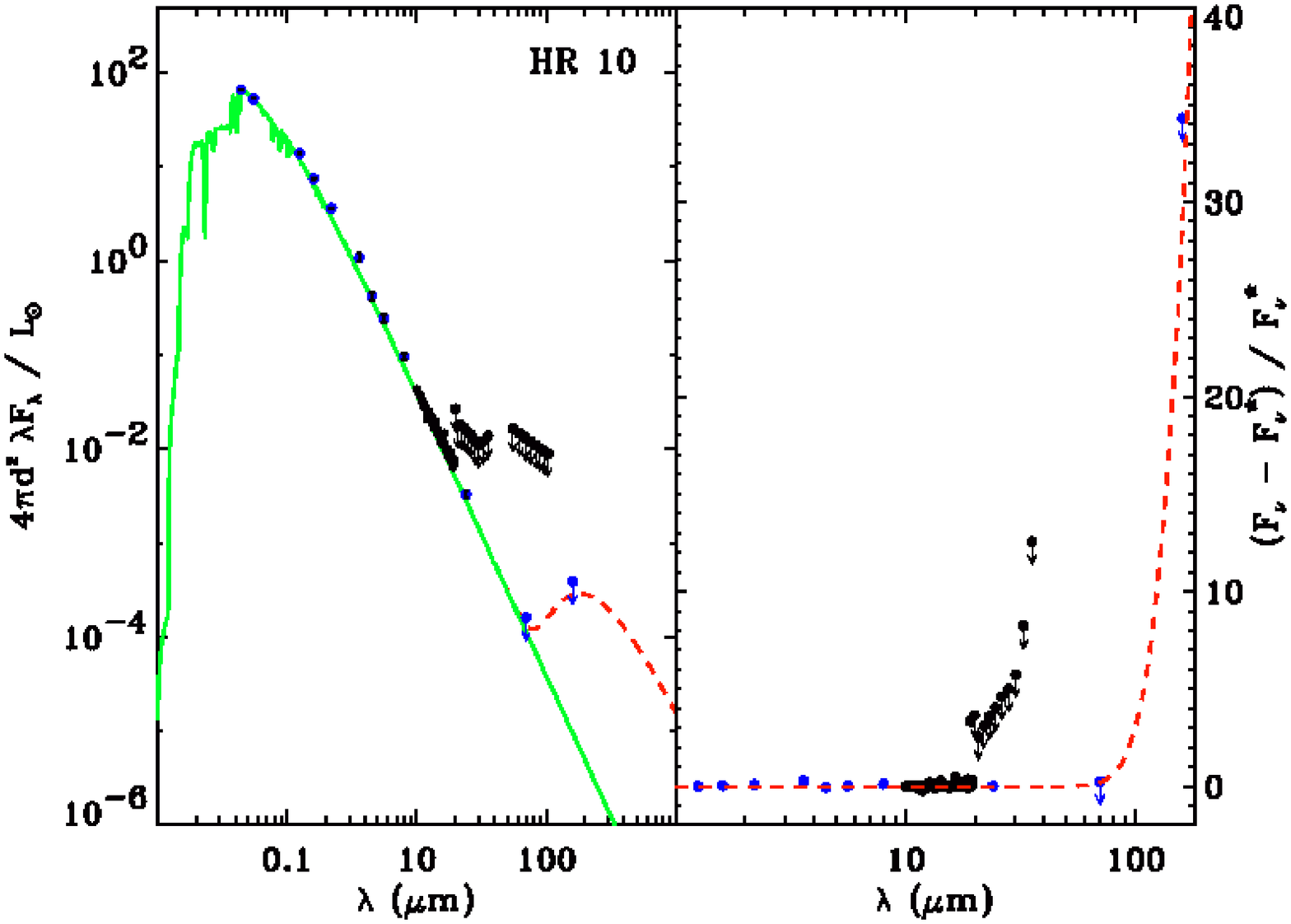}
\caption{SED ({\it left}) and (SED-star)/star ({\it right}) for HR10. Symbols
are the same as in Figure~\ref{fig:sed_alphaoph}.
IRS LH spectrum was not detected and a single upper limit equal to 
3$\times$RMS is plotted for each order. MIPS SED data are outside
the plotted region in the right panel.
\label{fig:sed_hr10} }
\end{figure}

The IR SEDs for all four sources are shown in the left panels of
Figures~\ref{fig:sed_alphaoph}--\ref{fig:sed_hr10}.  These SEDs
include $B$, $V$, $J$, $H$, and $K$ bands and IRAC and MIPS
photometric data (blue points) as well as IRS and MIPSSED spectra
(black lines).  $\alpha$ Oph and $\beta$ Car are detected at all bands
except 160 $\mu$m, while HR10 and HD85905 are not detected at 70 or
160 $\mu$m.  HR10 and HD85905 are also undetected in the IRS LH
module, and upper limits, equal to 3$\times$RMS are calculated for
each order.  The upper limits for MIPS 70 $\mu$m are more constraining
than are the upper limits for the LH $\sim19$--37 $\mu$m data, so the
latter are not used for fitting the excess.  All 4 sources appear
photospheric in all detected bands, so the upper limits for the 70
$\mu$m and/or 160 $\mu$m MIPS bands are used to place upper limits on
the temperature and amount of dust that may exist in disks around
these stars.  All optical and IR photometric measurements used in the
SEDs are given in Table~\ref{tab:obsflux}.

In order to model a debris disk around our primary targets, we must
estimate several stellar parameters, such as effective temperature
($T_{\star}$), luminosity ($L_{\star}$), radius ($R_{\star}$), gravity
($\log g$), mass ($M_{\star}$), and age.  The parameters used for our
models are listed in Table~\ref{tab:stparam}.  Two of our primary
targets, $\alpha$ Oph and $\beta$ Car, have excellent temperature and
radii measurements, and therefore well determined luminosities, since
they have been observed from the ultraviolet to the IR, and also have
radio angular diameter measurements and accurate distances
(\citeauthor{code76} \citeyear{code76}; \citeauthor{beeckmans77}
\citeyear{beeckmans77}; \citeauthor{malagnini86}
\citeyear{malagnini86}; \citeauthor*{richichi05}
\citeyear{richichi05}).  Stellar parameters for our other two primary
targets, HD85905 and HR10, are calculated using relations from
\citet*{napiwotzki93} and \citet{flower96}.  The spectral types were
used to estimate the gravity and stellar mass.  The age of these
systems was estimated using isochrones from \citet{bertelli94}.

The observed MIPS fluxes (and upper limits) 
and modeled stellar spectra are used to
constrain the amount of excess at dust temperatures comparable to the
Asteroid Belt ($T_{dust}=150$--250 K) and Kuiper Belt
($T_{dust}=30$--60 K), using the method described by
\citet{bryden_06}.  Assuming that the dust in the debris disk is well
represented by a single temperature, then the ratio of the observed
flux relative to the stellar flux can be used to calculate the ratio
of the total dust disk luminosity relative to the stellar luminosity
on the Rayleigh-Jeans tail of the stellar blackbody curve,
\begin{equation}
\frac{L_{dust}}{L_{\star}} = \frac{F_{dust}}{F_{\star}} \frac{kT_{dust}^4 (e^{{h{\nu}/{kT_{dust}}}}-1)}{{h}{\nu}{T_{\star}^3}}\;.
\end{equation}
Using temperatures of 100, 35, and 15 K, to correspond to blackbody
curves peaking at 24, 70 and 160 $\mu$m, we use the following
simplified expression to calculate $L_{dust}/L_{\star}$,
\begin{equation}
\label{eq:irsed}
\frac{L_{dust}}{L_{\star}} = \frac{F_{\lambda,dust}}{F_{\lambda,\star}} \left(\frac{5600 K}{T_{\star}}\right)^3 C_{\lambda}\;,
\end{equation}
where $C_{\lambda}$ is a constant that is dependent on the temperature
of the dust ($T_{dust}$) and wavelength and equal to
$3.7{\times}10^{-4}$, $1.5{\times}10^{-5}$, and $1.3{\times}10^{-6}$
for 24, 70 and 160 $\mu$m, respectively.  The flux ratios and
resulting $L_{dust}/L_{\star}$ estimates are listed in
Table~\ref{tab:dustdisk}.  Equation~\ref{eq:irsed} is used to
calculate upper limits for $L_{dust}/L_{\star}$ for regions of
temperatures corresponding to the 24, 70 and 160 $\mu$m photometry.
For the entire sample, we can limit $L_{dust}/L_{\star}$ to less than
$\sim$5000 times that of the Asteroid Belt \citep[$L_{dust}/L_{\star}
\sim 10^{-8}$ to $10^{-7}$;][]{dermott02}, by using the 24 $\mu$m
fluxes ($T\approx100$ K), and we can limit $L_{dust}/L_{\star}$ to
less than $\sim$18--30 times that of the Kuiper Belt
\citep[$L_{dust}/L_{\star} \sim 10^{-7}$ to $10^{-6}$;][]{stern96}, by
using the 70 $\mu$m fluxes ($T\approx35$ K).  For material at lower
temperatures than the Kuiper Belt, we can constrain
$L_{dust}/L_{\star}$ to less than $\sim$26--66 times that of the
Kuiper Belt, using the 160 $\mu$m fluxes ($T\approx15$ K).

The fractional excess emission ($F_{\nu}^{tot} - F_{\nu}^{star} /
F_{\nu}^{star}$) is plotted in the right panels of
Figures~\ref{fig:sed_alphaoph}--\ref{fig:sed_hr10}.  The stellar
photospheres are fit using NextGen model atmospheres
\citep*{hauschildt_99} matching the stellar parameters listed in
Table~\ref{tab:stparam} and scaled to match the J-band flux taken from
the literature, as the SEDs are clearly photospheric at 1.2 $\mu$m.
By making some assumptions about the distribution of the dust
contributing to the excess we can also calculate limits on the
luminosity of dust in these disks.  Once the stellar contribution has
been removed, a blackbody function is fit to the excess from
$\lambda=1$--160 $\mu$m.  As we are interested in calculating the
maximum possible excess in these fits, upper limits are ignored if
there is a detection or upper limit at longer wavelength with a
smaller fractional excess.  Fitting the excess with black body
function implies that the dust contributing to the excess lies in a
ring, of approximately constant temperature ($T_{dust}$), at a
corresponding distance, $D=(1/2) (T_{\star}/T_{dust})^2 R_{\star}$,
from the star.  The results of the fit, listed in
Table~\ref{tab:likechen}, are the temperature of the blackbody dust
($T_{dust}$), the solid angle subtended by the dust ($\Omega$) and the
reduced $\chi^2$ for each source.  The excess infrared luminosity is
calculated as $L_{IR} = 4 {\Omega} {\sigma} T_{dust}^4 d^2$, where $d$
is the distance in cm to each source and $\sigma$ is the
Stephan-Boltzmann constant.  All four sources possess fractional
luminosities, listed in Table~\ref{tab:likechen}, of
$L_{IR}/L_{\star}\le 5{\times}10^{-6}$, consistent with or less than
the fractional luminosities calculated above and similar to the least
luminous of known debris disks \citep{chen06}.


In order to place limits on the mass of dust surrounding these stars,
we need to make some assumptions about the dust properties
\citep[see][for a detailed description]{chen06}.  First, for
simplicity, we assume that the dust is composed primarily of
silicates, with a corresponding bulk density (${\rho}_s$) of 3.3 g
cm$^{-3}$.  (Note that changing the dust composition to include carbon
or silica grains would change the overall bulk density slightly to
${\rho}_s=3.5$ g cm$^{-3}$ and ${\rho}_s=2.3$ g cm$^{-3}$ for carbon
and silica grains, respectively.)  Next we assume that radiation
pressure removes grains smaller than $a_{min,0}$, thus setting the
minimum grain size.  Therefore, we can calculate the mass of small
grains with radii equal $a_{min,0}$, using the relation
$M_{dust}=(16/3) \pi (L_{IR}/L_{\star}) \rho_s D^2 a_{min,0}$
\citep[c.f.,][Equation 5]{chen06}.  This is a lower limit to the dust
surrounding these stars.  An upper limit to the mass of dust can be
found if we assume that the grain sizes follow the distribution $n_0
a^{-3.5}$ with a maximum radius of $a_{max} = 10$ cm and using the
relation $M_{10cm} = (4/3) \rho_s \sqrt{a_{min,0} a_{max}} d^2 \Omega$
\citep[c.f.,][Equation 6]{chen06}, where $d$ is the distance in cm to
the observed star.  As shown in Table~\ref{tab:likechen}, the masses
of material contributing to the measured excesses from these 4 disks
are not particularly small, ranging from $1{\times}10^{-3}$
$M_{\earth}$ to 2 $M_{\earth}$.  One must note, however, that this is
the mass of dust located at very large radii, $D\sim1000$-2400 AU from
the star.  The lack of significant excess at shorter wavelengths, and
thus smaller radii, suggests that much of the inner regions of these
systems have been effectively cleared, which would increase the
difficulty of feeding the variable gas material that is observed
around three of these stars, $\beta$ Car, HD85905, and HR10 (see
Sec~\ref{sec:var}).

\section{Constraints on Circumstellar Bulk Disk Gas \label{sec:ddg}}

We searched for several IR bulk gas phase atomic lines including
\ion{Ne}{2}, \ion{Ne}{3}, \ion{Fe}{1}, \ion{Fe}{2}, \ion{S}{1},
\ion{Si}{2}, as well as 4 ro-vibrational transitions of molecular
hydrogen, H$_2$ S(0)-S(4). Unfortunately, none were detected.  Upper
limits are presented in Tables~\ref{tab:gas}--\ref{tab:hgas}
($3{\times}RMS{\times}FWHM$, where $FWHM$ is the expected full width
at half maximum of an atomic emission line).  Upper limits for the
masses of gas in these disks are calculated from the H$_2$ S(0) and H$_2$ S(1) 
line upper limits, assuming local thermal equilibrium (LTE) and using
excitation temperatures of $T_{ex}=50$ K and 100 K, which are
appropriate if the dust and gas are roughly cospatial.  The total column
density of H$_2$ gas can be computed as
\begin{equation}
N_{tot}=\frac{F_{ul}\;\lambda_{ul}}{A_{ul}\;h\;c\;{\chi}_{u}}
\,\frac{4{\pi}}{\Omega} \;\;[{\rm cm}^{-2}]\;, 
\end{equation}
where $\lambda_{ul}$, $F_{ul}$, and $A_{ul}$ are the wavelength in cm,
observed flux upper limit in erg s$^{-1}$ cm$^{-2}$, and Einstein A
coefficient in s$^{-1}$ for the transition, $\Omega$ is the beam solid
angle in steradians, and $\chi_u$ is the fractional column density of
H$_2$ in the upper level,
\begin{equation}
\chi_{u}=\frac{N_u}{N_{tot}}=\frac{(2J+1)\;g{_N}}{Z(T_{ex})}\,e^{-E_{u}/kT_{ex}}\;.
\end{equation}
For H$_2$ S(0), $A_{ul}=2.94{\times}10^{-11}$ s$^{-1}$ and 
$E_{u}/k=510$ K and for H$_2$ S(1), 
$A_{ul}=4.76{\times}10^{-10}$ s$^{-1}$ and $E_{u}/k=1015$ K  
\citep*{wolniewicz_98}.  As transitions are restricted to
either ortho (parallel nuclear spin) or para (antiparallel nuclear
spin) states, if we assume ortho/para ${\approx}$ 3, then $g_{N} = 1$
for para transitions, such as H$_2$ S(0), and $g_{N} = 3$ for ortho
transitions, such as H$_2$ S(1).  Finally, the total mass of H$_2$ in
the disk can be calculated,
\begin{equation}
M_{tot}=N_{tot} \;m_{H_2}\; {\Omega} \;{d^2} \;\;[{\rm g}]\;,
\end{equation}
where, $m_{H_2}$ is the molecular weight of H$_2$ in grams and $d$ is
the distance to the disk in centimeters.  The upper limits for the
total column densities of H$_2$ are listed in the last four columns of
Table~\ref{tab:hgas}.  The cold gas mass constraints are
$<$2--100~$M_{\oplus}$ for $T=100$~K and $<$200--1$\times$10$^{6}$ 
$M_{\oplus}$ for $T=50$~K.  Note that the H$_2$ S(1) line is
located in a particularly noisy region of the SH spectrum, and thus
upper limits are much higher than for H$_2$ S(0).

The atomic sulfur line (\ion{S}{1}) at 25.23 $\mu$m may be a more
sensitive tracer of gas mass in low mass disks than
H$_2$. \citet{gorti04} find that for disks around G and K stars, with
gas masses of $10^{-3}$--1 $M_J$ and dust masses of
$10^{-7}$--$10^{-4}$ $M_J$, the strength of the \ion{S}{1} emission
line can be up to 1000 times that of the H$_2$~S(0) line, if both
lines are optically thin.  Using a line strength for H$_2$~S(0) of
1000 times less than the observed \ion{S}{1} upper limits, we find
that the gas mass could be $\sim$500 times less than the values
calculated from H$_2$~S(0) for $\alpha$ Oph and $\beta$ Car.
No estimates can be made for HD85905 and HR10, since these sources
were not detected in the LH IRS module, which contains both
the  \ion{S}{1} and H$_2$~S(0) lines.

\section{Discussion}

\subsection{Circumstellar or Interstellar?\label{sec:cirorism}}

Due to the similarity in spectral profile, a single spectrum is often
not sufficient to distinguish whether the absorption is caused by
circumstellar or interstellar material, or both.  Several techniques
have been used to determine the source of absorption. (1) Observations
of stars in close angular proximity and similar distance to the
primary target can be used to reconstruct the interstellar medium
along the line of sight to the primary target
\citep[e.g.,][]{crawford01}. (2) Repeated observations of the primary
target can be used to search for short term absorption variability
that is not observed in the large-scale structures of the interstellar
medium \citep[e.g,][]{lagrangehenri90hr10,petterson99}.  (3)
Observations of transitions, such as metastable lines, that are not
observed in the relatively low density interstellar medium 
can be used to indicate the presence of circumstellar material
\citep[e.g.,][]{kondo85,hobbs88}.

We present results using techniques (1) and (2).  The monitoring
campaign to search for temporal variability and mini-surveys of stars
in close angular proximity to our primary targets, clearly indicate
that 3 of our 4 targets are surrounded by circumstellar material.
Temporal variability is detected in our observations of $\beta$ Car,
HD85905, and HR10, confirming detections of variation in these objects
by \citet{lagrangehenri90hr10}, \citet{welsh98}, and \citet{hempel03}.
In addition, our survey of the ISM in the direction of these targets
indicate that little to none of the absorption can be attributed to
the interstellar medium.  Although it has been speculated that the
anomalously high absorption toward $\alpha$ Oph could be caused by
circumstellar material, we use nearby stars to firmly identify the
interstellar material that is responsible for the absorption toward
$\alpha$ Oph, confirming a similar study of more distant stars by
\citet{crawford01}.  Future work will entail looking for absorption
from metastable lines for our three targets that show evidence for
circumstellar gas.

\subsection{$\beta$ Pic-like Debris Disk or Be Star-like Stellar Wind Disk?}

The origin of the circumstellar gas in edge-on systems that show
absorption line variability is a long-standing question.  The systems
studied in this work may qualify as either weak debris disk systems
that currently have variable gas located very close to star but very
little dust, or weak winded, rapidly rotating, early type stars that
expel gas and form disks similar to classical Be stars.

The objects studied are relatively mature systems, older than $\beta$
Pic (which is $\sim$12 Myr), but roughly contemporaneous with Vega,
0.4 Gyr, and $\epsilon$~Eri, 0.6 Gyr \citep{zuckerman01}.  Although
most stars at this age have cleared their stellar systems of
primordial disk material, several are still in the evolutionary
transition period where they have retained a significant amount of
secondary dust and gas in their circumstellar surroundings.

The prototypical debris disks mentioned above all have IR excesses,
further evidence that the circumstellar material is processed from a
protostellar disk.  In addition, the one system that is oriented
edge-on ($\beta$ Pic) also shows gas absorption.  Our targets are all
A stars similar to $\beta$ Pic, and 3 of the 4 show gas absorption at
levels lower or comparable to $\beta$ Pic (e.g., $N_{\rm
CaII}($HR10$)/N_{\rm CaII}(\beta$ Pic$) \sim 0.65$).  However, the
fractional luminosities caused by an infrared excess consistent with
upper limits of the SEDs of our targets are lower than $\beta$ Pic by
more than 2 to 3 orders of magnitude \citep{backman93}.  No stable gas
component located at the stellar radial velocity is detected in our
targets, which since for $\beta$ Pic, the stable gas appears to be
associated with the bulk dust disk \citep{brandeker04}, is consistent
with our nondetection of any infrared excess.  However, there remains
the difficulty of feeding a variable gas component, most probably by
multitudes of star-grazing planetismal small bodies, without creating
an observable secondary dust disk through collisions of the same
bodies.

On the other hand, rapidly rotating B stars with strong radiatively
driven winds deposit a significant amount of gas into their
circumstellar environments.  These B stars often have strong emission
lines (e.g., hydrogen), and hence are classified as Be stars.  The ``Be''
phenomenon has also been observed in some early A stars and late O
stars, but peaks at spectral types B1--B2 \citep{porter03}.  A stars
can power weak radiatively driven winds, but the mass loss rates are
significantly smaller, $\dot{M} < 10^{-16} M_{\odot}$~yr$^{-1}$, and
only metals are expelled \citep{babel95}.  Over the ages of the stars
studied here, even with such a weak wind, enough mass can be delivered
to the circumstellar environment to be consistent with our
observations, although it does need to be retained relatively close
the star.  Our stars do not show any hydrogen emission lines in their
optical or infrared spectra, but as rapidly rotating early-type stars,
they may be able to produce an irregular circumstellar disk from
stellar winds.

Due to the similarity in signatures of gas disks in $\beta$ Pic-like
debris disks and Be star-like stellar wind disks, it is important to
keep in mind that it is difficult to distinguish the two based on gas
absorption lines alone.  In a study of rapidly rotating (i.e.,
edge-on) A stars similar to our sample, \citet{abt97} find that
$\sim$25\% of their stars show \ion{Ti}{2} absorption.  This is a
similar ratio as found for A stars with debris disks at 24 $\mu$m;
$\sim$32\% \citep{su06}.  \citet{abt97} argue that the observed
material cannot be remnants of star formation because they do not
observe absorption at all 3 epochs (spanning a total of 22 years) in 3
of their 7 \ion{Ti}{2} absorption stars.  However, it is plausible
that the mechanism causing variability in debris disks, such as
$\beta$ Pic, can be dramatic enough to result in nondetections of gas
absorption, particularly in weak sources, as 2 of their 3 variable
\ion{Ti}{2} absorption stars are.  Although the similarity of
detection fraction in these two studies may be a coincidence, it would
be interesting to search for IR excesses around these \ion{Ti}{2}
absorbers to look for any remnants of protostellar dust.

Detections of dust around our stars that have variable circumstellar
gas absorption would have strengthened their identification as debris
disk systems rather than stellar outflow disks.  However, IR excess
nondetections leave the origin of the observed gas disks an open
question.  The upper limits on the fractional IR luminosity could
still be consistent with a debris disk, albeit with much less dust
than debris disks like $\beta$ Pic, but still comparable to other {\it
Spitzer} debris disks \citep{chen06}.  At the same time, if there is
actually little or no dust in these systems, it is quite possible that
the origin of the variable circumstellar gas disk is stellar winds.
Expanding the sample and further monitoring of the gas content in
edge-on systems will help resolve this issue.

\section{Conclusions}

We present {\it Spitzer} infrared photometry and spectroscopy together
with high resolution optical spectra of 4 nearby stars that have
variable or anomalous optical absorption suspected to be due to
circumstellar material.  Our findings include:

\begin{enumerate}

\item The optical atomic absorption transitions of \ion{Ca}{2} and
\ion{Na}{1} were monitored toward all 4 stars at high spectral
resolution.  The observational baseline was more than 2.8 years.
Absorption line variability was detected in 3 of 4 targets, $\beta$
Car showed variability in \ion{Na}{1} while strong \ion{Ca}{2}
variability was detected toward HD85905 and HR10.  Our observations
add to previous studies of these same targets by other researchers,
which now extends the observed baseline to more than 20 years.

\item Night-to-night variability is detected toward HD85905 and HR10.
Although similar to the short term variability detected $\beta$ Pic,
the magnitude and frequency of the variations are lower toward HD85905
and HR10.

\item The fraction of the circumstellar absorption that is redshifted
relative to the radial velocity of the star ranges from 0--100\%.
Unlike $\beta$ Pic, the distribution of variable absorption toward our
targets is not heavy skewed to the red.

\item Mini-surveys (5--7 stars) of the LISM were conducted within
10$^{\circ}$ of each primary target.  We restricted our sample to
stars as close in distance as possible to our primary targets in order
to avoid contamination by more distant interstellar material.

\item In the direction of $\alpha$ Oph, we firmly identified the LISM
material that causes the anomalously high absorption seen, and thereby
show that circumstellar material is not responsible for the observed
absorption.  In particular, HR6594, the nearest star to $\alpha$ Oph
and only 35.5\,pc away, shows comparable absorption in \ion{Ca}{2} and
\ion{Na}{1}.  Absorption levels drop off rapidly indicating a small
and possibly filamentary LISM structure in that direction.  The lack
of variability and the extremely constraining IR excess measurements
support the lack of circumstellar material around $\alpha$ Oph.

\item The LISM in the direction of the other 3 targets, $\beta$ Car,
HD85905, and HR10, is responsible for little to none of the observed
absorption.  Only the constant \ion{Na}{1} feature see toward HD85905,
may be caused by material in the Local Bubble shell, and unrelated to
the circumstellar material around HD85905.

\item The \ion{Ca}{2} to \ion{Na}{1} ratio is measured for all stars.
Only HR10 shows an extremely high ratio, consistent with some of the
high values seen toward $\beta$ Pic.  The other targets show levels
that are high, but not inconsistent with LISM and circumstellar
measurements.  Unless $N($\ion{Ca}{2}$)/N($\ion{Na}{1}$) \gg 10$, the
variation in the interstellar ratio make it difficult to use this
ratio alone to determine if the absorbing material is circumstellar or
interstellar.  In this respect, the observed targets differ
significantly from $\beta$ Pic, which shows a strong IR excess and
stable gas absorption component.

\item We search for IR excesses with {\it Spitzer} in all 4 stars that
have shown variable or anomalous optical absorption.  We do not detect
any significant IR excesses in IRAC or MIPS photometry or IRS
spectroscopy, in any of the targets.  This is consistent with no
detection of a stable gas component at rest in the stellar reference
frame.

\item Sensitive measurements of the IR SEDs provide strong constraints
on the maximum possible dust luminosities (i.e., consistent with the
{\it Spitzer} upper limits at the longest IR wavelengths) of these
systems.  Fractional luminosity upper limits range from 1.8 to $5.4
\times 10^{-6}$, and are several orders of magnitude lower than
measured for $\beta$ Pic, despite that the gas absorption line column
densities are only slightly lower than those observed toward $\beta$
Pic.

\item No molecular hydrogen lines are detected in the IRS spectra, nor
are any atomic transitions detected.  Limits on the integrated line
fluxes for important transitions are provided.

\item We estimate upper limits to the mass of the variable gas
component causing the optical atomic absorption, that range from 0.4
to 20 $\times 10^{-8} M_{\oplus}$.  Combined with the nondetection and
tight constraints on any dust in these systems, the source of the
variable gas component remains an open question.  If evaporation of
small star-grazing objects are responsible for the variable gas
absorption, they are not contributing significantly to any dusty
debris disk.

\end{enumerate}

\acknowledgements 

S.R. acknowledges support provided by NASA through Hubble Fellowship
grant HST-HF-01190.01 awarded by the Space Telescope Science
Institute, which is operated by the Association of Universities for
Research in Astronomy, Inc., for NASA, under contract NAS 5-26555.
J.E.K-S is supported by the Spitzer Fellowship Program, provided by
NASA through contract 1256316, issued by the Jet Propulsion
Laboratory, California Institute of Technology, under NASA contract
1407.  This work is based in part on observations made with the {\emph
Spitzer Space Telescope}, which is operated by the Jet Propulsion
Laboratory, California Institute of Technology under a contract with
NASA. Support for this work was provided by NASA through an award
issued by JPL/Caltech.  We thank David Doss at McDonald Observatory
and Stuart Ryder at the Anglo-Australian Observatory for their
valuable assistance in acquiring the high resolution optical spectra.
We are grateful for the insightful comments and discussions with
Christine Chen, Aki Roberge, and Paul Harvey.  The helpful comments by
the anonymous referee were much appreciated.  This research has made
use of NASA's Astrophysics Data System Bibliographic Services.  This
research has made use of the SIMBAD database, operated at CDS,
Strasbourg, France.

{\it Facilities:} \facility{Spitzer (IRAC, MIPS, MIPSSED, IRS)},
\facility{Smith (CS12, CS21, CS23)}, \facility{AAT (UHRF)}


\begin{thebibliography}{98}
\expandafter\ifx\csname natexlab\endcsname\relax\def\natexlab#1{#1}\fi

\bibitem[{{Abt} \& {Morrell}(1995)}]{abt95}
{Abt}, H.~A., \& {Morrell}, N.~I. 1995, \apjs, 99, 135

\bibitem[{{Abt} {et~al.}(1997){Abt}, {Tan}, \& {Zhou}}]{abt97}
{Abt}, H.~A., {Tan}, H., \& {Zhou}, H. 1997, \apj, 487, 365

\bibitem[{{Alonso} {et~al.}(1994){Alonso}, {Arribas}, \&
  {Martinez-Roger}}]{alonso94}
{Alonso}, A., {Arribas}, S., \& {Martinez-Roger}, C. 1994, \aaps, 107, 365

\bibitem[{{Aumann}(1985)}]{aumann85}
{Aumann}, H.~H. 1985, \pasp, 97, 885

\bibitem[{{Babel}(1995)}]{babel95}
{Babel}, J. 1995, \aap, 301, 823

\bibitem[{{Backman} \& {Paresce}(1993)}]{backman93}
{Backman}, D.~E., \& {Paresce}, F. 1993, in Protostars and Planets III, ed.
  E.~H. {Levy} \& J.~I. {Lunine}, 1253--1304

\bibitem[{{Barrado y Navascues}(1998)}]{barradoynavascues98}
{Barrado y Navascues}, D. 1998, \aap, 339, 831

\bibitem[{{Beeckmans}(1977)}]{beeckmans77}
{Beeckmans}, F. 1977, \aap, 60, 1

\bibitem[{{Beichman} {et~al.}(2006)}]{beichman06}
{Beichman}, C.~A., {et~al.} 2006, \apj, 639, 1166

\bibitem[{{Bertelli} {et~al.}(1994){Bertelli}, {Bressan}, {Chiosi}, {Fagotto},
  \& {Nasi}}]{bertelli94}
{Bertelli}, G., {Bressan}, A., {Chiosi}, C., {Fagotto}, F., \& {Nasi}, E. 1994,
  \aaps, 106, 275

\bibitem[{{Bertin} {et~al.}(1993){Bertin}, {Lallement}, {Ferlet}, \&
  {Vidal-Madjar}}]{bertin93}
{Bertin}, P., {Lallement}, R., {Ferlet}, R., \& {Vidal-Madjar}, A. 1993, \aap,
  278, 549

\bibitem[{{Beust}(1994)}]{beust94}
{Beust}, H. 1994, in Circumstellar Dust Disks and Planet Formation, ed.
  R.~{Ferlet} \& A.~{Vidal-Madjar} (Gif sur Yvette Cedex: Editions
  Fronti\`{e}res), 35

\bibitem[{{Beust} \& {Morbidelli}(2000)}]{beust00}
{Beust}, H., \& {Morbidelli}, A. 2000, Icarus, 143, 170

\bibitem[{{Bevington} \& {Robinson}(1992)}]{bevington92}
{Bevington}, P.~R., \& {Robinson}, D.~K. 1992, {Data Reduction and Error
  Analysis for the Physical Sciences}, 2nd edn. (New York: McGraw-Hill)

\bibitem[{{Bouchet} {et~al.}(1991){Bouchet}, {Schmider}, \&
  {Manfroid}}]{bouchet91}
{Bouchet}, P., {Schmider}, F.~X., \& {Manfroid}, J. 1991, \aaps, 91, 409

\bibitem[{{Brandeker} {et~al.}(2004){Brandeker}, {Liseau}, {Olofsson}, \&
  {Fridlund}}]{brandeker04}
{Brandeker}, A., {Liseau}, R., {Olofsson}, G., \& {Fridlund}, M. 2004, \aap,
  413, 681

\bibitem[{{Bryden} {et~al.}(2006)}]{bryden_06}
{Bryden}, G., {et~al.} 2006, \apj, 636, 1098

\bibitem[{{Chen} {et~al.}(2005)}]{chen05}
{Chen}, C.~H., {et~al.} 2005, \apj, 634, 1372

\bibitem[{{Chen} {et~al.}(2006)}]{chen06}
---. 2006, \apjs, 166, 351

\bibitem[{{Cheng} {et~al.}(1992){Cheng}, {Bruhweiler}, {Kondo}, \&
  {Grady}}]{cheng92}
{Cheng}, K.-P., {Bruhweiler}, F.~C., {Kondo}, Y., \& {Grady}, C.~A. 1992,
  \apjl, 396, L83

\bibitem[{{Code} {et~al.}(1976){Code}, {Bless}, {Davis}, \& {Brown}}]{code76}
{Code}, A.~D., {Bless}, R.~C., {Davis}, J., \& {Brown}, R.~H. 1976, \apj, 203,
  417

\bibitem[{{Crawford}(2001)}]{crawford01}
{Crawford}, I.~A. 2001, \mnras, 327, 841

\bibitem[{{Crawford} {et~al.}(1998){Crawford}, {Beust}, \&
  {Lagrange}}]{crawford98}
{Crawford}, I.~A., {Beust}, H., \& {Lagrange}, A.-M. 1998, \mnras, 294, L31

\bibitem[{{Crawford} \& {Dunkin}(1995)}]{crawford95}
{Crawford}, I.~A., \& {Dunkin}, S.~K. 1995, \mnras, 273, 219

\bibitem[{{Crawford} {et~al.}(1994){Crawford}, {Spyromilio}, {Barlow}, {Diego},
  \& {Lagrange}}]{crawford94}
{Crawford}, I.~A., {Spyromilio}, J., {Barlow}, M.~J., {Diego}, F., \&
  {Lagrange}, A.~M. 1994, \mnras, 266, L65

\bibitem[{{Dermott} {et~al.}(2002){Dermott}, {Durda}, {Grogan}, \&
  {Kehoe}}]{dermott02}
{Dermott}, S.~F., {Durda}, D.~D., {Grogan}, K., \& {Kehoe}, T.~J.~J. 2002, in
  Asteroids III, ed. W.~F. {Bootke Jr,} {et~al.} (Tucson: Univ. Arizona Press),
  423

\bibitem[{{Diego}(1993)}]{diego93}
{Diego}, F. 1993, \ao, 32, 6284

\bibitem[{{Diego} {et~al.}(1995)}]{diego95}
{Diego}, F., {et~al.} 1995, \mnras, 272, 323

\bibitem[{{Dole} {et~al.}(2004)}]{dole04}
{Dole}, H., {et~al.} 2004, \apjs, 154, 93

\bibitem[{{Fazio} {et~al.}(2004)}]{fazio04}
{Fazio}, G.~G., {et~al.} 2004, \apjs, 154, 10

\bibitem[{{Ferlet} {et~al.}(1987){Ferlet}, {Vidal-Madjar}, \&
  {Hobbs}}]{ferlet87}
{Ferlet}, R., {Vidal-Madjar}, A., \& {Hobbs}, L.~M. 1987, \aap, 185, 267

\bibitem[{{Flower}(1996)}]{flower96}
{Flower}, P.~J. 1996, \apj, 469, 355

\bibitem[{{Freudling} {et~al.}(1995){Freudling}, {Lagrange}, {Vidal-Madjar},
  {Ferlet}, \& {Forveille}}]{freudling95}
{Freudling}, W., {Lagrange}, A.-M., {Vidal-Madjar}, A., {Ferlet}, R., \&
  {Forveille}, T. 1995, \aap, 301, 231

\bibitem[{{Gomes} {et~al.}(2005){Gomes}, {Levison}, {Tsiganis}, \&
  {Morbidelli}}]{gomes05}
{Gomes}, R., {Levison}, H.~F., {Tsiganis}, K., \& {Morbidelli}, A. 2005, \nat,
  435, 466

\bibitem[{{Gorti} \& {Hollenbach}(2004)}]{gorti04}
{Gorti}, U., \& {Hollenbach}, D. 2004, \apj, 613, 424

\bibitem[{{Grenier} {et~al.}(1999){Grenier}, {Burnage}, {Faraggiana},
  {Gerbaldi}, {Delmas}, {G{\'o}mez}, {Sabas}, \& {Sharif}}]{grenier99}
{Grenier}, S., {Burnage}, R., {Faraggiana}, R., {Gerbaldi}, M., {Delmas}, F.,
  {G{\'o}mez}, A.~E., {Sabas}, V., \& {Sharif}, L. 1999, \aaps, 135, 503

\bibitem[{{Harvey} {et~al.}(2004){Harvey}, {Cieza}, {Spiesman}, \& {c2d
  Team}}]{harvey04}
{Harvey}, P., {Cieza}, L., {Spiesman}, W., \& {c2d Team}. 2004, American
  Astronomical Society Meeting Abstracts, 204, 4132

\bibitem[{{Hauschildt} {et~al.}(1999){Hauschildt}, {Allard}, \&
  {Baron}}]{hauschildt_99}
{Hauschildt}, P.~H., {Allard}, F., \& {Baron}, E. 1999, \apj, 512, 377

\bibitem[{{Heap} {et~al.}(2000){Heap}, {Lindler}, {Lanz}, {Cornett}, {Hubeny},
  {Maran}, \& {Woodgate}}]{heap00}
{Heap}, S.~R., {Lindler}, D.~J., {Lanz}, T.~M., {Cornett}, R.~H., {Hubeny}, I.,
  {Maran}, S.~P., \& {Woodgate}, B. 2000, \apj, 539, 435

\bibitem[{{Hempel} \& {Schmitt}(2003)}]{hempel03}
{Hempel}, M., \& {Schmitt}, J.~H.~M.~M. 2003, \aap, 408, 971

\bibitem[{{Hobbs}(1986)}]{hobbs86}
{Hobbs}, L.~M. 1986, \apj, 308, 854

\bibitem[{{Hobbs} {et~al.}(1985){Hobbs}, {Vidal-Madjar}, {Ferlet}, {Albert}, \&
  {Gry}}]{hobbs85}
{Hobbs}, L.~M., {Vidal-Madjar}, A., {Ferlet}, R., {Albert}, C.~E., \& {Gry}, C.
  1985, \apjl, 293, L29

\bibitem[{{Hobbs} {et~al.}(1988){Hobbs}, {Welty}, {Lagrange-Henri}, {Ferlet},
  \& {Vidal-Madjar}}]{hobbs88}
{Hobbs}, L.~M., {Welty}, D.~E., {Lagrange-Henri}, A.~M., {Ferlet}, R., \&
  {Vidal-Madjar}, A. 1988, \apjl, 334, L41

\bibitem[{{Holweger} {et~al.}(1999){Holweger}, {Hempel}, \&
  {Kamp}}]{holweger99}
{Holweger}, H., {Hempel}, M., \& {Kamp}, I. 1999, \aap, 350, 603

\bibitem[{{Houck} {et~al.}(2004)}]{houck04}
{Houck}, J.~R., {et~al.} 2004, \apjs, 154, 18

\bibitem[{{Jenkins} \& {Tripp}(2001)}]{jenkins01}
{Jenkins}, E.~B., \& {Tripp}, T.~M. 2001, \apjs, 137, 297

\bibitem[{{Jones} {et~al.}(2000){Jones}, {Balogh}, \& {Horbury}}]{jones00}
{Jones}, G.~H., {Balogh}, A., \& {Horbury}, T.~S. 2000, \nat, 404, 574

\bibitem[{{Kalas} {et~al.}(2004){Kalas}, {Liu}, \& {Matthews}}]{kalas04}
{Kalas}, P., {Liu}, M.~C., \& {Matthews}, B.~C. 2004, Science, 303, 1990

\bibitem[{{Kondo} \& {Bruhweiler}(1985)}]{kondo85}
{Kondo}, Y., \& {Bruhweiler}, F.~C. 1985, \apjl, 291, L1

\bibitem[{{Lagrange} {et~al.}(2000){Lagrange}, {Backman}, \&
  {Artymowicz}}]{lagrange00}
{Lagrange}, A.-M., {Backman}, D.~E., \& {Artymowicz}, P. 2000, in Protostars
  and Planets IV, ed. V.~{Mannings}, A.~P. {Boss}, \& S.~S. {Russell} (Tucson:
  Univ. Arizona Press), 639

\bibitem[{{Lagrange-Henri} {et~al.}(1990{\natexlab{a}}){Lagrange-Henri},
  {Beust}, {Ferlet}, {Vidal-Madjar}, \& {Hobbs}}]{lagrangehenri90hr10}
{Lagrange-Henri}, A.~M., {Beust}, H., {Ferlet}, R., {Vidal-Madjar}, A., \&
  {Hobbs}, L.~M. 1990{\natexlab{a}}, \aap, 227, L13

\bibitem[{{Lagrange-Henri} {et~al.}(1990{\natexlab{b}}){Lagrange-Henri},
  {Ferlet}, {Vidal-Madjar}, {Beust}, {Gry}, \& {Lallement}}]{lagrangehenri90}
{Lagrange-Henri}, A.~M., {Ferlet}, R., {Vidal-Madjar}, A., {Beust}, H., {Gry},
  C., \& {Lallement}, R. 1990{\natexlab{b}}, \aaps, 85, 1089

\bibitem[{{Lallement} {et~al.}(1993){Lallement}, {Bertin}, {Chassefiere}, \&
  {Scott}}]{lallement93}
{Lallement}, R., {Bertin}, P., {Chassefiere}, E., \& {Scott}, N. 1993, \aap,
  271, 734

\bibitem[{{Lallement} {et~al.}(1986){Lallement}, {Vidal-Madjar}, \&
  {Ferlet}}]{lallement86}
{Lallement}, R., {Vidal-Madjar}, A., \& {Ferlet}, R. 1986, \aap, 168, 225

\bibitem[{{Lallement} {et~al.}(2003){Lallement}, {Welsh}, {Vergely}, {Crifo},
  \& {Sfeir}}]{lallement03}
{Lallement}, R., {Welsh}, B.~Y., {Vergely}, J.~L., {Crifo}, F., \& {Sfeir}, D.
  2003, \aap, 411, 447

\bibitem[{{Levato} \& {Malaroda}(1970)}]{levato70}
{Levato}, H., \& {Malaroda}, S. 1970, \pasp, 82, 741

\bibitem[{{Makovoz} \& {Khan}(2005)}]{mopex}
{Makovoz}, D., \& {Khan}, I. 2005, in Astronomical Society of the Pacific
  Conference Series, ed. P.~{Shopbell}, M.~{Britton}, \& R.~{Ebert}, 81

\bibitem[{{Malagnini} {et~al.}(1986){Malagnini}, {Morossi}, {Rossi}, \&
  {Kurucz}}]{malagnini86}
{Malagnini}, M.~L., {Morossi}, C., {Rossi}, L., \& {Kurucz}, R.~L. 1986, \aap,
  162, 140

\bibitem[{{Mann} {et~al.}(2006){Mann}, {K{\"o}hler}, {Kimura}, {Cechowski}, \&
  {Minato}}]{mann06}
{Mann}, I., {K{\"o}hler}, M., {Kimura}, H., {Cechowski}, A., \& {Minato}, T.
  2006, \aapr, 13, 159

\bibitem[{{Napiwotzki} {et~al.}(1993){Napiwotzki}, {Schoenberner}, \&
  {Wenske}}]{napiwotzki93}
{Napiwotzki}, R., {Schoenberner}, D., \& {Wenske}, V. 1993, \aap, 268, 653

\bibitem[{{Petterson} \& {Tobin}(1999)}]{petterson99}
{Petterson}, O.~K.~L., \& {Tobin}, W. 1999, \mnras, 304, 733

\bibitem[{{Porter} \& {Rivinius}(2003)}]{porter03}
{Porter}, J.~M., \& {Rivinius}, T. 2003, \pasp, 115, 1153

\bibitem[{{Redfield}(2006)}]{redfield06}
{Redfield}, S. 2006, in ASP Conf. Ser. 352, New Horizons in Astronomy, Frank N.
  Bash Symposium 2005, ed. S.~J. {Kannappan}, S.~{Redfield}, J.~E.
  {Kessler-Silacci}, M.~{Landriau}, \& N.~{Drory} (San Francisco: ASP), 79

\bibitem[{{Redfield}(2007)}]{redfield07hd32297}
{Redfield}, S. 2007, \apjl, 656, L97

\bibitem[{{Redfield} \& {Linsky}(2001)}]{redfield01}
{Redfield}, S., \& {Linsky}, J.~L. 2001, \apj, 551, 413

\bibitem[{{Redfield} \& {Linsky}(2002)}]{redfield02}
---. 2002, \apjs, 139, 439

\bibitem[{{Redfield} \& {Linsky}(2004)}]{redfield04tt}
---. 2004, \apj, 613, 1004

\bibitem[{{Richichi} {et~al.}(2005){Richichi}, {Percheron}, \&
  {Khristoforova}}]{richichi05}
{Richichi}, A., {Percheron}, I., \& {Khristoforova}, M. 2005, \aap, 431, 773

\bibitem[{{Rieke} {et~al.}(2004)}]{rieke04}
{Rieke}, G.~H., {et~al.} 2004, \apjs, 154, 25

\bibitem[{{Roberge} {et~al.}(2002){Roberge}, {Feldman}, {Lecavelier des
  Etangs}, {Vidal-Madjar}, {Deleuil}, {Bouret}, {Ferlet}, \&
  {Moos}}]{roberge02}
{Roberge}, A., {Feldman}, P.~D., {Lecavelier des Etangs}, A., {Vidal-Madjar},
  A., {Deleuil}, M., {Bouret}, J.-C., {Ferlet}, R., \& {Moos}, H.~W. 2002,
  \apj, 568, 343

\bibitem[{{Roberge} {et~al.}(2006){Roberge}, {Feldman}, {Weinberger},
  {Deleuil}, \& {Bouret}}]{roberge06}
{Roberge}, A., {Feldman}, P.~D., {Weinberger}, A.~J., {Deleuil}, M., \&
  {Bouret}, J.-C. 2006, \nat, 441, 724

\bibitem[{{Roberge} {et~al.}(2005){Roberge}, {Weinberger}, {Redfield}, \&
  {Feldman}}]{roberge05}
{Roberge}, A., {Weinberger}, A.~J., {Redfield}, S., \& {Feldman}, P.~D. 2005,
  \apjl, 626, L105

\bibitem[{{Savage} \& {Sembach}(1991)}]{savage91}
{Savage}, B.~D., \& {Sembach}, K.~R. 1991, \apj, 379, 245

\bibitem[{{Savage} \& {Sembach}(1996)}]{savage96}
---. 1996, \araa, 34, 279

\bibitem[{{Schechter} {et~al.}(1993){Schechter}, {Mateo}, \&
  {Saha}}]{schechter93}
{Schechter}, P.~L., {Mateo}, M., \& {Saha}, A. 1993, \pasp, 105, 1342

\bibitem[{{Schneider} {et~al.}(2005){Schneider}, {Silverstone}, \&
  {Hines}}]{schneider05}
{Schneider}, G., {Silverstone}, M.~D., \& {Hines}, D.~C. 2005, \apjl, 629, L117

\bibitem[{{Silverstone} {et~al.}(2006)}]{silverstone06}
{Silverstone}, M.~D., {et~al.} 2006, \apj, 639, 1138

\bibitem[{{Slettebak}(1975)}]{slettebak75}
{Slettebak}, A. 1975, \apj, 197, 137

\bibitem[{{Slettebak}(1988)}]{slettebak88}
---. 1988, \pasp, 100, 770

\bibitem[{{Smith} \& {Terrile}(1984)}]{smith84}
{Smith}, B.~A., \& {Terrile}, R.~J. 1984, Science, 226, 1421

\bibitem[{{\it Spitzer} Science Center(2006)}]{mipsdatahandbook}
{\it Spitzer} Science Center. 2006, {\it Spitzer Space Telescope} Multiband
  Imaging Photometer for Spitzer (MIPS) Data Handbook Version 3.2.1 (Pasadena:
  SSC), http://ssc.spitzer.caltech.edu/mips

\bibitem[{{Stern}(1996)}]{stern96}
{Stern}, S.~A. 1996, \aap, 310, 999

\bibitem[{{Su} {et~al.}(2006)}]{su06}
{Su}, K.~Y.~L., {et~al.} 2006, ArXiv Astrophysics e-prints

\bibitem[{{Th{\'e}bault} \& {Beust}(2001)}]{thebault01}
{Th{\'e}bault}, P., \& {Beust}, H. 2001, \aap, 376, 621

\bibitem[{{Tody}(1993)}]{tody93}
{Tody}, D. 1993, in ASP Conf. Ser. 52: Astronomical Data Analysis Software and
  Systems II, ed. R.~J. {Hanisch}, R.~J.~V. {Brissenden}, \& J.~{Barnes} (San
  Francisco: ASP), 173

\bibitem[{{Tull}(1972)}]{tull72}
{Tull}, R.~G. 1972, in Proc. ESO/CERN Conference on Auxiliary Instrumentation
  for Large Telescopes (Geneva: ESO), 259

\bibitem[{{Tull} {et~al.}(1995){Tull}, {MacQueen}, {Sneden}, \&
  {Lambert}}]{tull95}
{Tull}, R.~G., {MacQueen}, P.~J., {Sneden}, C., \& {Lambert}, D.~L. 1995,
  \pasp, 107, 251

\bibitem[{{Vidal-Madjar} {et~al.}(1986){Vidal-Madjar}, {Ferlet}, {Hobbs},
  {Gry}, \& {Albert}}]{vidalmadjar86}
{Vidal-Madjar}, A., {Ferlet}, R., {Hobbs}, L.~M., {Gry}, C., \& {Albert}, C.~E.
  1986, \aap, 167, 325

\bibitem[{{Vidal-Madjar} {et~al.}(1998){Vidal-Madjar}, {Lecavelier des Etangs},
  \& {Ferlet}}]{vidalmadjar98}
{Vidal-Madjar}, A., {Lecavelier des Etangs}, A., \& {Ferlet}, R. 1998, \planss,
  46, 629

\bibitem[{{Vidal-Madjar} {et~al.}(1994)}]{vidalmadjar94}
{Vidal-Madjar}, A., {et~al.} 1994, \aap, 290, 245

\bibitem[{{Welsh} {et~al.}(1998){Welsh}, {Craig}, {Crawford}, \&
  {Price}}]{welsh98}
{Welsh}, B.~Y., {Craig}, N., {Crawford}, I.~A., \& {Price}, R.~J. 1998, \aap,
  338, 674

\bibitem[{{Welsh} {et~al.}(1997){Welsh}, {Craig}, {Jelinsky}, \&
  {Sasseen}}]{welsh97}
{Welsh}, B.~Y., {Craig}, N., {Jelinsky}, S., \& {Sasseen}, T. 1997, \aap, 321,
  888

\bibitem[{{Welsh} {et~al.}(1994){Welsh}, {Craig}, {Vedder}, \&
  {Vallerga}}]{welsh94}
{Welsh}, B.~Y., {Craig}, N., {Vedder}, P.~W., \& {Vallerga}, J.~V. 1994, \apj,
  437, 638

\bibitem[{{Welsh} {et~al.}(1999){Welsh}, {Sfeir}, {Sirk}, \&
  {Lallement}}]{welsh99}
{Welsh}, B.~Y., {Sfeir}, D.~M., {Sirk}, M.~M., \& {Lallement}, R. 1999, \aap,
  352, 308

\bibitem[{{Welty} {et~al.}(1994){Welty}, {Hobbs}, \& {Kulkarni}}]{welty94}
{Welty}, D.~E., {Hobbs}, L.~M., \& {Kulkarni}, V.~P. 1994, \apj, 436, 152

\bibitem[{{Welty} {et~al.}(1996){Welty}, {Morton}, \& {Hobbs}}]{welty96}
{Welty}, D.~E., {Morton}, D.~C., \& {Hobbs}, L.~M. 1996, \apjs, 106, 533

\bibitem[{{Werner} {et~al.}(2004)}]{werner04}
{Werner}, M.~W., {et~al.} 2004, \apjs, 154, 1

\bibitem[{{Whipple}(1987)}]{whipple87}
{Whipple}, F.~L. 1987, \aap, 187, 852

\bibitem[{{Wolniewicz} {et~al.}(1998){Wolniewicz}, {Simbotin}, \&
  {Dalgarno}}]{wolniewicz_98}
{Wolniewicz}, L., {Simbotin}, I., \& {Dalgarno}, A. 1998, \apjs, 115, 293

\bibitem[{{Zuckerman}(2001)}]{zuckerman01}
{Zuckerman}, B. 2001, \araa, 39, 549

\end{thebibliography}

\clearpage

\begin{deluxetable}{lllcllcclc}
\tablewidth{0pt}
\tabletypesize{\scriptsize}
\tablecaption{Stellar Properties of Primary Targets\tablenotemark{a} \label{tab:basics}}
\tablehead{ HD & Other & Spectral & $m_{\rm V}$ & $v_{\rm R}$ & $v \sin i$ & $l$ & $b$ & Distance\tablenotemark{b} & Dist. to LB Boundary\tablenotemark{c} \\
\# & Name & Type & (mag) & (km s$^{-1}$) & (km s$^{-1}$) & (deg) & (deg) & (pc) & (pc) }
\startdata
159561 & $\alpha$ Oph & A5III    & 2.10 & 12.6   & 228 & 35.9  &  22.6  & 14.32$^{+ 0.18}_{- 0.18}$ & $\sim$55\\
80007  & $\beta$ Car  & A2IV     & 1.70 & --5    & 145 & 286.0 & --14.4 & 34.08$^{+ 0.55}_{- 0.54}$ & $\sim$85\\
85905  & HR3921       & A2/A3III & 6.23 & 16\tablenotemark{d} & 264 & 257.8 &  24.5  & 140$^{+15}_{-13}$ & $\sim$120 \\
256    & HR10         & A2IV/V   & 6.23 & --10.8\tablenotemark{e} & 241 & 74.36 & --75.9 & 160$^{+25}_{-19}$ & $>$250 \\

\enddata
\tablenotetext{a}{All values from SIMBAD unless otherwise noted.}
\tablenotetext{b}{Distances calculated from {\it Hipparcos} parallaxes.}
\tablenotetext{c}{\citet{lallement03}}
\tablenotetext{d}{This work.}
\tablenotetext{e}{\citet{grenier99}}

\end{deluxetable}

\begin{deluxetable}{llllllcrr}
\tablewidth{0pt}
\tabletypesize{\scriptsize}
\tablecaption{Observational Parameters for Primary Stars\label{tab:optobs}}
\tablehead{ HD & Other & Date & Telescope\tablenotemark{a} & Instrument\tablenotemark{b} & Ion & $v_{\rm atm}$\tablenotemark{c} & Exposure & $S/N$  \\
\# & Name & & & & & (km s$^{-1}$) & (s) & }
\startdata
159561 & $\alpha$ Oph & 2003 Oct 6  & McD2.7 & CS12 & CaII & \nodata & 3600 & 42 \\
159561 & $\alpha$ Oph & 2004 Apr 7  & McD2.7 & CS12 & CaII & \nodata & 2400 & 36 \\
159561 & $\alpha$ Oph & 2004 Aug 27 & McD2.7 & CS21 & CaII & \nodata & 1800 & 227 \\
159561 & $\alpha$ Oph & 2004 Aug 27 & McD2.7 & CS21 & CaII & \nodata & 1800 & 112 \\
159561 & $\alpha$ Oph & 2004 Aug 28 & McD2.7 & CS21 & CaII & \nodata & 1800 & 263 \\
159561 & $\alpha$ Oph & 2004 Aug 28 & McD2.7 & CS21 & CaII & \nodata & 1800 & 136 \\
159561 & $\alpha$ Oph & 2004 Aug 29 & McD2.7 & CS21 & CaII & \nodata & 1800 & 115 \\
159561 & $\alpha$ Oph & 2004 Aug 30 & McD2.7 & CS21 & CaII & \nodata & 1800 & 156 \\
\enddata
\tablenotetext{a}{McD2.7: the Harlan J. Smith 2.7m Telescope at McDonald Observatory; AAT3.9: the Anglo-Australian 3.9m Telescope at the Anglo-Australian Observatory.}
\tablenotetext{b}{CS12: Coud\'{e} double-pass Spectrometer ($R\,\sim\,$500,000); CS21: Cross-Dispersed Echelle Spectrometer (2D Coud\'{e}) Focus 1 ($R\,\sim\,$240,000); CS23: Cross-Dispersed Echelle Spectrometer (2D Coud\'{e}) Focus 3 ($R\,\sim\,$60,000); UHRF: Ultra High Resolution Facility ($R\,\sim\,$1,000,000).}
\tablenotetext{c}{Projected velocity of the Earth's atmosphere at time of observation.  Occasionally, atmospheric NaI is detected in high $S/N$ spectra and can contaminate the circumstellar or interstellar absorption feature.  No atmospheric lines fall near the CaII feature.}
\end{deluxetable}  

\begin{deluxetable}{llllllcrr}
\tablewidth{0pt}
\tabletypesize{\scriptsize}
\tablecaption{Observational Parameters for Stars in Close Proximity to Primary Targets\label{tab:optobsnab}}
\tablehead{ HD & Other & Date & Telescope\tablenotemark{a} & Instrument\tablenotemark{b} & Ion & $v_{\rm atm}$\tablenotemark{c} & Exposure & $S/N$  \\
\# & Name & & & & & (km s$^{-1}$) & (s) & }
\startdata
165777 & 72 Oph   & 2004 Aug 27 & McD2.7 & CS21 & CaII & \nodata & 2700 & 147 \\
165777 & 72 Oph   & 2004 Oct 20 & McD2.7 & CS21 & NaI  & --22.4 & 1200 & 59 \\
165777 & 72 Oph   & 2005 Jun 15 & AAT3.9 & UHRF & CaII & \nodata & 3300 & 23 \\
165777 & 72 Oph   & 2005 Sep 9  & McD2.7 & CS12 & CaII & \nodata & 7200 & 19 \\
160910 & HR6594   & 2004 Aug 28 & McD2.7 & CS21 & CaII & \nodata & 1800 & 28 \\
160910 & HR6594   & 2005 Jul 16 & McD2.7 & CS21 & NaI  & --11.5 & 1200 & 54 \\
159332 & HR6541   & 2005 Jul 17 & McD2.7 & CS21 & NaI  & --12.3 & 1200 & 50 \\
154494 & 60 Her   & 2004 Oct 20 & McD2.7 & CS21 & CaII & \nodata & 1200 & 16 \\
\enddata
\tablenotetext{a}{McD2.7: the Harlan J. Smith 2.7m Telescope at McDonald Observatory; AAT3.9: the Anglo-Australian 3.9m Telescope at the Anglo-Australian Observatory.}
\tablenotetext{b}{CS12: Coud\'{e} double-pass Spectrometer ($R\,\sim\,$500,000); CS21: Cross-Dispersed Echelle Spectrometer (2D Coud\'{e}) Focus 1 ($R\,\sim\,$240,000); CS23: Cross-Dispersed Echelle Spectrometer (2D Coud\'{e}) Focus 3 ($R\,\sim\,$60,000); UHRF: Ultra High Resolution Facility ($R\,\sim\,$1,000,000).}
\tablenotetext{c}{Projected velocity of the Earth's atmosphere at time of observation.  Occasionally, atmospheric NaI is detected in high $S/N$ spectra and can contaminate the circumstellar or interstellar absorption feature.  No atmospheric lines fall near the CaII feature.}
\end{deluxetable}  

\begin{deluxetable}{lcccccccc}
\tablewidth{0pt}
\tabletypesize{\tiny}
\tablecaption{{\emph Spitzer} Observations \label{tab:obs}}
\tablehead{\colhead{}      & \colhead{IRAC}  & \colhead{} & \colhead{MIPS}  & \colhead{} & \colhead{IRS} & \colhead{} & \colhead{MIPSSED} & \colhead{} \\
\colhead{Source}      & \colhead{AOR Key} & \colhead{Date} & \colhead{AOR Key} & \colhead{Date} & \colhead{AOR Key} & \colhead{Date}  & \colhead{AOR Key} & \colhead{Date}} 
\startdata
$\alpha$ Oph	&11026688 &2004 Sep 03  &11026176            &2005 Apr 09               &11026944 &2004 Sep 28  &11026432 &2005 Sep 26 \\
$\beta$ Car	&11027712 &2005 Jan 20  &11027200, 13658624  &2005 Jan 28, 2005 May 20  &11027968 &2005 Jan 13  &11027456 &2005 Sep 26 \\
HD85905	        &11029760 &2004 Dec 17  &11029248, 13658880  &2004 Dec 04, 2005 May 20  &11030016 &2004 Dec 09  &11029504 &2005 Sep 26 \\
HR10		&11028736 &2004 Nov 25  &11028224, 13658368  &2004 Dec 05, 2005 Jun 23  &11028992 &2005 Jun 30  &11028480 &2005 Sep 26 \\

\enddata
\end{deluxetable}

\begin{deluxetable}{llllccll}
\tablewidth{0pt}
\tabletypesize{\scriptsize}
\tablecaption{High Resolution Absorption Line Measurement History \label{tab:longterm}}
\tablehead{ HD & Other & Ion & Date & $\langle v \rangle_N$\tablenotemark{a} & $\log N$ & Reference \\
\# & Name & & & (km s$^{-1}$) & (cm$^{-2}$) &}
\startdata
159561 & $\alpha$ Oph& CaII & 2006 Jul 06 & $-25.7 \pm 2.3$ & $11.49 \pm 0.04$ & this work\\
159561 & $\alpha$ Oph& CaII & 2006 Feb 14 & $-26.1 \pm 2.7$ & $11.55 \pm 0.01$ & this work\\
159561 & $\alpha$ Oph& CaII & 2006 Feb 13 & $-25.7 \pm 2.1$ & $11.44 \pm 0.02$ & this work\\
159561 & $\alpha$ Oph& CaII & 2005 Sep 14 & $-26.1 \pm 2.7$ & $11.57 \pm 0.01$ & this work\\
159561 & $\alpha$ Oph& CaII & 2005 Sep 11 & $-25.8 \pm 2.6$ & $11.51 \pm 0.01$ & this work\\
159561 & $\alpha$ Oph& CaII & 2005 Aug 18 & $-25.9 \pm 2.5$ & $11.50 \pm 0.02$ & this work\\
159561 & $\alpha$ Oph& CaII & 2005 Jul 18 & $-25.5 \pm 2.1$ & $11.52 \pm 0.01$ & this work\\
159561 & $\alpha$ Oph& CaII & 2005 Jul 15 & $-26.0 \pm 2.4$ & $11.53 \pm 0.01$ & this work\\
\enddata
\tablenotetext{a}{Column density weighted mean velocity, and an ``error'' which is the weighted average variance, essentially the range of velocities in which significant absorption is present.  The error in determining the velocity of any given absorption is typically 0.1--2.0 km~s$^{-1}$, depending on the spectral resolution of the observations.}
\end{deluxetable}  

\begin{deluxetable}{lllcllcclcc}
\tablewidth{0pt}
\tabletypesize{\scriptsize}
\tablecaption{Stellar Parameters for Targets Near $\alpha$ Oph\tablenotemark{a}\label{tab:basicaophnab}}
\tablehead{ HD & Other & Spectral & $m_{\rm V}$ & $v_{\rm R}$ & $v \sin i$ & $l$ & $b$ & Distance\tablenotemark{b} & $\Delta\theta$\tablenotemark{c} & $\Delta r_{\rm POS}$\tablenotemark{d}\\
\# & Name & Type & (mag) & (km s$^{-1}$) & (km s$^{-1}$) & (deg) & (deg) & (pc) & (deg) & (pc) }
\startdata
165777 & 72 Oph   & A4IVs & 3.72 & --23.9 & 65  & 36.6 & 14.1 & 25.38$^{+ 0.55}_{- 0.52}$ & 8.53 & 2.1 \\
160910 & HR6594   & F4Vw  & 5.55 & --43.7 & 30  & 40.0 & 22.4 & 35.5$^{+1.1}_{-1.0}$ & 3.79 & 0.9 \\
159332 & HR6541   & F6V   & 5.66 & --58.8 & 5   & 42.5 & 25.5 & 36.7$^{+1.0}_{-1.0}$ & 6.69 & 1.7 \\
154494 & 60 Her   & A4IV  & 4.88 & --4.2  & 117 & 32.8 & 29.2 & 44.1$^{+1.4}_{-1.4}$ & 7.16 & 1.8 \\
154228 & HR6341   & A1V   & 5.92 & --31.7 & 43  & 33.5 & 30.0 & 80.8$^{+4.8}_{-4.3}$ & 7.71 & 1.9 \\

\enddata
\tablenotetext{a}{All values from SIMBAD unless otherwise noted.}
\tablenotetext{b}{Distances calculated from {\it Hipparcos} parallaxes.}
\tablenotetext{c}{Angular distance from direction of $\alpha$ Oph.}
\tablenotetext{d}{Physical separation in the plane of the sky from $\alpha$ Oph, at the distance of $\alpha$ Oph.}
\end{deluxetable}

\begin{deluxetable}{lllcllcclcc}
\tablewidth{0pt}
\tabletypesize{\scriptsize}
\tablecaption{Stellar Parameters for Targets Near $\beta$ Car\tablenotemark{a}\label{tab:basicbcarnab}}
\tablehead{ HD & Other & Spectral & $m_{\rm V}$ & $v_{\rm R}$ & $v \sin i$ & $l$ & $b$ & Distance\tablenotemark{b} & $\Delta\theta$\tablenotemark{c} & $\Delta r_{\rm POS}$\tablenotemark{d}\\
\# & Name & Type & (mag) & (km s$^{-1}$) & (km s$^{-1}$) & (deg) & (deg) & (pc) & (deg) & (pc) }
\startdata
71243  & $\alpha$ Cha & F5III   & 4.07 & --13.7 & 0  & 289.9 & --21.7 & 19.46$^{+0.19}_{-0.18}$ & 8.19 & 2.8 \\
78045  & $\alpha$ Vol & A2.5IVm & 4.00 & 4.9    & 34 & 282.7 & --13.0 & 38.11$^{+0.76}_{-0.73}$ & 3.50 & 2.1 \\
83944  & m Car        & B9IV-V  & 4.51 & 20     & 68 & 281.9 & --6.6  & 68.6$^{+2.3}_{-2.1}$ & 8.78 & 5.3 \\
74405  & $\theta$ Vol & A0V     & 5.19 & 13     & \nodata & 284.7 & --17.1 & 73.3$^{+2.8}_{-2.6}$ & 2.98 & 1.8 \\
76728  & c Car        & B8.5II  & 3.80 & 25     & 0\tablenotemark{e} & 277.7 & --10.0 & 95.7$^{+4.5}_{-4.1}$ & 9.23 & 5.5 \\

\enddata
\tablenotetext{a}{All values from SIMBAD unless otherwise noted.}
\tablenotetext{b}{Distances calculated from {\it Hipparcos} parallaxes.}
\tablenotetext{c}{Angular distance from direction of $\beta$ Car.}
\tablenotetext{d}{Physical separation in the plane of the sky from $\beta$ Car, at the distance of the closest partner.}
\tablenotetext{e}{\citet{levato70}}
\end{deluxetable}  

\begin{deluxetable}{lllcllcclcc}
\tablewidth{0pt}
\tabletypesize{\scriptsize}
\tablecaption{Stellar Parameters for Targets Near HD85905\tablenotemark{a}\label{tab:basichd85905nab}}
\tablehead{ HD & Other & Spectral & $m_{\rm V}$ & $v_{\rm R}$ & $v \sin i$ & $l$ & $b$ & Distance\tablenotemark{b} & $\Delta\theta$\tablenotemark{c} & $\Delta r_{\rm POS}$\tablenotemark{d}\\
\# & Name & Type & (mag) & (km s$^{-1}$) & (km s$^{-1}$) & (deg) & (deg) & (pc) & (deg) & (pc)}
\startdata
87504  & $\upsilon^2$ Hya & B8V    & 4.59 & 28   & 50  & 252.5 & 33.1 & 85.0$^{+6.1}_{-5.3}$ & 9.77 & 14.6 \\
87130  & PP Hya           & A3III  & 6.83 & \nodata & \nodata & 257.8 & 27.3 & 108.1$^{+12.1}_{-9.9}$ & 2.80 & 5.3 \\
84367  & $\theta$ Ant     & A8V+...& 4.79 & 24.0 & \nodata & 259.9 & 19.1 & 117.8$^{+11.1}_{-9.3}$ & 5.77 & 11.9 \\
83953  & I Hya            & B5V    & 4.75 & 25.9 & 240 & 256.3 & 21.6 & 152$^{+17}_{-14}$ & 3.21 & 8\\
83754  & $\kappa$ Hya     & B4IV/V & 5.05 & 18   & 150 & 248.7 & 27.8 & 158$^{+27}_{-20}$ & 8.81 & 22 \\
86066  & HIP48683         & A2V    & 7.43 & \nodata & \nodata & 255.9 & 26.7 & 172$^{+26}_{-20}$ & 2.79 & 7\\

\enddata
\tablenotetext{a}{All values from SIMBAD unless otherwise noted.}
\tablenotetext{b}{Distances calculated from {\it Hipparcos} parallaxes.}
\tablenotetext{c}{Angular distance from direction of HD85905.}
\tablenotetext{d}{Physical separation in the plane of the sky from HD85905, at the distance of the closest partner.}
\end{deluxetable}  

\begin{deluxetable}{lllcllcclcc}
\tablewidth{0pt}
\tabletypesize{\scriptsize}
\tablecaption{Stellar Parameters for Targets Near HR10\tablenotemark{a}\label{tab:basichr10nab}}
\tablehead{ HD & Other & Spectral & $m_{\rm V}$ & $v_{\rm R}$ & $v \sin i$ & $l$ & $b$ & Distance\tablenotemark{b} & $\Delta\theta$\tablenotemark{c} & $\Delta r_{\rm POS}$\tablenotemark{d}\\
\# & Name & Type & (mag) & (km s$^{-1}$) & (km s$^{-1}$) & (deg) & (deg) & (pc) & (deg) & (pc)}
\startdata
222345 & $\omega^1$ Aqr & A7IV     & 4.99 & --2  & 105 & 67.0 & --68.9 & 41.0$^{+1.5}_{-1.4}$ & 7.34 & 5.3 \\
222661 & $\omega^2$ Aqr & B9V      & 4.48 & 3    & 142 & 67.7 & --69.6 & 47.3$^{+2.0}_{-1.9}$ & 6.60 & 5.5 \\
225132 & 2 Cet          & B9IVn    & 4.54 & --5  & 200 & 72.1 & --75.3 & 69.9$^{+4.8}_{-4.2}$ & 0.83 & 1.0 \\
221565 & b$^3$ Aqr      & A0V      & 4.71 & 15   & 180 & 47.9 & --71.1 & 98.2$^{+7.8}_{-6.7}$ & 8.81 & 15.2 \\
222847 & i$^1$ Aqr      & B9V      & 5.24 & 14   & 310 & 59.2 & --72.2 & 101.7$^{+9.9}_{-8.3}$ & 5.55 & 9.9 \\
1064   & HR51           & B9V      & 5.76 & 16.9 & 124 & 95.1 & --70.4 & 106.4$^{+9.6}_{-8.2}$ & 8.06 & 15.1 \\
223466 & HR9026         & A3V      & 6.42 & 16.5 & 70  & 36.9 & --76.1 & 121$^{+15}_{-12}$ & 8.92 & 19 \\
1256   & HR61           & B6III/IV & 6.49 & 19   & 150 & 72.1 & --79.5 & 187$^{+36}_{-26}$ & 3.63 & 10 \\

\enddata
\tablenotetext{a}{All values from SIMBAD unless otherwise noted.}
\tablenotetext{b}{Distances calculated from {\it Hipparcos} parallaxes.}
\tablenotetext{c}{Angular distance from direction of HR10.}
\tablenotetext{d}{Physical separation in the plane of the sky from HR10, at the distance of the closest partner.}

\end{deluxetable}

\begin{deluxetable}{lllccccc}
\tablewidth{0pt}
\tabletypesize{\scriptsize}
\tablecaption{Absorption Line Properties \label{tab:aodmeas}}
\tablehead{ HD & Other & $\langle v \rangle_{\rm CaII}$\tablenotemark{a} & $\log N_{\rm CaII}$\tablenotemark{b} & $\langle v \rangle_{\rm NaI}$\tablenotemark{a} & $\log N_{\rm NaI}$\tablenotemark{b} & CaII/NaI \\
\# & Name & (km s$^{-1}$) & (cm$^{-2}$) & (km s$^{-1}$) & (cm$^{-2}$) & }
\startdata
159561 & $\alpha$ Oph & $-25.8 \pm 0.3$ & $11.53 \pm 0.03$ & $-26.6 \pm 0.6$ & $10.59 \pm 0.04$ & $8.7 \pm 1.0$ \\
\hline
165777 & 72 Oph & $-30.6 \pm 5.7$ & $10.03 \pm 0.47$ & \nodata         & $<$10.7 & $>$0.2 \\
160910 & HR6594 & $-25.0 \pm 3.2$ & $11.50 \pm 0.10$ & $-23.6 \pm 1.2$ & $10.77 \pm 0.06$ & $5.4^{+1.6}_{-1.3}$ \\
159332 & HR6541 & \nodata         & \nodata          & \nodata         & $<$10.4 & \nodata \\
154494 & 60 Her & $-20.3 \pm 2.8$ & $10.62 \pm 0.11$ & \nodata         & $<$10.7 & $>$0.8 \\
154228 & HR6341 & $-18.2 \pm 1.5$ & $10.78 \pm 0.16$ & $-18.1 \pm 2.7$ & $10.78 \pm 0.02$ & $1.0^{+0.4}_{-0.3}$ \\
\hline
80007 & $\beta$ Car & $2.4 \pm 0.9$ & $10.47 \pm 0.07$ & $3.1 \pm 7.5$ & $9.88 \pm 0.07$ & $3.9^{+1.0}_{-0.8}$ \\
\hline
71243 & $\alpha$ Cha & \nodata        & $<$11.4          & \nodata        & $<$11.1 & \nodata \\
\enddata
\tablenotetext{a}{Column density weighted mean velocity, and an ``error'' which is the weighted average variance, essentially the range of velocities in which significant absorption is present.  The error in determining the velocity of any given absorption is typically 0.1--2.0 km~s$^{-1}$, depending on the spectral resolution of the observations.}
\tablenotetext{b}{Total column density using apparent optical depth technique.}
\end{deluxetable}

\begin{deluxetable}{lcccccccccccc}
\rotate
\tablewidth{0pt}
\tabletypesize{\tiny}
\tablecaption{IR Flux Measurements \label{tab:obsflux}}
\tablehead{\colhead{} & \colhead{$B$} & \colhead{$V$} & \colhead{$J$} & \colhead{$H$} & \colhead{$K$} & \colhead{$F_{3.6}$}      & \colhead{$F_{4.5}$}  & \colhead{$F_{5.8}$} & \colhead{$F_{8.0}$}  & \colhead{$F_{24}$} & \colhead{$F_{70}$} & \colhead{$F_{160}$} \\
\colhead{Source} & \colhead{(mag)} & \colhead{(mag)} & \colhead{(mag)} & \colhead{(mag)} & \colhead{(mag)} & \colhead{(mJy)} & \colhead{(mJy)} & \colhead{(mJy)} & \colhead{(mJy)} & \colhead{(mJy)} & \colhead{(mJy)} & \colhead{(mJy)} }
\startdata
$\alpha$ Oph &2.22(3) &2.07(3) &1.75(4) &1.67(3) &1.64(3) &sat.        &sat.       &2.5E+4(2) &1.5E+4(1) &1.4E+3(1) &1.5E+2(3) &$<$5.9E+2 \\
$\beta$ Car  &1.68(1) &1.68(2) &1.55(2) &1.55(1) &1.53(1) &sat.        &sat.       &2.9E+4(3) &1.5E+4(2) &1.6E+3(2) &1.9E+2(4) &$<$6.0E+2 \\
HD85905      &6.27(1) &6.23(1) &6.05(2) &6.04(3) &6.04(2) &1.2E+03(1)  &6.6E+02(7) &4.9E+2(5) &2.6E+2(3) &2.7E+1(3) &$<$4.7E+0 &$<$2.3E+1 \\
HR10         &6.33(1) &6.23(1) &5.86(2) &5.83(3) &5.75(3) &1.6E+03(2)  &8.0E+02(8) &5.7E+2(6) &3.2E+2(3) &3.3E+1(3) &$<$4.7E+0 &$<$2.6E+1 \\
\enddata
\tablecomments{The numbers in parentheses note the error in the last significant digit.}
\tablerefs{$B$ and $V$ photometric data are from SIMBAD.  Near-IR $J$, $H$, $K$ fluxes are from the 2 Micron All Sky Survey (2MASS) for HD85905 and HR10, from \citet*{alonso94} for $\alpha$ Oph, and \citet*{bouchet91} for $\beta$ Car.}
\end{deluxetable}

\begin{deluxetable}{lcccccc}
\tablewidth{0pt}
\tabletypesize{\footnotesize}
\tablecaption{Estimated Stellar Parameters \label{tab:stparam}}
\tablehead{\colhead{} & \colhead{$T_{\star}$} & \colhead{$L_{\star}$} & \colhead{$R_{\star}$} & \colhead{} & \colhead{$M_{\star}$} & \colhead{$\log$ Age} \\
\colhead{Source} & \colhead{(K)} & \colhead{($L_{\sun}$)} & \colhead{($R_{\sun}$)} & \colhead{$\log{g}$} & \colhead{($M_{\sun}$)} & \colhead{(yr)} }
\startdata
$\alpha$ Oph  &7980	 &22.6      &2.5	&3.8     &2.7	  &9.0 \\
$\beta$ Car   &9280	 &220.5     &5.8	&3.6     &3.6	  &8.6 \\
HD85905       &8840	 &51.8      &3.1 	&3.3     &3.6	  &8.8 \\
HR10          &8370	 &63.8      &3.8	&4.1     &2.5	  &8.9 \\
\enddata
\end{deluxetable}

\begin{deluxetable}{lccccccccc}
\tablewidth{0pt}
\tabletypesize{\scriptsize}
\tablecaption{Constraints on Excess Luminosity from Photometry \label{tab:dustdisk}}
\tablehead{\colhead{} & \multicolumn{3}{c}{24$\mu$m} & \multicolumn{3}{c}{70$\mu$m} & \multicolumn{3}{c}{160$\mu$m} \\ 
\colhead{Source} & \colhead{$F_{\star}$} & \colhead{$F_{MIPS}$/$F_{\star}$} & \colhead{$L_{dust}$/$L_{\star}$} & \colhead{$F_{\star}$} & \colhead{$F_{MIPS}$/$F_{\star}$} & \colhead{$L_{dust}$/$L_{\star}$} & \colhead{$F_{\star}$} & \colhead{$F_{MIPS}$/$F_{\star}$} & \colhead{$L_{dust}$/$L_{\star}$} }
\startdata
$\alpha$ Oph   &1.5       &0.95    &2.5E-4       &0.17       &0.89      &9.4E-06     &3.4E-2     &$<$18       &$<$1.6E-05 \\      
$\beta$ Car    &1.7       &0.92    &2.1E-4       &0.20       &0.97      &8.8E-06     &3.9E-2     &$<$16       &$<$1.2E-05 \\    
HD85905        &2.8E-2    &0.97    &2.3E-4       &3.2E-3     &$<$1.5    &$<$1.4E-05  &6.3E-4     &$<$38       &$<$3.1E-05 \\    
HR10           &3.3E-2    &1.01    &2.5E-4       &3.8E-3     &$<$1.2    &$<$1.2E-05  &7.4E-4     &$<$35       &$<$3.1E-05 \\      
\enddata
\end{deluxetable}

\begin{deluxetable}{lccccccccccccc}
\tablewidth{0pt}
\tabletypesize{\footnotesize}
\tablecaption{Constraints on Excess Luminosity from Complete SED\tablenotemark{a} \label{tab:likechen}}
\tablehead{ \colhead{} &\colhead{$T_{dust}$} &\colhead{$\Omega$}  &\colhead{BB Min.} &\colhead{} &\colhead{$D$} &\colhead{$<{a}>$} &\colhead{$M_{dust}$} &\colhead{$M_{10cm}$} \\
\colhead{Source} &\colhead{(K)} &\colhead{($10^{-14}$ sr$^{-1})$}  &\colhead{$\chi^{2}$} &\colhead{$L_{IR}/L_{\star}$} &\colhead{(AU)} &\colhead{($\mu$m)} &\colhead{($M_{\earth}$)} &\colhead{($M_{\earth}$)} }
\startdata
$\alpha$ Oph  &21$\pm$2   &300$\pm$200   & 8.8   &2.9E-06      &900       &3.2    &6.7E-4      &0.15   \\   
$\beta$ Car   &22$\pm$2   &200$\pm$200   & 5.3   &1.7E-06      &2400      &24     &2.0E-2      & 1.7   \\   
HD85905       &22$\pm$5   &9$\pm$7       & 23    &4.8E-06      &1100      &5.5    &3.2E-3      &0.55   \\   
HR10          &20$\pm$4   &17$\pm$6      & 18    &6.1E-06      &1600      &9.8    &1.4E-2      &1.8   \\   
\enddata
\tablenotetext{a}{Although no significant excess flux was detected, tight constraints on the fractional luminosity and other dust properties are made for the maximum hypothetical excess that is still consistent with our IR upper limits.}
\end{deluxetable}

\begin{deluxetable}{@{\extracolsep{\fill}}lccccccc}
\tablewidth{0pt}
\tabletypesize{\tiny}
\tablecaption{Constraints on Atomic Line Fluxes from IRS \label{tab:gas}}
\tablehead{ \colhead{ } &\colhead{Ne II} &\colhead{Ne III} &\colhead{Fe I} &\colhead{Fe II} &\colhead{Fe II} &\colhead{S I} &\colhead{Si II} \\
\colhead{Source} &\colhead{12.814 $\mu$m} &\colhead{15.56 $\mu$m} &\colhead{24.042 $\mu$m} &\colhead{17.936 $\mu$m} &\colhead{25.988 $\mu$m} &\colhead{25.249 $\mu$m} &\colhead{34.820 $\mu$m}  \\
\colhead{ } &\colhead{(erg s$^{-1}$cm$^{-2}$)}  &\colhead{(erg s$^{-1}$cm$^{-2}$)}  &\colhead{(erg s$^{-1}$cm$^{-2}$)}  &\colhead{(erg s$^{-1}$cm$^{-2}$)}  &\colhead{(erg s$^{-1}$cm$^{-2}$)}  &\colhead{(erg s$^{-1}$cm$^{-2}$)}  &\colhead{(erg s$^{-1}$cm$^{-2}$)} }
\startdata
$\alpha$ Oph  &$<$2.24E-13   &$<$8.62E-14   &$<$1.61E-14   &$<$4.69E-14   &$<$1.08E-14   &$<$1.12E-14   &$<$4.73E-15  \\
$\beta$ Car   &$<$2.49E-13   &$<$9.54E-14   &$<$1.67E-14   &$<$5.19E-14   &$<$1.16E-14   &$<$1.18E-14   &$<$5.12E-15  \\
HD85905       &$<$4.38E-15   &$<$1.77E-15   &...           &$<$1.11E-15   &...           &...           &...          \\
HR10          &$<$5.67E-15   &$<$2.34E-15   &...           &$<$1.53E-15   &...           &...           &...          \\
\enddata
\end{deluxetable}

\begin{deluxetable}{@{\extracolsep{\fill}}lcccccccc}
\tablewidth{0pt}
\tabletypesize{\scriptsize}
\tablecaption{Constraints on Molecular Hydrogen Line Fluxes from IRS \label{tab:hgas}}
\tablehead{ \colhead{ } &\colhead{H$_2$ S(0)} &\colhead{H$_2$ S(1)} &\colhead{H$_2$ S(2)} &\colhead{H$_2$ S(3)} &\colhead{50 K} &\colhead{100 K}  &\colhead{50 K} &\colhead{100 K} \\
\colhead{Source} &\colhead{28.221 $\mu$m} &\colhead{17.035 $\mu$m} &\colhead{12.279 $\mu$m} &\colhead{9.665 $\mu$m} &\colhead{$M_{H_2}$ S(0)} &\colhead{$M_{H_2}$ S(0)}  &\colhead{$M_{H_2}$ S(1)} &\colhead{$M_{H_2}$ S(1)}  \\
\colhead{ } &\colhead{(erg s$^{-1}$cm$^{-2}$)}  &\colhead{(erg s$^{-1}$cm$^{-2}$)}  &\colhead{(erg s$^{-1}$cm$^{-2}$)}  &\colhead{(erg s$^{-1}$cm$^{-2}$)}  &\colhead{(M$_{\earth}$)} &\colhead{(M$_{\earth}$)}  &\colhead{(M$_{\earth}$)} &\colhead{(M$_{\earth}$)} }
\startdata
$\alpha$ Oph  &$<$8.76E-15   &$<$5.86E-14   &$<$2.60E-13   &$<$3.91E-13   &$<$200        &$<$2      &$<$3E+05       &$<$20  \\
$\beta$ Car   &$<$8.57E-15   &$<$6.51E-14   &$<$2.90E-13   &$<$4.36E-13   &$<$1000       &$<$10     &$<$2E+06       &$<$100 \\
HD85905       &$<$1.45E-15   &$<$1.26E-15   &$<$5.08E-15   &$<$7.22E-15   &...           &...       &$<$5E+05       &$<$40  \\
HR10          &$<$1.07E-15   &$<$1.75E-15   &$<$6.54E-15   &$<$8.89E-15   &...           &...       &$<$1E+06       &$<$80  \\
\enddata
\end{deluxetable}


\clearpage

\clearpage

\clearpage

\clearpage

\clearpage

\clearpage

\clearpage
 
\end{document}